\documentclass[prd,onecolumn,floatfix,
preprintnumbers,nofootinbib]{revtex4}
\usepackage{amsmath,amssymb,amsfonts}

\begin{document}

\input epsf.tex    


\input psfig.sty

\title{Explosion Mechanisms of Core-Collapse Supernovae}



\author{Hans-Thomas Janka}
\affiliation{Max Planck Institute for Astrophysics, Karl-Schwarzschild-Str.~1,
D-85748 Garching, Germany; email: thj@mpa-garching.mpg.de}

\begin{abstract}
Supernova theory, numerical and analytic, has made 
remarkable progress in the past decade. 
This progress was made possible
by more sophisticated simulation tools, especially for
neutrino transport, improved microphysics, and deeper 
insights into the role of hydrodynamic instabilities. 
Violent, large-scale nonradial mass motions are
generic in supernova cores. The neutrino-heating
mechanism, aided by nonradial flows, drives 
explosions, albeit low-energy ones, of ONeMg-core and some
Fe-core progenitors. The characteristics of the neutrino
emission from new-born neutron stars were revised, new
features of the gravitational-wave signals were discovered,
our notion of supernova nucleosynthesis was shattered, and our
understanding of pulsar kicks and explosion asymmetries was
significantly improved. But simulations also suggest that 
neutrino-powered explosions might not explain the most
energetic supernovae and hypernovae, which seem to
demand magnetorotational driving. Now that modeling is being
advanced from two to three dimensions, more realism, new
perspectives, and hopefully answers to long-standing
questions are coming into reach.
\end{abstract}

\preprint{Accepted by Annual Review of Nuclear and Particle Science}


\maketitle

Keywords: 
massive stars --- neutrinos --- hydrodynamics --- 
magnetic fields --- neutron stars --- black holes

\section{INTRODUCTION: ROOTS AND QUESTIONS}
\label{sec:intro}

When, why, and how can the catastrophic infall of the core of 
a massive star be reversed to trigger the powerful ejection of 
the stellar mantle and envelope in a supernova (SN) explosion? 
This fundamental problem of stellar astrophysics has been a 
matter of intense research since the crucial role of SNe
for the synthesis of heavy elements and for the dissemination of
the nuclear burning products of stars had been
recognized by Burbidge et al.~\cite{Burbidgeetal1957}. 
The latter authors also noticed that nuclear statistical
equilibrium in the hot, dense core of evolved stars 
(at $T \gtrsim 7\times 10^9$\,K) favors iron
dissociation to alpha particles, and they concluded that the
huge demand of energy (about 1.7\,MeV per nucleon or 
$1.7\times 10^{18}\,$erg per gram) must be supplied by gravitational
binding energy, leading to a contraction of the stellar core
and ultimately to a dynamical implosion on a timescale of
less than a second, $t_\mathrm{coll}\sim 0.21/\sqrt{\rho_8}$\,s,
when the average density $\rho_8 \equiv \rho/(10^8\mathrm{g/cm^3})$
exceeds unity. This groundbreaking insight is in line with
Baade \& Zwicky's earlier idea that SNe could represent the transition
of ordinary stars to neutron stars (NSs)~\cite{BaadeZwicky1934}.

Already in 1960 Hoyle \& Fowler~\cite{HoyleFowler1960} proposed
the two basic scenarios of stellar death: thermonuclear runaway
at degenerate conditions (which, as we know now, drives the 
destruction of white dwarf stars in Type~Ia SNe) and
the implosion of stellar cores (associated with what is called
core-collapse supernovae (CCSNe) of Types II, Ib/c, and 
hypernovae\footnote{Observationally, SNe~II exhibit strong
H-Balmer lines in their early spectra, whereas SNe~I show no 
H-lines. In SNe~Ia there are Si-lines, 
in SNe~Ib no Si- but He-lines,
and in SNe~Ic none of these, indicating explosions of stars that
had lost their hydrogen envelope or both the outer hydrogen and 
helium shells before collapse.
More sub-classes have been introduced, some of them 
motivated only by recent discoveries: 
SNe~II-P and II-L are discriminated by a plateau
phase or linear decay of their lightcurves after the peak,
IIb events have only thin H-shells left, and spectra of IIa and
IIn cases possess signatures of a dense circumstellar medium.}).
They hypothesized (following~\cite{Burbidgeetal1957})
that the gravitational compression of the core raises the 
temperature such that thermonuclear fuel could be ignited to
release the energy for triggering the ejection of the outer parts
of the star. They also mentioned simulations by Colgate \&
Johnson~\cite{ColgateJohnson1960,ColgateGrasbergerWhite1961},
in which the ``bounce'' of a forming NS launched a spherical shock 
wave that reversed the infall of the overlying stellar shells
to make them gravitationally unbound. Colgate \&
White~\cite{ColgateWhite1966} realized that
gravitational binding energy of order
$E_\mathrm{b}\sim GM_\mathrm{ns}^2/R_\mathrm{ns} > 10^{53}$\,erg,
which is released when the core of a star collapses to a NS,
is converted to neutrino emission and provides
a huge energy reservoir for powering the SN blast wave. They
argued correctly that in stellar layers pulled inward at
supersonic speed along with the imploding core, thermonuclear
combustion is unable to initiate an outward acceleration. Instead
they proposed that a fraction of the intense neutrino flux may
get absorbed in the mantle of the star to cause the explosion.

More than four decades of theoretical and numerical modeling 
work, spearheaded by early pioneers of the field like Dave Arnett,
Jim Wilson, Hans Bethe, Gerry Brown, Steve Bruenn, Wolfgang Hillebrandt, 
Jim Lattimer, and David Schramm, have helped to sharpen
our picture of the diverse physical ingredients and processes that 
play a role in the core of dying stars, among them 
magnetohydrodynamic (MHD) effects, fluid instabilities and turbulent 
flows, the finite-temperature
equation of state (EoS) of NS matter, neutrino transport and 
neutrino-matter interactions, and general relativistic gravity.
While the bounce-shock mechanism is not supported by any modern
simulation with state-of-the-art treatment of the physics, the
``delayed neutrino-heating mechanism'' as discussed by Bethe and
Wilson~\cite{BetheWilson1985} and aided by violent, nonradial 
mass motions in the collapsing stellar 
core~\cite{Herantetal1994,Burrowsetal1995,JankaMueller1995,JankaMueller1996}, 
has advanced to the
widely favored scenario for powering the majority of SNe.

The momentum behind the quest 
for solving the puzzle of the SN mechanism originates from important 
questions at the interface of astrophysics and nuclear, particle,
and gravitational physics, for example:
\begin{itemize}
\item
What is the link between the properties 
of SNe and their progenitor stars? 
\item
Which stars collapse to black 
holes (BHs) instead of NSs, which fraction of stellar collapses
do not yield explosions? 
\item
What are the birth properties of the compact remnants, i.e.\ their
masses, spins, magnetic fields, and recoil velocities?
\item
How can the high velocities of young pulsars be explained?
Is any exotic physics necessary?
\item
What characteristics does the neutrino burst from a SN have 
and what does it tell us about neutrino properties and the
extreme conditions in the newly formed NS?
\item
What is the gravitational-wave signature of
a stellar collapse event and which information can we extract
about the dynamical processes in the SN core? 
\item
What is the nucleosynthetic role of massive star explosions
in the chemogalactic history?
\item
Are SNe the long-sought sources of r-process 
elements, in particular also of the lanthanides,
the third abundance peak, and actinides? 
\item
What is the population-integrated energetic footprint
left by SN explosions in the dynamical evolution of galaxies?
\end{itemize}
In the following sections we will review the known types of 
stellar collapse events (Sect.~\ref{sec:progenitors}), 
the ingredients and current status of 
numerical modeling (Sect.~\ref{sec:numericalmodeling}),
the mechanisms by which massive stars might explode
(Sect.~\ref{sec:mechanisms}), and the signatures of the explosion
mechanism that might serve for observational diagnostics
(Sects.~\ref{sec:coresignals} and \ref{sec:remnants}).
We will provide an update of recent developments as follow-up
and supplement of previous reports that have approached the
topic from different 
perspectives~\cite{Fryer2004,Mezzacappa2005,WoosleyJanka2005,WoosleyBloom2006,Kotakeetal2006,Jankaetal2007,Ott2009,Thielemannetal2011}.


\begin{figure}
\centerline{\psfig{figure=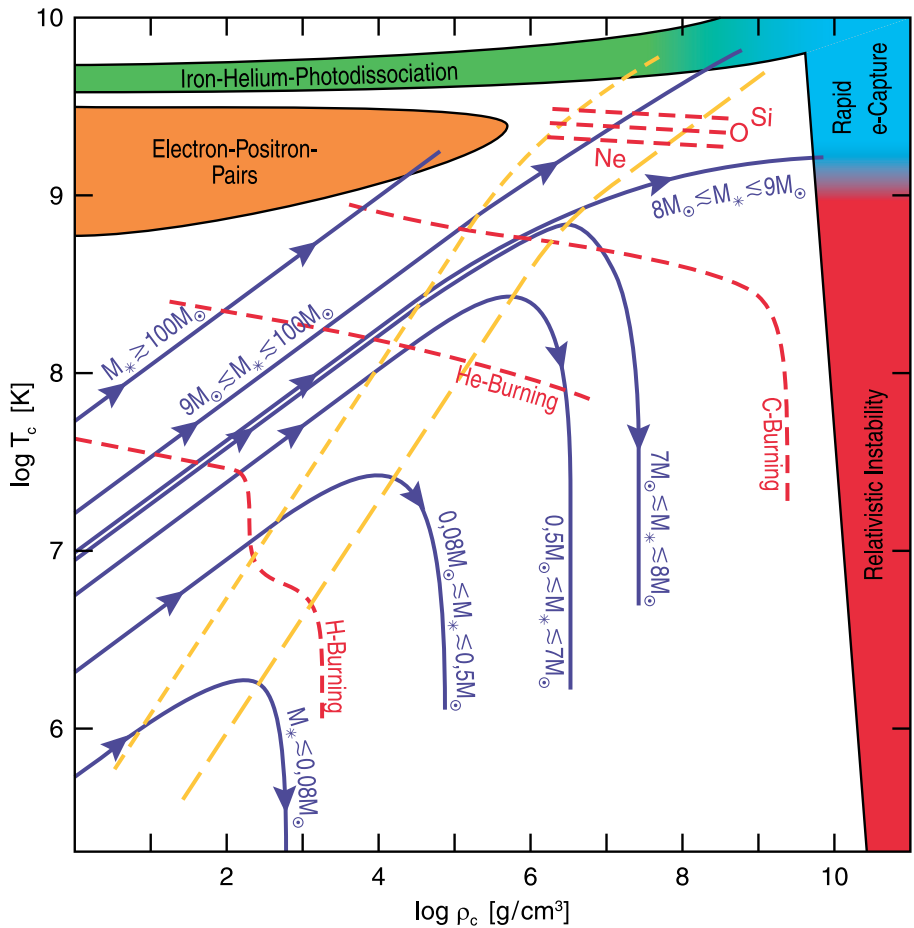,width=12truecm}}
\caption{{\small 
Stellar death regions with schematic stellar evolution
tracks in the plane of central density ($\rho_\mathrm{c}$) and
central temperature ($T_\mathrm{c}$). Colored death regions
are labeled by the instability process causing the collapse of
the stellar core, and the blue tracks are labeled by the
corresponding rough birth-mass range of objects
reaching the different stages of central burning
(indicated by red dashed lines). Yellow diagonal lines mark
the beginning of degeneracy (short-dashed) and strong degeneracy
(long-dashed) of the electron plasma. Note that realistic
stellar tracks exhibit wiggles and loops when the ignition
of the next burning stage is reached and the stellar core 
adjusts to the new energy source (see Ref.~\cite{Wheeler1990}.)}}
\label{jankafig1}
\end{figure}

\begin{figure}
\centerline{\psfig{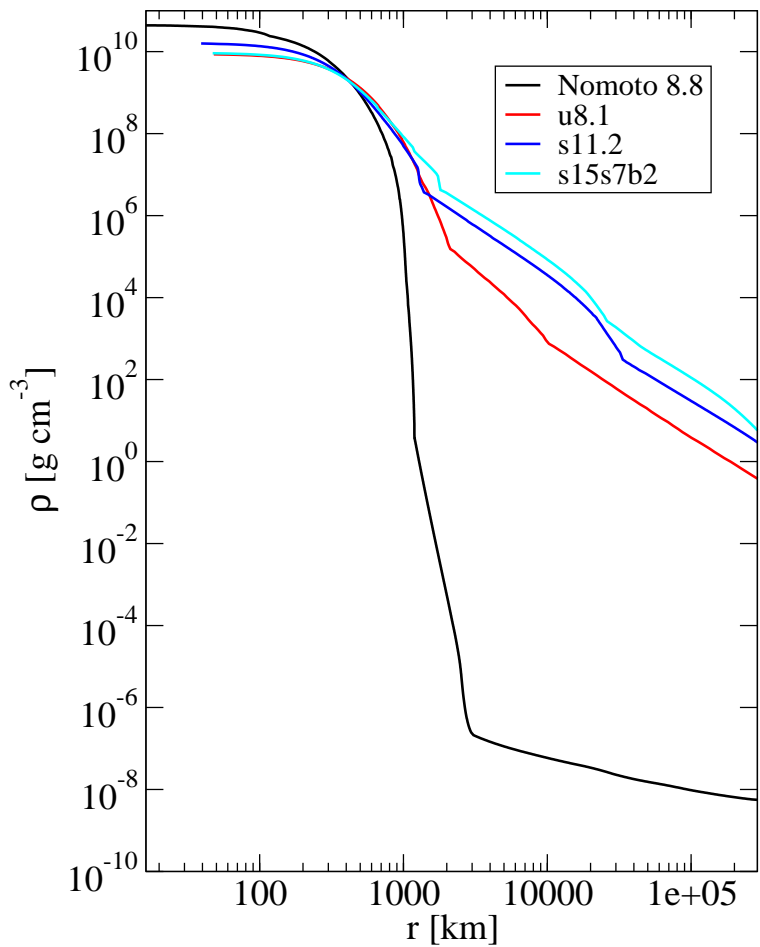}}
\caption{{\small 
Core-density profiles of different SN progenitors at the 
onset of gravitational collapse. The black line corresponds
to the ONeMg core of an 8.8\,$M_\odot$ star~\cite{Nomoto1984},
the other three are SN progenitors with iron cores: an
8.1\,$M_\odot$ ultra metal-poor ($10^{-4}$ solar metallicity)
star (A.~Heger, 
private communication) and 11.2\,$M_\odot$~\cite{Woosleyetal2002}
and 15\,$M_\odot$~\cite{WoosleyWeaver1995} solar-metallicity stars.
The steps and kinks in the curves correspond to composition-shell
interfaces (Fe/Si and O/C for the 11.2 and 15\,$M_\odot$ models
and inner and outer boundaries of a C-O-Ne-layer for the 
8.1\,$M_\odot$ case).}}
\label{jankafig2}
\end{figure}

\section{ROUTES TO STELLAR CORE COLLAPSE}
\label{sec:progenitors}

Massive stars possess finite lifetimes of millions to tens of
millions of years, which are mainly determined
by the period the star spends on the main sequence (MS) during central
hydrostatic hydrogen burning. The evolution time of stars scales
approximately like $t_\mathrm{evol}\approx 7.3\times 10^9\,\mathrm{yr}\,
(M_\ast/M_\odot)/(L_\ast/L_\odot)$ with the stellar mass $M_\ast$
and luminosity $L_\ast/L_\odot \approx (M_\ast/M_\odot)^{3.5}$
(where $M_\odot = 1.989\times 10^{33}$\,g and 
$L_\odot = 3.85\times 10^{33}$\,erg/s are solar mass and luminosity).
When hydrogen in the stellar core gets exhausted and the star
leaves the MS, its evolution speeds up considerably because the
efficiency of energy production in the higher stages of
nuclear burning decreases and concurrently energy losses through
neutrino-antineutrino pairs rise dramatically. This is the case
in particular when the central temperature of the star climbs to
$T_\mathrm{c} \sim 10^9$\,K, at which time $e^+e^-$ pairs become
abundant and the energy drain in $\nu\bar\nu$ pairs accelerates
with $T_\mathrm{c}^9$. At this time neutrino losses exceed the
radiation losses of the star and the evolution of the helium core
decouples from that of the stellar envelope.

The energy drain happens at the expense of gravitational 
binding, leading to continuous contraction of the stellar core,
which is slowed down only temporarily by the periods of nuclear
burning. As long as nondegenerate particles dominate the pressure
of the stellar plasma, hydrostatic equilibrium requires that the
central temperature, $T_\mathrm{c}$, and central density,
$\rho_\mathrm{c}$, roughly follow the proportionality
\begin{equation}
\frac{T_\mathrm{c}^3}{\rho_\mathrm{c}} \propto M_\mathrm{c}^2
\sim \mathrm{const}\ .
\label{eq:rhoctc}
\end{equation}
According to this relation more massive stars with bigger He-cores
(larger $M_\mathrm{c}$) are hotter (Fig.~\ref{jankafig1}). 
For sufficiently high central
temperature, nuclear fuel can ignite in the next burning 
stage, building up heavier and more stable elements in their inner 
core. If, however, the stellar interior enters the regime of 
electron degeneracy before\footnote{Fermions approach the degeneracy
when their Fermi energy begins to exceed the thermal energy
$k_\mathrm{B}T$, i.e.\ at $T_8\sim 4\rho_5^{2/3}$ for nonrelativistic 
electrons and at $T_{10}\sim \rho_8^{1/3}$ for relativistic ones with
$T_x \equiv T/(10^x\,\mathrm{K})$ and $\rho_y \equiv
\rho/(10^y\,\mathrm{g\,cm}^{-3})$.} (yellow, short-dashed line 
in Fig.~\ref{jankafig1}) it ends as a white dwarf,
being stabilized by lepton degeneracy pressure and cooling at
essentially fixed density. 

Stars beyond certain birth-mass limits can reach the ``death zones''
in the upper and right parts of Fig.~\ref{jankafig1}, where the 
stellar core becomes gravitationally unstable. Contraction, and 
in the case of a runaway process finally collapse, sets in when 
the effective adiabatic index drops below the critical value 
of $4/3$ for mechanical stability (the actual value is slightly 
decreased by rotation and increased by general relativistic gravity).

Three different processes can initiate the implosion of stellar
cores in three areas of the $\rho_\mathrm{c}$-$T_\mathrm{c}$-plane
indicated by different colors in Fig.~\ref{jankafig1}, playing a
role in different kinds of CC events.

\begin{figure}
\centerline{\psfig{figure=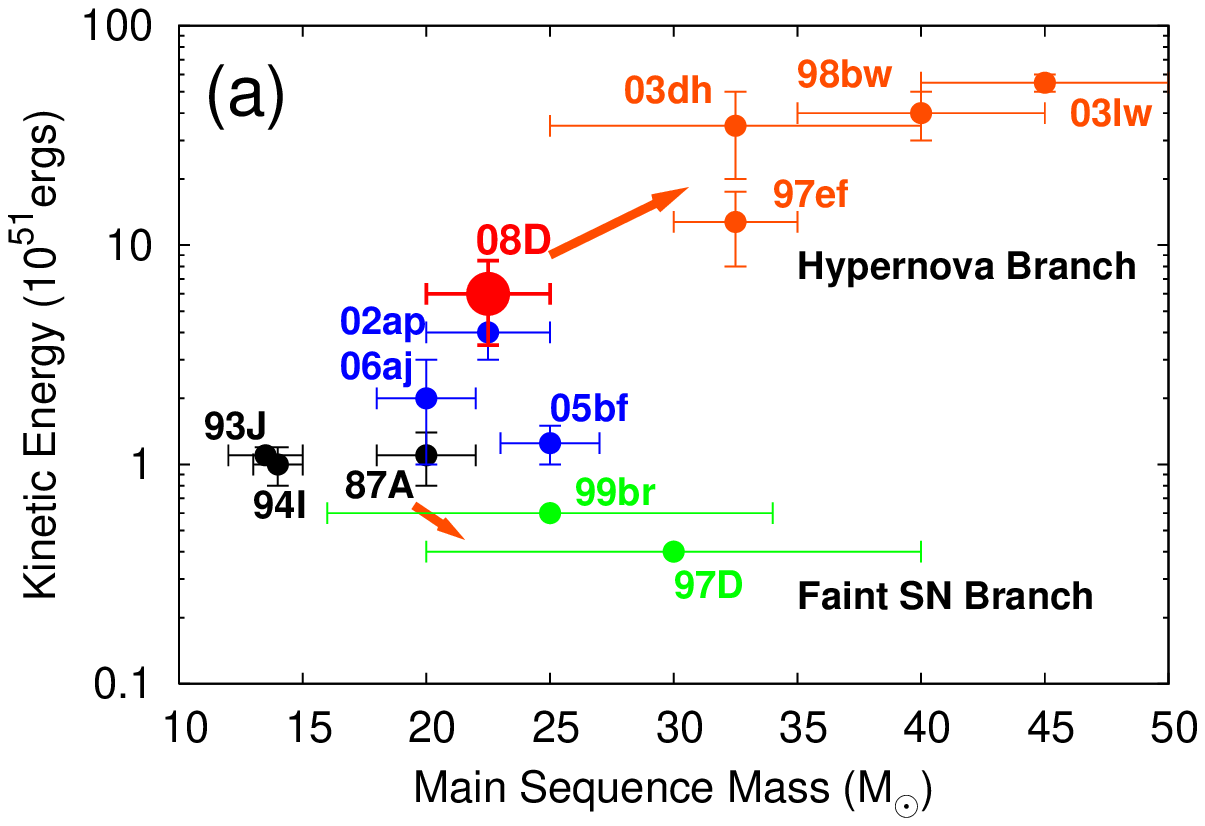,width=7truecm}
            \psfig{figure=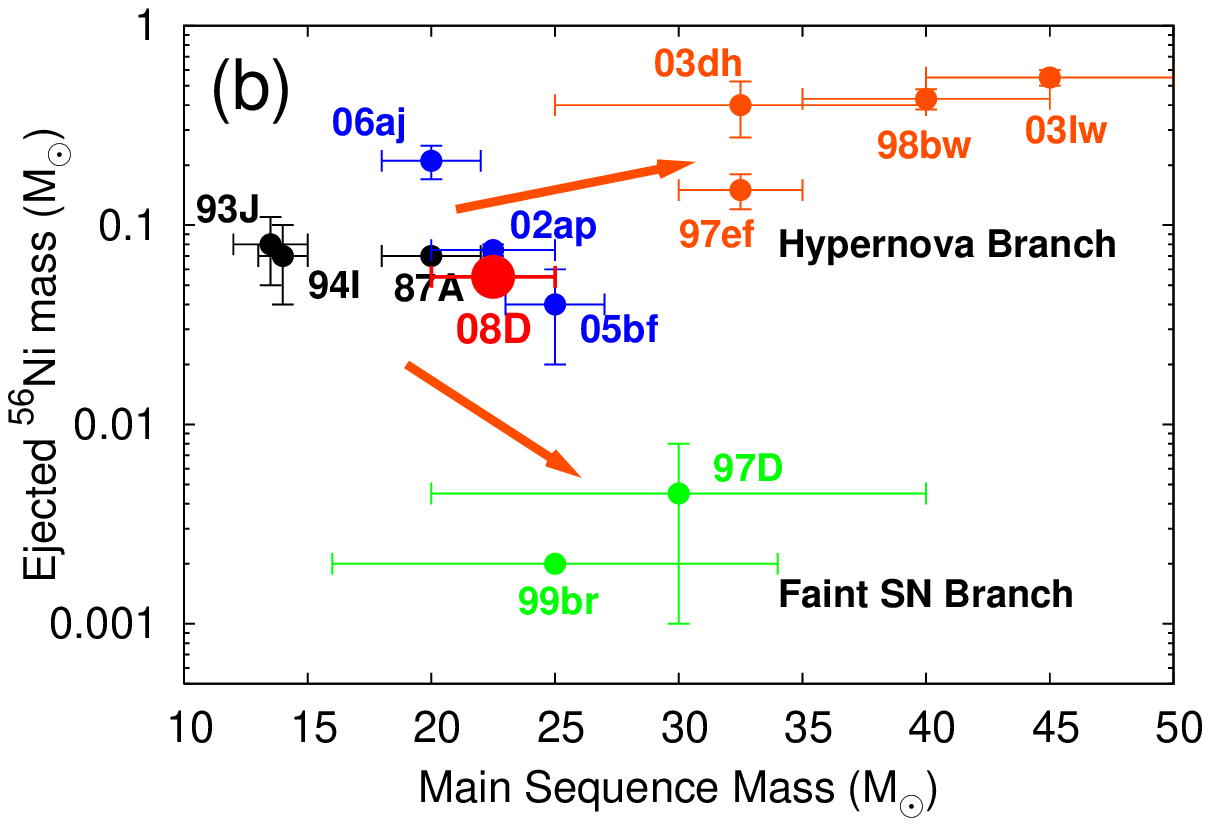,width=7truecm}}
\centerline{\psfig{figure=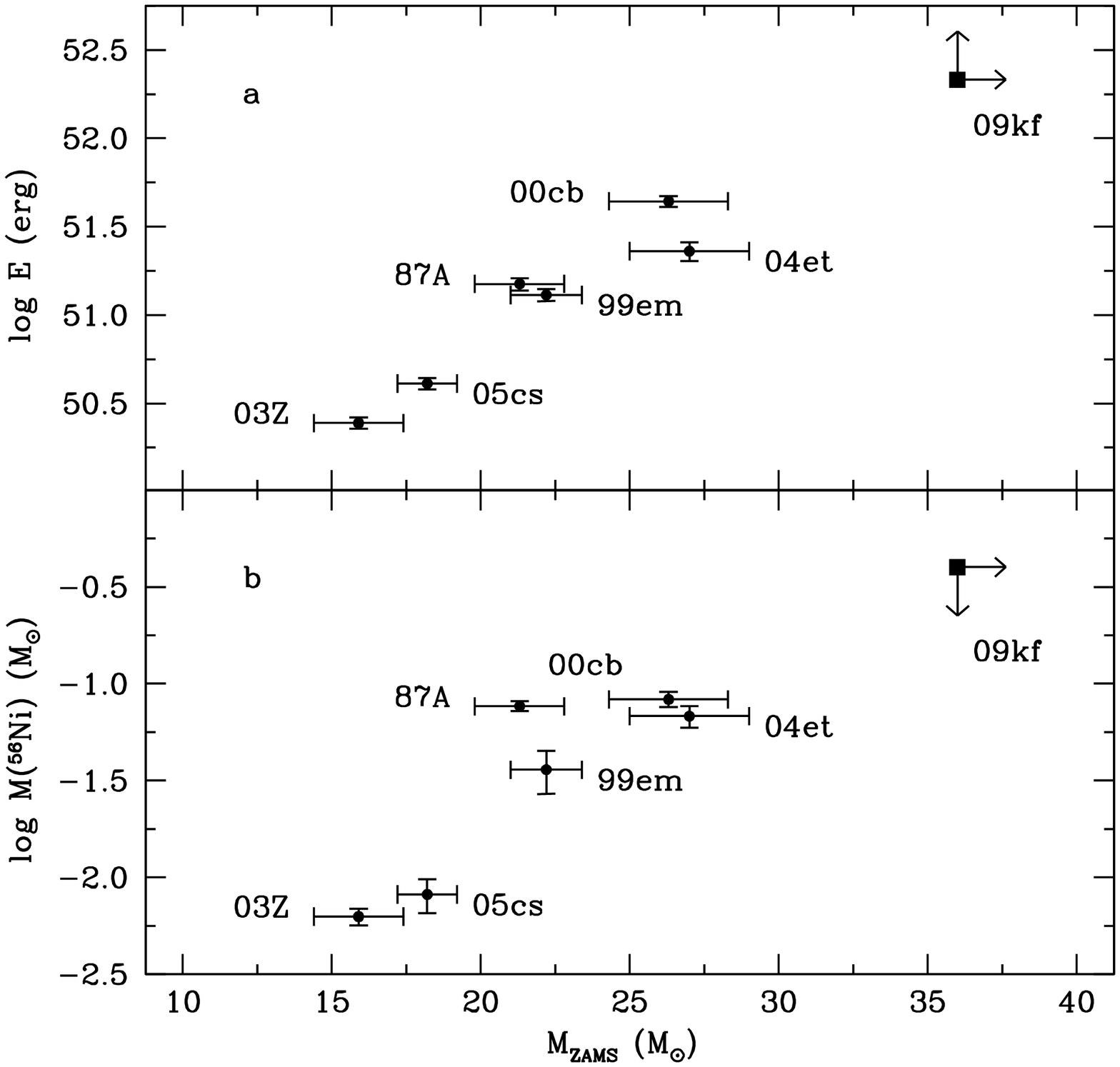,width=7truecm}
            \psfig{figure=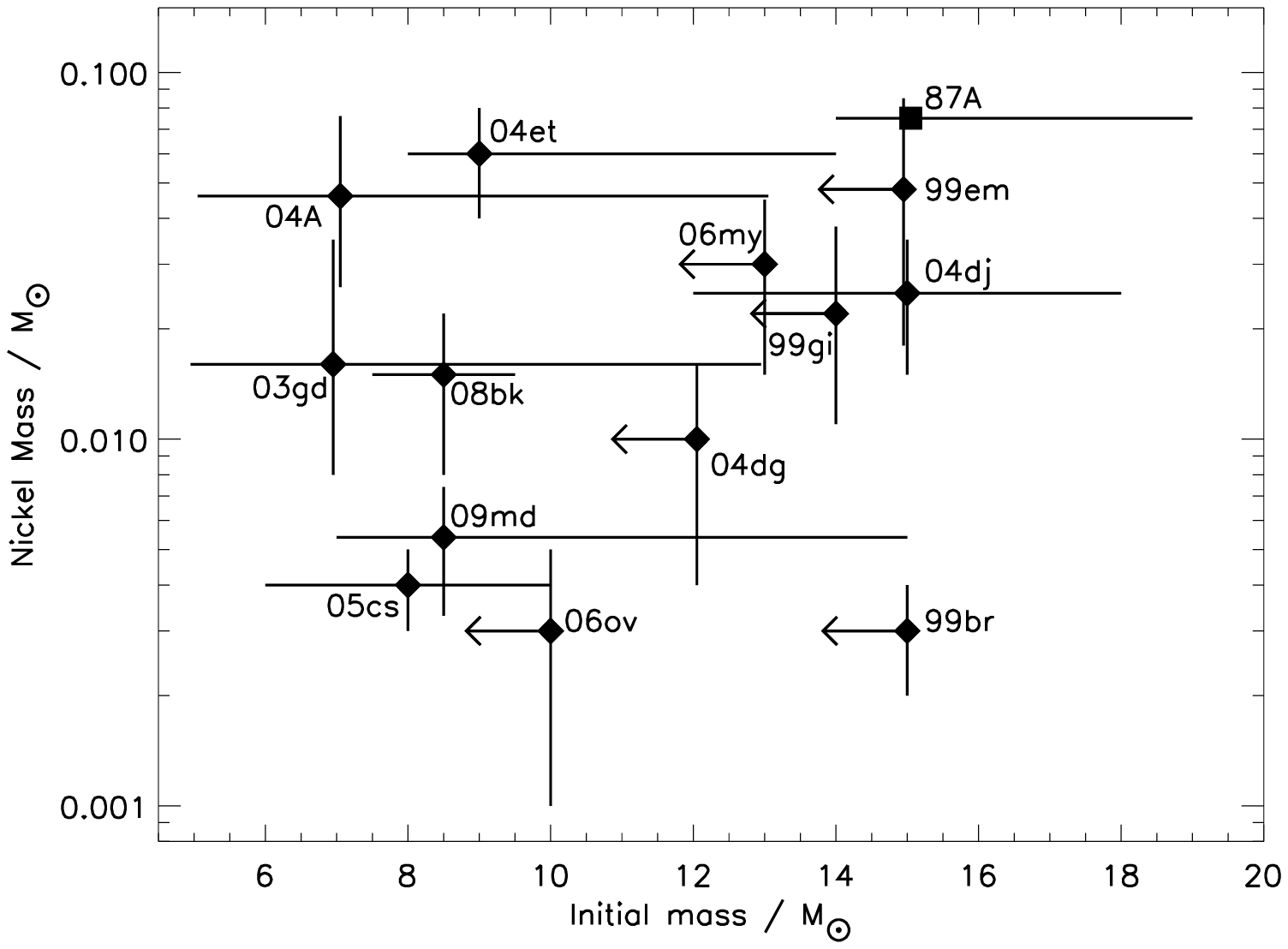,width=7truecm}}
\caption{{\small 
Kinetic energies and ejected nickel masses for stellar
explosions versus initial (zero-age main sequence; ZAMS) masses 
from different authors. While explosion properties are
deduced from comparing observations with lightcurve and spectra 
calculations based on (spherically symmetric) models, the
ZAMS masses are estimated by linking ejecta masses to initial
masses through stellar evolution models with mass-loss assumptions
({\em upper panels}~\cite{Tanakaetal2009} and {\em lower left
panel}~\cite{UtrobinChugai2011}; reproduced by permission of the
authors and AAS and \copyright ESO, respectively) or by inferring ZAMS
masses or upper limits from computed stellar evolution histories that 
account for the properties of discovered SN progenitors or their
stellar environments (i.e., coeval star clusters, 
host galaxies~\cite{Smarttetal2009,Smartt2009};
{\em lower right panel}, kindly provided by
John Eldridge and Stephen Smartt). Due to theoretical
uncertainties this leads to largely different mass determinations for
some cases (SN~1987A, SN~1999br, SN~1999em, SN~2004et, SN~2005cs). 
For masses $\gtrsim$25\,$M_\odot$
Tanaka et al.~\cite{Tanakaetal2009} discriminated a very energetic
and bright HN branch from a low-energy, faint SN branch. The objects
with little nickel production of the latter branch, however, have 
also been interpreted as weak explosions (possibly ECSNe) 
near the lower mass limit for SN progenitors.}}
\label{jankafig3}
\end{figure}

\subsection{Electron-Capture Supernovae}
\label{sec:ECSNe}

The lowest-mass progenitors of CCSNe develop oxygen-neon-magnesium
(ONeMg) cores through C-burning~\cite{Nomoto1984,Nomoto1987,Poelarendsetal2008}
but reach $e$-degeneracy before hydrostatic Ne-burning can be ignited.
Due to low reaction thresholds of Ne and Mg, the increasing electron 
Fermi energy enables $e$-captures (right upper corner of 
Fig.~\ref{jankafig1}), triggering gravitational 
collapse and resulting in an electron-capture SN (ECSN).
Solar-metallicity stars\footnote{The metallicity $Z$ is the total
mass fraction of chemical elements heavier than helium in the matter
the star was formed of. The solar value has been determined to be
0.016.} with 9--9.25\,$M_\odot$ are estimated to
have that destiny~\cite{Poelarendsetal2008}, but the mass window is
expected to shift and widen for lower metallicities~\cite{Pumoetal2009} 
and in binary systems with mass loss or 
transfer~\cite{PodsiadlowskiLangerPoelarendsetal2004} 
so that ECSNe could contribute even 20--30\% of all
SNe~\cite{Wanajoetal2009,WanajoJankaMueller2011}.

Because of the extremely steep density decline in a thin C-O-shell
($\sim$0.1\,$M_\odot$ between about $3\times 10^4$\,g/cm$^3$ and
$4\times 10^8$\,g/cm$^3$) at the edge of the O-Ne core
(Fig.~\ref{jankafig2}), these stars have special explosion properties 
(Sect.~\ref{sec:neutrinomechanism}). They eject little carbon and 
oxygen and very little nickel, their 
SNe will therefore be relatively faint.
The Crab remnant of SN~1054 is thought to be the relic of such an 
explosion~\cite{Nomotoetal1982,Hillebrandt1982}, and an increasing
number of 
dim events like SN~1997D, 1999br, 2005cs, 2008S and other recently
observed transient sources are discovered as possible candidates.

\subsection{Iron-Core Supernovae}
\label{sec:FeCSNe}

Massive stars that ignite hydrostatic Ne-burning form an iron
core. The latter becomes gravitational unstable when the nuclear
statistical equilibrium (NSE) at temperatures around $10^{10}$\,K
($k_\mathrm{B}T\sim 1$\,MeV) favors the dissociation of iron-group
nuclei to $\alpha$-particles and a growing number of free nucleons
(upper region in Fig.~\ref{jankafig1}).
With the onset of contraction and increasing density and electron
chemical potential, also $e$-captures on nuclei (and some free
protons) speed up and accelerate the implosion. The dynamical
collapse is abruptly stopped only when nuclear densities 
($\rho\gtrsim 2.7\times 10^{14}$\,g/cm$^3$) are reached,
and the phase transition to homogeneous nuclear matter leads to
a sudden increase of the effective adiabatic index
due to repulsive short-range forces between nucleons.

When the overshooting inner core rebounds and crashes
supersonically into the subsequently infalling layers, sound waves
steepen into a shock front that ultimately leads to the disruption
of the star in the SN explosion. However, different from ONeMg 
cores, the much flatter density profile in and around Fe cores 
(Fig.~\ref{jankafig2}) leads to long-lasting, 
high mass accretion rates and large ram pressure of the infalling
shells. This impedes the outward propagation of the shock and
makes Fe-core progenitors harder to blow up than stars with
ONeMg core.
Although more massive stars exhibit a gross tendency to bigger
He-cores and shallower density decline, the variation with stellar
birth mass is not necessarily monotonic~\cite{Woosleyetal2002}.
The mechanism(s) by which ECSNe and Fe-core SNe succeed to explode
will be discussed in Sect.~\ref{sec:mechanisms}.
 
Stellar cores of pre-SN stars are expected to rotate 
relatively slowly, i.e.\ with average pre-collapse spin periods 
of tens of seconds or more. This is a consequence of angular 
momentum loss associated with mass loss phases (in particular when 
the star becomes a red giant), because magnetic torques from fields
generated by differential rotation in the star couple core and
envelope and thus transport angular
momentum efficiently out of the core~\cite{HegerWoosleySpruit2005}.
Stellar rotation is therefore not expected to play a crucial
role for the explosion mechanism of normal CCSNe 
(Sect.~\ref{sec:magneticmechanism}).

\subsection{Gamma-Ray Burst Supernovae}
\label{sec:GRBSNe}

Rapid stellar rotation, however, is thought to be crucial
in the case of gamma-ray burst (GRB) SNe
and hypernovae (HNe) (for a review, see~\cite{WoosleyBloom2006}). 
The latter originally obtained their name because
of exceptional brightness and thus high nickel 
production~\cite{Paczynski1998} but are now considered as stellar 
explosions with unusually high ejecta velocities (i.e., 
very broad spectral lines) and thus large kinetic 
energies~\cite{Iwamotoetal1998} (Fig.~\ref{jankafig3}). 
HNe are found to be associated with long-duration
($t_\mathrm{GRB} \gtrsim 2$\,s) GRBs, either observed
spectroscopically (e.g., SN~1998bw with GRB~980425,
SN~2003dh with GRB~030329, SN~2003lw with GRB~031203, SN~2006aj
with GRB~060218, SN~2010bh with GRB~100316D)
or as late lightcurve humps superimposed on the power-law
decline of the afterglow that follows the GRB.
 
GRBs with their extremely luminous high-energy radiation 
are understood as ultrarelativistic, collimated outflows
(``jets''). Strong global asymmetry is also suggested by line 
profiles (in particular double-peaked oxygen emission lines) 
seen in many HNe. Such 
events are interpreted as signatures of BH-forming stellar collapses
(``collapsars''~\cite{MacFadyenWoosley1999}), in which matter 
around a rapidly spinning BH is able to set free energy in neutrinos,
electromagnetic poynting flux, and mass outflow with an efficiency of
up to roughly 40\% of the rest-mass energy of accreted material,
$\dot E_\mathrm{acc} \lesssim 0.4\dot Mc^2
\sim 10^{54} (\dot M/M_\odot$/s)\,erg/s. Alternatively,
a nearly critically rotating NS, $t_\mathrm{rot} \sim 1\,$ms,
with ultrastrong dynamo-generated magnetic field, 
$\left\langle B\right\rangle \gtrsim 10^{15}$\,G,
(``millisecond magnetar'') is discussed as possible central 
engine of GRBs and HNe. The jet and stellar explosion could either 
be powered by rotational energy of the magnetar or by gravitational
and rotational energy of the accretion flow and BH. Both can be tapped
by magnetic fields through MHD effects 
(Sect.~\ref{sec:magneticmechanism}) and by neutrinos radiated 
from matter heated by magnetically generated viscous
dissipation~\cite{Pophametal1999}. The existence of expected
strong disk ``winds'' with the observed large Ni 
production~\cite{WoosleyBloom2006}, however, seems to be challenged
by MHD simulations~\cite{Fujimoto2008}.

The progenitors of collapsars and GRB-HNe are thought to possess a 
massive core, which forms a BH instead of exploding before. They
must be compact stars without an extended hydrogen envelope
in order to allow jets to emerge ultrarelativistically,
i.e., the crossing time of the jet must be shorter than the on-time
of the central engine: $R_\ast/c \lesssim t_\mathrm{engine}$. 
Moreover, the collapsing stellar core must contain a high specific
angular momentum, $j \gtrsim GM_\mathrm{BH}/c 
\gtrsim 10^{16}\,M_\mathrm{BH}/(3\,M_\odot)$\,cm$^2$/s, to 
either form a magnetar with the necessary huge reservoir of rotational
energy or to allow for a thick, massive accretion disk that remains 
long enough around the newly formed BH to efficiently release energy.

Such requirements favor rapidly rotating Wolf-Rayet stars as 
progenitors, but special initial conditions (a high birth spin) and
evolution paths avoiding combined mass and angular momentum loss,
or alternatively binary scenarios, are 
necessary~\cite{YoonLanger2005,WoosleyHeger2006}.
In the present-day universe HNe and GRB-SNe 
are rare (with a GRB/SN ratio of $\sim$1/1000, less than 1\% of all
SN~Ib/c produce GRBs), but BH-forming CC events and GRBs could be
very common in the early (metallicity less than $\sim$1/10 solar) 
universe. This theoretical expectation
is compatible with the fact that GRB-SNe are 
preferentially (but not exclusively) observed in low-metallicity
environments.

\subsection{Pair-Instability Supernovae}
\label{sec:PISNe}

Stars above about 100\,$M_\odot$ are very hot and encounter the pair
instability (in the upper left corner of Fig.~\ref{jankafig1}) after
central carbon burning (e.g., \cite{Hegeretal2003,Woosleyetal2002}
and refs.\ therein) at $T\sim 10^9$\,K. The gravitational instability
occurs because the formation of $e^+e^-$ pairs from high-energy 
photons converts thermal energy to rest-mass energy and thus reduces 
the adiabatic index of the EoS below $4/3$. 

In the stellar mass range between $\sim$100\,$M_\odot$ and 
$\sim$140\,$M_\odot$ and for $M_\ast\gtrsim 260\,M_\odot$ 
collapse to a BH is expected.
For intermediate masses the ignition of the 
still available thermonuclear fuel during the implosion is violent
enough to trigger the complete disruption of the star with an explosion 
energy up to more than 10$^{53}$\,erg and the production of up to
$\gtrsim$50\,$M_\odot$ of $^{56}$Ni 
\cite{Hegeretal2003,Woosleyetal2002}. While such 
``thermonuclear core-collapse SNe'' were originally
termed ``hypernovae'' by Woosley \& Weaver~\cite{WoosleyWeaver1982},
they are now commonly called pair-instability SNe (PISNe) or 
pair-capture SNe (PCSNe).
In the case of BH formation, in particular in the presence of 
rotation allowing for an accretion torus, huge amounts of energy
are released in neutrinos, roughly 
(0.01--0.03)$M_\ast c^2\sim 10^{55}$\,erg,
depending on the angular momentum~\cite{FryerWoosleyHeger2001}.

While some recently discovered ultra-bright SNe and transients 
(for example SN~2002ic, 2005gj, 2005ap, 
2006gy, 2007bi, 2008es, 2010gx) have been discussed as PISN 
candidates (e.g., \cite{SmithLiFoleyetal2007,GalYametal2009}), other 
explanations for the extreme luminosity than excessive Ni yields
have been proposed, 
e.g., interaction of explosion ejecta with a dense circumstellar
medium~\cite{WoosleyBlinnikovHeger2007} or additional
energy release by magnetar spin-down~\cite{KasenBildsten2010}.
The expected rate of PISNe is small, maybe one of 100--1000 normal
stellar core collapses, and presumably mostly associated 
with metal-poor host galaxies.
In the Milky Way possibly two dozen very massive hypergiants
like the evolved luminous blue variable star $\eta$ Carinae might end
their lives in such events. 

Although the nature of the stellar death events associated with 
ultra-bright transients and in particular the energy source of their
extraordinary luminosity are not at all clear and will remain a topic
of intense research and debate in the coming years, space limitations
demand to constrain the rest of this article mostly on the physics and
processes that are relevant for the far majority of ordinary CCSNe.

\begin{table}%
\def~{\hphantom{0}}
\caption{Neutrino reactions with stellar-medium particles and between 
      neutrinos in the Garching models. $N$ means either $n$ or $p$,
      $\nu \in \{\nu_e, \bar\nu_e, \nu_\mu, \bar\nu_\mu, \nu_\tau, \bar\nu_\tau\}$,
      and $\nu_x\in \{\nu_\mu, \bar\nu_\mu, \nu_\tau, \bar\nu_\tau\}$.
      In addition to ``inelastic'' nucleon recoil, thermal motions, 
      phase-space blocking,
      high-density $N$-$N$-correlations \cite{burrows_98} 
      and weak magnetism corrections \cite{horowitz_02}, also quenching 
      of the axial-vector coupling \cite{carter_02} and the reduction of
      the effective nucleon mass at high densities \cite{reddy_99} are 
      taken into account in the rates marked with a dagger ($^\dagger$).
      A prime indicates that the neutrino can exchange energy
      with the scattering target (non-conservative or ``inelastic'' scattering)}
\label{jankatab1}
\begin{tabular}{@{}llcc@{}}%
\toprule
  Process     & References       \\
\colrule
 Beta-Processes  &     \\
      $\nu_e +  n \ \rightleftharpoons\ e^- +  p$         & \cite{burrows_98}$^\dagger$\\
      $\bar{\nu}_e + p \ \rightleftharpoons\ e^+ + n$      & \cite{burrows_98}$^\dagger$\\
      $\nu_e + (A,Z) \ \rightleftharpoons\ e^- + (A,Z+1)$  & \cite{langanke_03}\\
 Scattering Reactions &  \\
      $\nu + (A,Z) \ \rightleftharpoons\ \nu' + (A,Z)$     & \cite{horowitz_97} (ion-ion correlations)\\
                                                           & \cite{langanke_08} (inelastic contribution)\\
      $\nu + N \ \rightleftharpoons\ \nu' + N$             & \cite{burrows_98}$^\dagger$\\
      $\nu + e^\pm \ \rightleftharpoons\ \nu' + e^\pm$     & \cite{mezzacappa_93} \\
 (``Thermal'') Pair Production &  \\ 
      $\nu + \bar{\nu} \ \rightleftharpoons\ e^- + e^+ $   & \cite{bruenn_85,pons_98} \\
 Nucleon-Nucleon Bremsstrahlung &  \\
      $\nu + \bar{\nu} + N+N \ \rightleftharpoons\ N+N $   & \cite{hannestad_98} \\
 Reactions between Neutrinos &  \\
      $\nu_{\mu,\tau}+ \bar{\nu}_{\mu,\tau} \ \rightleftharpoons\ \nu_{e} + \bar{\nu}_{e}$
                                                           & \cite{buras_03} \\
      $\nu_x + \{\nu_e,\bar\nu_e\} \ \rightleftharpoons\ \nu'_x + \{\nu'_e,\bar\nu'_e \}$
                                                           & \cite{buras_03} \\
\botrule
\end{tabular}
\end{table}

\section{NUMERICAL MODELING AND PHYSICS INGREDIENTS}
\label{sec:numericalmodeling}

While over two decades that followed the poineering work by 
Colgate \& White~\cite{ColgateWhite1966}, Arnett~\cite{Arnett1966},
and Wilson~\cite{Wilson1971}, SN modeling 
was constrained to spherically symmetric (1D) simulations with few
exceptions only~\cite{LeBlancWilson1970,Smarretal1981,Symbalisty1984,Moenchmeyeretal1991}, the situation
has radically changed in the post-SN~1987A era. Detailed observations
of this nearest SN in the era of modern astronomy
revealed that large-scale mixing processes had transported
radioactive nuclei with velocities up to $\sim$4000\,km/s from the 
deep core far into the hydrogen envelope of the exploding star,
suggesting that spherical symmetry was broken already during the very
first moments of the blast~\cite{Arnettetal1989,HillebrandtHoeflich1989}.
Moreover, two-dimensional (2D) simulations in the early 1990s demonstrated
that violent convective overturn takes place in the neutrino-heating layer
between the gain radius and stalled accretion 
shock~\cite{Herantetal1994,Burrowsetal1995,JankaMueller1995,JankaMueller1996}. 
This raised hopes that buoyant energy transport to the shock could
crucially support the delayed neutrino-heating mechanism and finally 
ensure robust explosions after 1D models had turned out to be 
successful only with special assumptions that could not withstand
closer and more detailed analysis. For example, neutron-finger 
instability inside the nascent NS was proposed to enhance the 
neutrino luminosities and thus neutrino heating~\cite{WilsonMayle1988}
but is disfavored because lepton equilibration between
fingers and surroundings was shown to proceed faster than thermal
equilibration~\cite{BruennDineva1996,Mezzacappa2005}. 

In the following sections more recent developments and the present 
status of numerical approaches will be briefly summarized. While
three-dimensional (3D), general relativistic (magneto-)hydrodynamic
simulations including microphysical EoS and sophisticated,
energy-dependent neutrino transport are the ultimate, brave objective, 
only first steps have so far been achieved, approaching the 
goal from different directions. Mastering this grand computational
challenge will require highly parallelized codes with
excellent scaling capability on tens of thousands of processor cores
to achieve sustained performance on the hundreds of teraflop/s to
petaflop/s level. Still, one 3D model calculation will take a
wall-clock time of several weeks to months.

\subsection{Hydrodynamics and Gravity}

To date fully self-consistent modeling of stellar 
collapse and explosion in 3D has been achieved only
by Fryer and 
collaborators~\cite{FryerWarren2002,FryerWarren2004,FryerYoung2007},
yet only by sacrificing many aspects which are 
important for quantitatively reliable and 
conclusive results concerning the SN mechanism.
In particular, Newtonian gravity and a grey,
flux-limited neutrino diffusion (FLD) scheme~\cite{Herantetal1994}
were applied 
in combination with a smoothed particle hydrodynamics (SPH) method,
which permits economical calculations in 3D with relatively low 
resolution but is noisy and diffusive. Good resolution and an
accurate representation of the hydrodynamical quantities,
however, are essential to treat the growth of fluid
instabilities from initial seeds in the SN
core~\cite{Schecketal2008,SatoFoglizzoFromang2009},
and general relativity as well as a multi-group description of neutrino
transport including velocity-dependent observer corrections were found
to cause important differences in 1D~\cite{Bruennetal2001,Lentzetal2011}
and 2D simulations~\cite{Burasetal2006A,MuellerJankaMarek2012}.

Other groups, using mesh-based discretization schemes for solving
the hydrodynamics, have so far studied only more constrained problems
in 3D than~\cite{FryerWarren2002,FryerWarren2004,FryerYoung2007}, 
mostly also making even more 
radical approximations of the relevant (micro)physics. For example,
in refs.~\cite{BlondinMezzacappa2007,Fernandez2010} 
the development of a nonradial hydrodynamic instability 
of the accretion shock in a collapsing stellar core, the 
so-called standing accretion shock
instability (SASI;~\cite{Blondinetal2003}), was investigated 
for a steady-state flow through outer and inner grid boundaries 
with an ideal-gas EoS and parametrized neutrino-cooling terms.
In refs.~\cite{Iwakamietal2008,Iwakamietal2009} a similar 
accretion setup was studied with a microphysical EoS and
additional simple neutrino-heating terms for prescribed
luminosities and spectra (neutrino ``lightbulb'' approximation, 
NLA), which enabled neutrino-driven convection.
Using the NLA, refs.~\cite{Nordhausetal2010,Hankeetal2011} 
investigated the onset of an explosion in ``realistic'' 
collapsing stellar cores systematically by varying the 
driving neutrino luminosity to explore the dependence
on the dimension (1D, 2D or 3D) of the simulation.
Gravitational-wave (GW) signals from infall, core bounce, and 
early postbounce ($\sim$100\,ms) phases were computed with
3D general relativity (GR) 
for NS and BH formation~\cite{Ottetal2007,Ottetal2011} and
with 3D Newtonian hydrodynamics and an effective
general relativistic potential (developed as an approximation
of GR gravity in~\cite{RamppJanka2002,Mareketal2006})
for NS formation~\cite{Scheideggeretal2010}, making various
crude simplifications of the neutrino effects and partly even
of the EoS of the stellar plasma.

A grey description of the neutrino transport outside of an
excised high-density core of the proto-neutron star (PNS) 
according to ref.~\cite{Schecketal2006} was applied for 
exploring NS kicks, neutrino emission asymmetries, and GW 
signal characteristics by long-time 3D simulations of SN 
explosions in
refs.~\cite{Wongwathanaratetal2010,MuellerJankaWongwathanarat2011}.
Very first results of Newtonian 3D calculations with more
detailed multi-group (MG) transport treatments have already been
put out, using ``ray-by-ray'' (RbR) MGFLD~\cite{Bruennetal2009}
or an implementation of the ``isotropic diffusion source    
approximation'' (IDSA, see Sect.~\ref{sec:neutrinotransport})
for $\nu_e$ and $\bar\nu_e$ with~\cite{TakiwakiKotakeSuwa2011}
or without~\cite{Liebendoerferetal2010} a RbR approach,
coupled to a trapping treatment for heavy-lepton neutrinos.

\subsection{Neutrino Transport}
\label{sec:neutrinotransport}

Over the past decade sophisticated multi-energy group
solvers for three-flavor neutrino transport including
energy-bin coupling and velocity-dependent terms (corrections
due to the motion of the stellar plasma) have been developed
and applied to all stages of stellar core collapse and the 
transition to explosion in 1D calculations. On the one hand 
this was achieved by direct integration
of the Boltzmann transport equation (BTE) with a 
discrete-ordinate (S$_\mathrm{N}$) method in GR
simulations~\cite{YamadaJankaSuzuki1999,Liebendoerferetal2004},
on the other hand by integrating the set of two-moment equations
of the BTE for neutrino number, energy, and momentum using a
variable Eddington-factor closure obtained from convergent
iteration with a model (i.e.\ simplified) Boltzmann equation.
The latter approach was developed for
Newtonian~\cite{Burrowsetal2000,RamppJanka2002}
as well as GR simulations~\cite{MuellerJankaDimmelmeier2010}.
It was also generalized for multi-dimensional applications by
adopting a ``ray-by-ray plus'' (RbR+) 
approximation~\cite{RamppJanka2002,Burasetal2006A}, in which
spherical transport problems are solved on each angular bin
of a 2D or 3D polar coordinate grid. This approximation
implies that the neutrino
intensity is assumed to be axially symmetric around the radial
direction and the neutrino flux is considered to be purely radial. 
The ``plus'' suffix signals, however, that neutrino
pressure gradients and the lateral advection of neutrinos with
fluid flows are taken into account in the optically
thick regime to prevent artificial hydrodynamic 
instabilities~\cite{Burasetal2006A}.

All the published 1D and 2D SN models of the Garching group, 
e.g.\ in
refs.~\cite{Burasetal2006A,Burasetal2006B,Kitauraetal2006,MarekJanka2009,Huedepohletal2010},
include the full, state-of-the-art set of neutrino interactions 
listed in Table~\ref{jankatab1}. Recently, 1D results based on a 
similarly refined treatment of the neutrino processes have been
put out by the Oak Ridge group~\cite{Lentzetal2011}.

Truely multi-dimensional, energy-dependent transport schemes for 
radiation-hydrodynamics with neutrinos have so far been used
extensively in 2D Newtonian simulations only by the 
Arizona-Hebrew-Princeton collaboration, which applied a MGFLD method
(e.g., in refs.~\cite{Burrowsetal2006,Burrowsetal2007,Burrowsetal2007B})
and an S$_\mathrm{N}$ solver for a multi-angle (MA) treatment
treatment~\cite{Ottetal2008,Brandtetal2011},
however without energy-bin coupling and without properly accounting
for effects associated with fluid motions.
This are severe shortcomings~\cite{Lentzetal2011,Burasetal2006A}, 
which are avoided in more elaborate 2D Newtonian 
implementations of MGFLD~\cite{SwestyMyra2009}
and of a two-moment closure scheme for the coupled set of neutrino 
energy and momentum equations~\cite{ObergaulingerJanka2011}.
An alternative approach is the ``isotropic diffusion source
approximation'' (IDSA;~\cite{LiebendoerferWhitehouseFischer2009}),
in which the neutrino distribution function is decomposed into 
trapped and streaming particle components, whose separate evolution 
equations are coupled by a diffusion source term. This method was
simplified to a RbR version for $\nu_e$ and $\bar\nu_e$ and
only a subset of neutrino processes in 2D~\cite{Suwaetal2010} and
3D~\cite{TakiwakiKotakeSuwa2011} SN simulations. A more detailed 
comparison and critical assessment of presently employed transport 
treatments can be found in ref.~\cite{Lentzetal2011}.

Pointing into the future, routes towards 3D time-dependent
neutrino transport in radiation-hydrodynamics calculations
have been outlined in the form of a rigorous solution of the
6+1 dimensional (three spatial dimensions, energy and
two direction angles for the radiation momentum, plus time)
BTE by an S$_\mathrm{N}$ discretization 
scheme~\cite{SumiyoshiYamada2012}, by spectral
methods~\cite{BonazzolaVasset2011}, and in GR by a truncated
moment formalism~\cite{Shibataetal2011}, but observer corrections 
due to fluid motion, relativistic effects, nonlinear,
energy-coupling interaction 
kernels, and high parallization efficieny are major challenges.

Direct comparisons of multi-dimensional SN calculations by 
different groups with different codes and approximations
have not been carried out so far, in contrast to the 1D
case~\cite{LiebendoerferRamppetal2005,MuellerJankaDimmelmeier2010},
and will be a formidable task for the coming years.
However, FLD was shown to 
underestimate angular variations of the radiated neutrinos
and to sphericize the radiation field compared to a multi-angle
(S$_\mathrm{N}$) treatment~\cite{Ottetal2008}, although 
fundamental changes of the hydrodynamic evolution were not
observed despite higher neutrino-heating rates with the 
S$_\mathrm{N}$ code. 
On the contrary, the RbR approximation generically
sharpens angular variations since all fluxes are radial. Local
emission maxima (``hot spots'') in the neutrinospheric region
therefore send radiation only in radial direction. Nevertheless,
because nonspherical accretion flows in the SN core exhibit 
unsteady behavior in space and time (see, e.g.,
\cite{MuellerJankaWongwathanarat2011}) ``variational averaging''
can be expected to diminish any dynamical consequences of local
emission peaks~\cite{Burasetal2006A}.

Despite undeniable weaknesses, the complexity and computational 
intensity of neutrino-hydrodynamics in full generality will make
the use of simplifications unavoidable still for some time.

\begin{figure}
\centerline{\psfig{figure=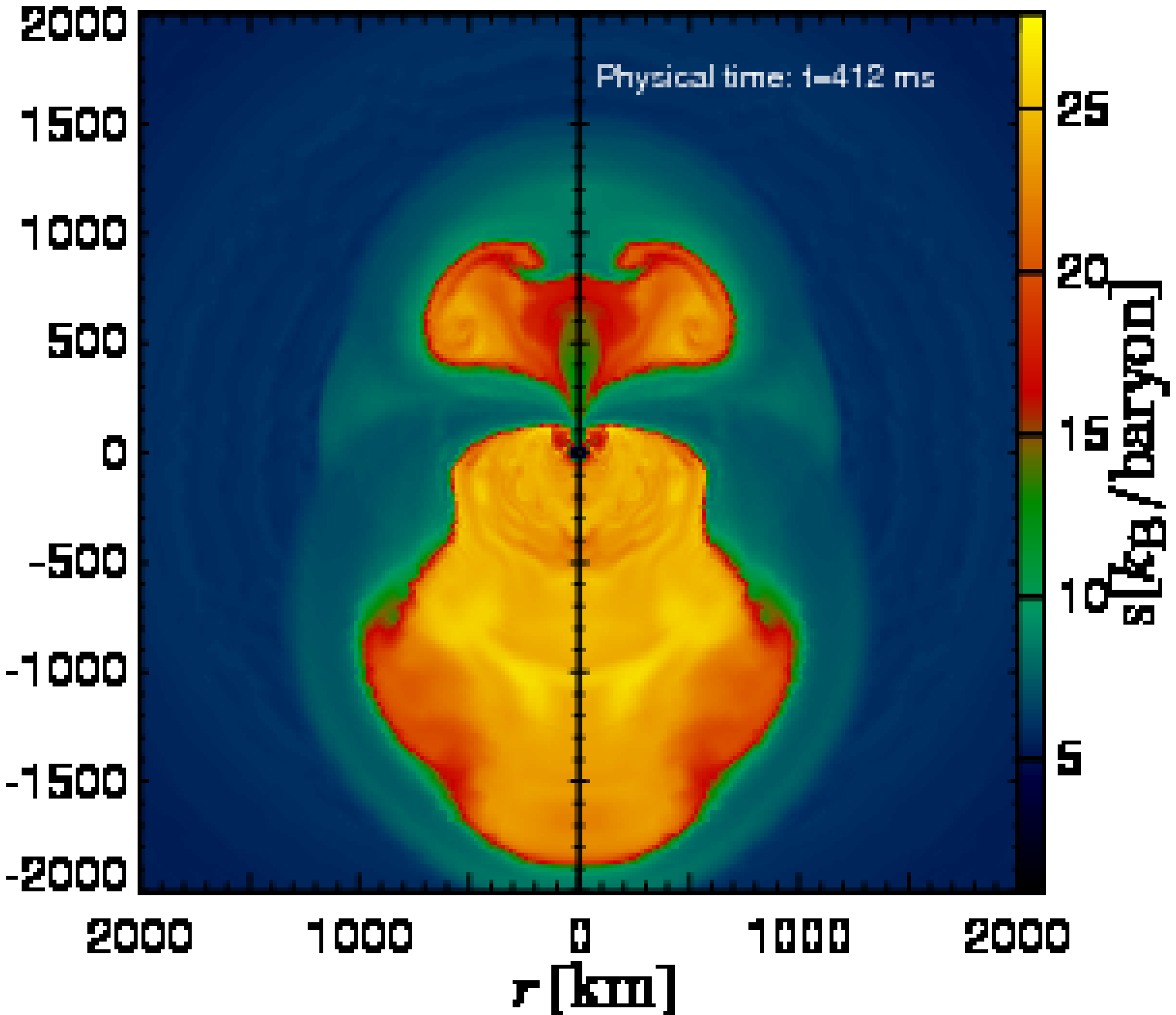,height=5truecm}
            \psfig{figure=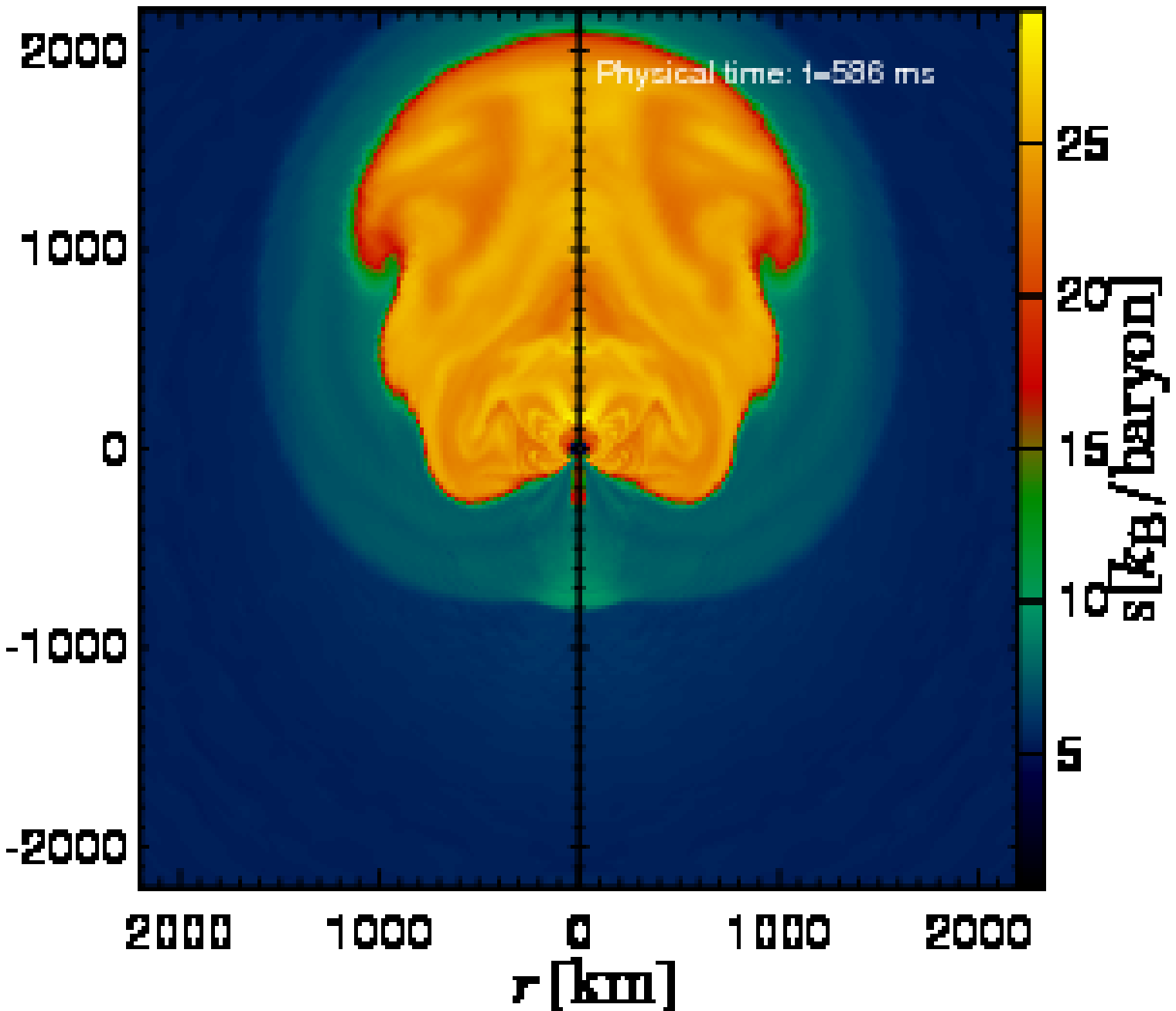,height=5truecm} 
            \psfig{figure=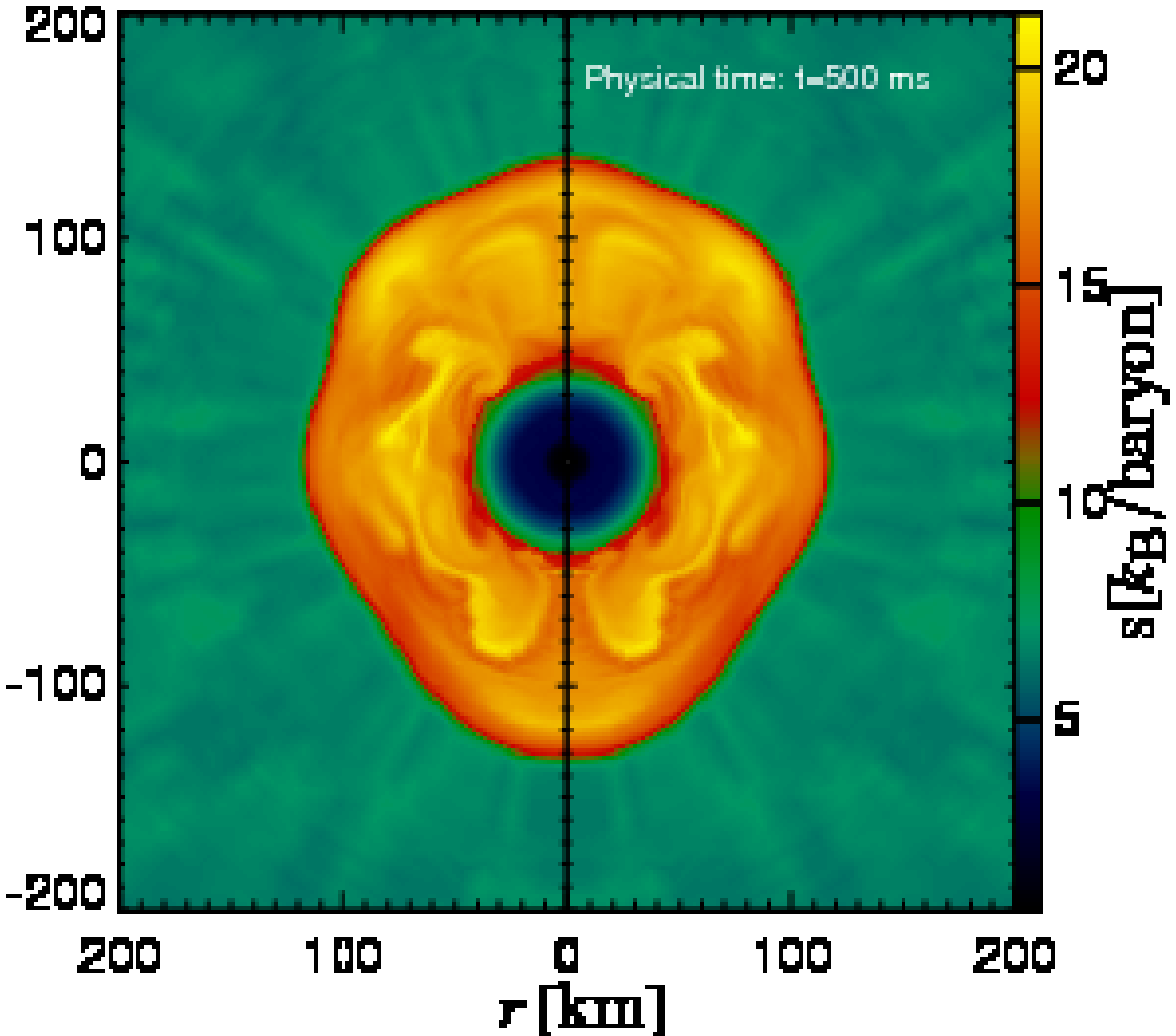,height=5truecm}}
\vspace{1.0truecm}
\centerline{\psfig{figure=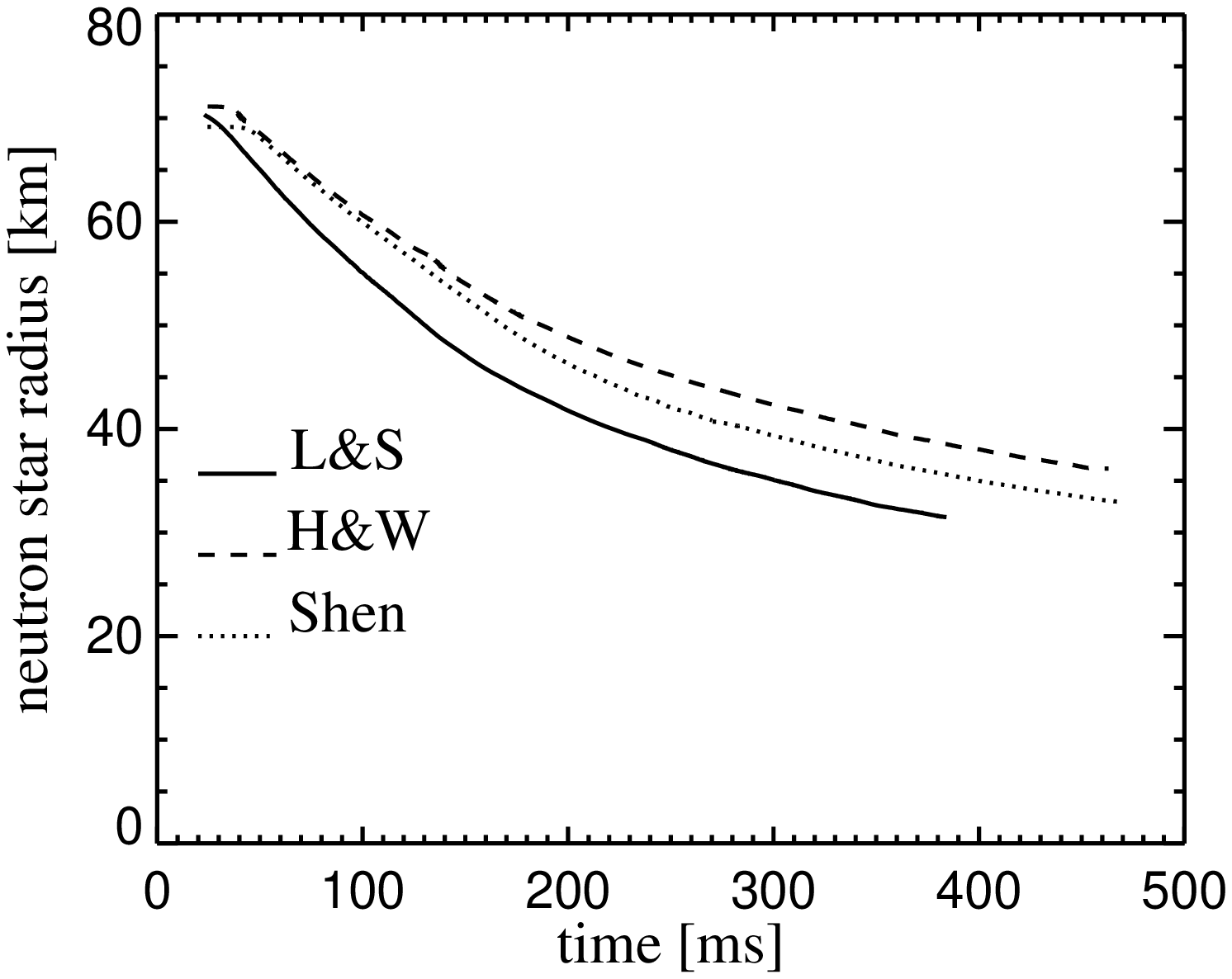,height=5truecm}\hspace{1.0truecm}
            \psfig{figure=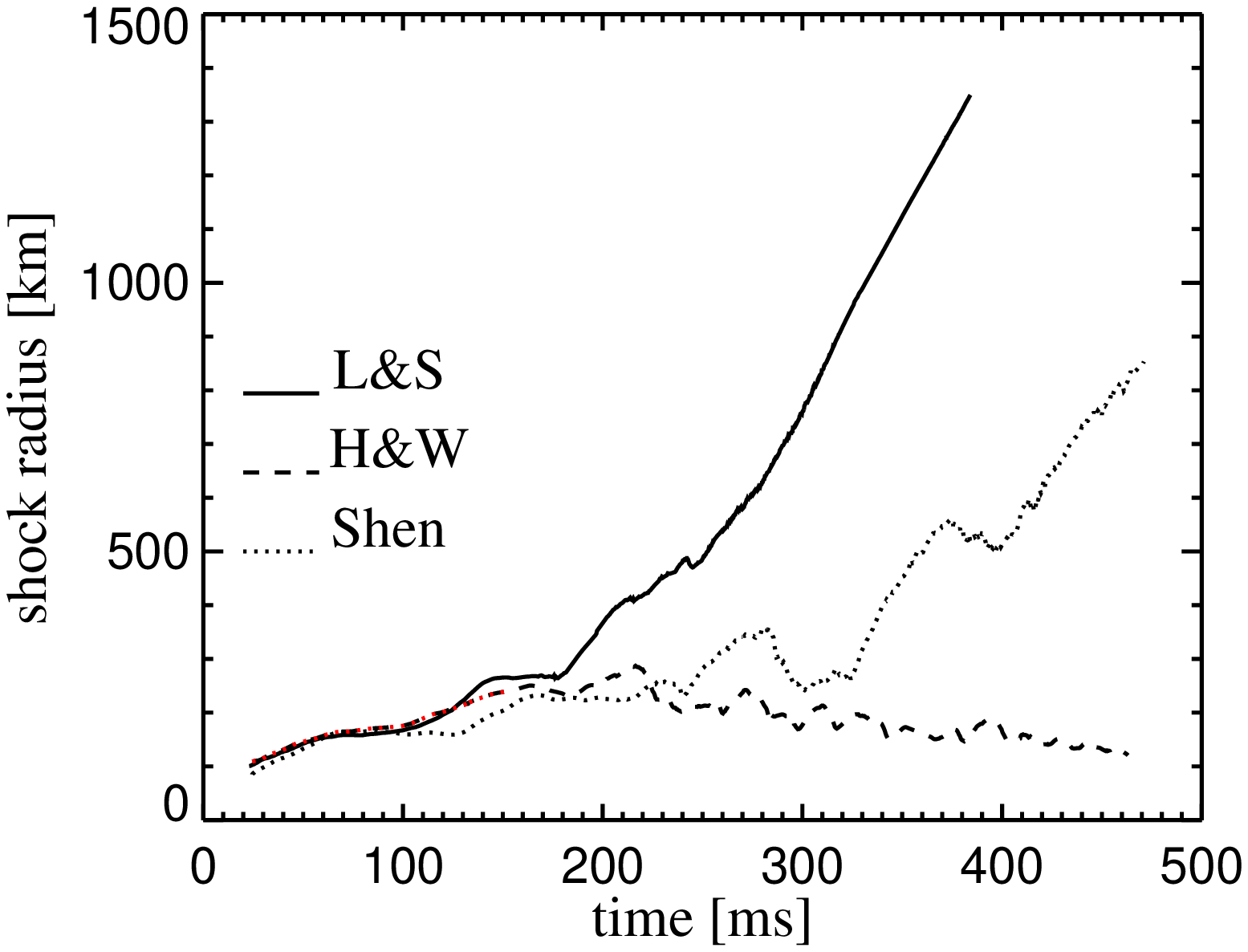,height=5truecm}}
\caption{{\small 
Two-dimensional SN simulations~\cite{MarekJanka2012}
of an 11.2\,$M_\odot$ star~\cite{Woosleyetal2002} for three
different nuclear EoSs. The {\em upper panels} show cross-sectional
entropy distributions at 412\,ms after bounce for the LS180-EoS
({\em left}), at 586\,ms p.b.\ for the STOS-EoS ({\em middle}), 
and at 500\,ms p.b.\
for the Hillebrandt \& Wolff EoS~\cite{HillebrandtWolff1984}.
The last is the stiffest EoS of the set. It leads to the 
slowest contraction of the PNS ({\em bottom left}) and because
of weaker neutrino heating and less vigorous hydrodynamic mass
motions does not yield an explosion within the simulated time
as visible in the evolution of the average shock radius
({\em bottom right}).}}
\label{jankafig4}
\end{figure}

\subsection{Equation of State and Composition of Stellar Plasma}
\label{sec:microphysics}

The nuclear and subnuclear EoS is an extremely important
ingredient for SN modeling. Unfortunately, our knowledge of,
in particular, the supranuclear regime is incomplete,
although information from nuclear 
theory and experiment~\cite{Hebeleretal2010} 
as well as astrophysical observations, for example by the
recently discovered 1.97\,$M_\odot$ binary
millisecond pulsar PSR J1614-2230~\cite{Demorestetal2010},
is rapidly growing and is beginning to set serious constraints on the
possible existence of bigger mass regions with exotic phases
in NS interiors~\cite{LattimerPrakash2010,SteinerLattimerBrown2010}.

The two EoSs for hot NS matter widely used for stellar core
collapse in the past decade are those of Lattimer \& 
Swesty~\cite{LattimerSwesty1991} and 
Shen et al.~\cite{Shenetal1998}. They
include nucleons and nuclei, electrons and positrons, and photons.
The former is based on a compressible liquid-drop 
model~\cite{LattimerPethicketal1985} with a Skyrme force for
nucleon interactions. The transition to homogeneous nuclear 
matter is established by a Maxwell construction.
Most of the SN simulations by the Garching group were performed
with a version (LS180-EoS) with an incompressibility
modulus of bulk nuclear matter of $K = 180\,$MeV and a value of
29.3\,MeV for the symmetry energy parameter. 
The Shen et al.\ EoS (STOS-EoS) employs a relativistic mean
field model with parameter settings that reproduce characteristic
properties of heavy nuclei. It is extended with the
Thomas-Fermi spherical-cell approximation to describe 
homogeneous matter as well as inhomogeneous conditions.
Its bulk incompressibility and symmetry energy have values of
281\,MeV and 36.9\,MeV, respectively. 

These EoSs describe the nuclear composition as a mix of free
nucleons, alpha particles, and a representative heavy nucleus,
whose mass and charge numbers $(A,Z)$ depend on density, 
temperature, and neutronization of the matter. Although largely
different $(A,Z)$ are returned by both EoSs during the infall
stage and affect, for example neutrino trapping through coherent 
neutrino-nuclei scatterings, 
1D simulations yield basically the same 
behavior. Quantitative differences occur only on a modest level 
of 5--25\% in quantities characterizing collapse, bounce, and
early postbounce evolution, e.g.\ in the central lepton fraction
at neutrino trapping, position of shock formation, peak luminosity
of the $\nu_e$ burst, and maximum radius to which the shock 
expands before it retreats
again~\cite{Jankaetal2007,Jankaetal2005NuclPhysA758,Lentzetal2010,Hempeletal2011}. 
This outcome is even more astonishing in view of the appreciably
different adiabatic index
$\Gamma = (\partial\ln P/\partial\ln\rho)_s$ 
($P$, $\rho$, and $s$ are pressure, density, and entropy per
nucleon, respectively) for both EoSs around nuclear density
($\Gamma_\mathrm{LS}\sim 2.2$, $\Gamma_\mathrm{STOS}\sim 2.9$)
and the correspondingly different maximum compression
and rebound behavior at bounce. Once again {\em Mazurek's Law}
applies, according to which any change of the microphysics is
moderated in its effects on collapsing stellar cores by a 
strong feedback between the EoS, weak interactions, neutrino
transport, and hydrodynamics~\cite{LattimerPrakash2000}.

In order to achieve a more elaborate treatment of the nuclear
composition in the shock-heated region below neutrinospheric 
densities after bounce and to connect smoothly to the
chemical abundances of the progenitor star, the Garching simulations
employ at $\rho < 10^{11}$\,g/cm$^3$ a Boltzmann-gas NSE description 
with typically two dozen nuclear species,
and in the non-NSE regime at $T\lesssim 5\times 10^9$\,K 
a nuclear ``flashing'' treatment~\cite{RamppJanka2002} or, 
alternatively available now, a small reaction network for 
nuclear burning.

With a maximum gravitational mass of 1.83\,$M_\odot$ for 
cold NSs in weak equilibrium, the LS180-EoS is not compatible
with PSR J1614-2230. Moreover, an incompressibility
of $K = 180\,$MeV seems in conflict with the experimentally
favored value of $K \sim 240\,$MeV for symmetric nuclear
matter~\cite{ShlomoKolomietzColo2006,Piekarewicz2010}. While
the STOS-EoS ($M_\mathrm{max}^\mathrm{STOS}\approx 2.22\,M_\odot$)
fulfills both constraints, its radius of
$\sim$15\,km for a 1.4\,$M_\odot$ NS does not match
the best NS radius estimate from the currently most 
comprehensive evaluation of astrophysical data,
$R_\mathrm{ns}\sim 11$--12.5\,km for 
$M_\mathrm{ns} = 1.4\,M_\odot$~\cite{SteinerLattimerBrown2010}.
This estimate overlaps with the range of $\sim$10--14\,km deduced 
from theoretical considerations~\cite{Hebeleretal2010}, which
in turn agrees with a NS radius of $\sim$12\,km for the LS180-EoS.

The properties of cold, neutronized NSs, however, are not 
necessarily conclusive for the conditions in the
hot SN-core environment. Indeed, for different versions of the
LS-EoS with $K = 180, 220, 375$\,MeV (the last two being 
compatible with PSR J1614-2230) the structure of hot PNSs 
well below the maximum mass, which is a relevant aspect for the
early postbounce evolution of collapsing stellar cores,
shows only smaller differences. Correspondingly, 1D CC simulations
with these EoS versions revealed only minor differences until hundreds
of ms after bounce~\cite{Myraetal1994,Thompsonetal2003,Lentzetal2010}.
During the later PNS cooling phase and in particular when mass 
accretion brings the PNS close to the mass limit, differences in the 
stiffness and the symmetry energy of the EoS can have important
consequences, e.g.\ for the time when BH formation 
occurs~\cite{Hempeletal2011} or for convective activity in the PNS
and its influence on the neutrino emission~\cite{Robertsetal2011}.
Moreover, 2D simulations showed~\cite{MarekJanka2009,MarekJanka2012}  
that the explosion of 11.2\,$M_\odot$ and 
15\,$M_\odot$ progenitors depends sensitively on the
radius evolution of the PNS in the first few 100\,ms after bounce,
i.e., the {\em radius contraction of the PNS} 
(in contrast to the final radius of the NS), because
a more rapidly shrinking remnant radiates neutrinos with higher
fluxes and energies~\cite{Jankaetal2005NuclPhysA758,Hempeletal2011}, 
thus enhancing neutrino heating and in particular also enabling
more violent hydrodynamic instabilities (Fig.~\ref{jankafig4}).

A variety of new non-zero temperature EoSs for SN studies have 
been put out recently
\cite{HempelSchaffnerBielich2010,ShenHorowitzTeige2011,ShenHorowitzOConnor2011,Shenetal2011}. With modern 1D SN codes being available, these
EoSs have been (or will be) channelled through an ``industrial''
testing pipeline, confirming the modest influence of differences
near or above nuclear-matter density on the early shock evolution
in 1D as reported from previous studies above~\cite{Hempeletal2011}. 
Also a refined description of the nuclear
composition~\cite{HempelSchaffnerBielich2010,Hempeletal2011} did
not manifest itself in a big impact on infall and shock formation.
Future studies, also in 2D and 3D, will have to show whether
so far ignored light nuclei ($^2$H, $^3$H, $^3$He, Li) 
besides $^4$He~\cite{SumiyoshiRoepke2008,Typeletal2010,Hempeletal2011b}
will have any relevant effects on the SN 
mechanism~\cite{SumiyoshiRoepke2008,Hempeletal2011} or the
neutrino-driven wind from the cooling PNS~\cite{Arconesetal2008}.

\begin{figure}
\centerline{\hspace{-0.5truecm}
            \psfig{figure=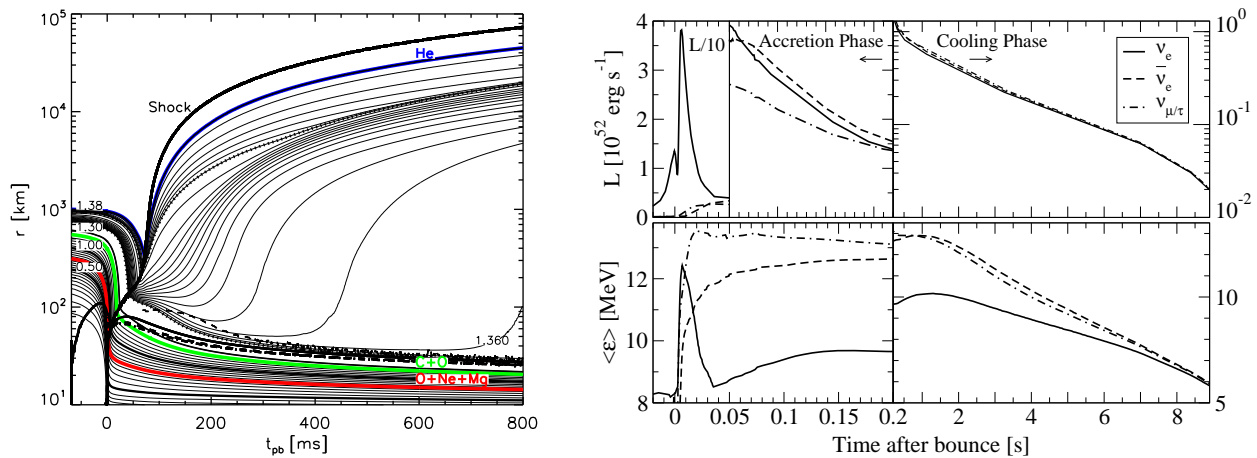,height=6.0truecm}
            \hspace{0.5truecm}
            \psfig{figure=JankaFig5b.eps,height=6.0truecm}}
\caption{{\small
{\em Left panel:} Neutrino-powered ECSN of an 
8.8\,$M_\odot$ star with ONeMg core~\cite{Nomoto1984,Nomoto1987}
visualized by mass-shell trajectories of a 1D simulation 
(from~\cite{Kitauraetal2006}). The SN shock
(bold, outgoing line) expands for $\sim$50\,ms as accretion
shock (the downstream velocities are negative) before it 
accelerates by reaching the steep density gradient
at the edge of the core. Neutrino heating subsequently 
drives a baryonic ``wind'' off the PNS surface. Colored
lines mark the inner boundaries of the Mg-rich layer in
the O-Ne-Mg core (red; at $\sim$0.72\,$M_\odot$), C-O shell
(green; at $\sim$1.23\,$M_\odot$), and He-shell
(blue; at $\sim$1.38\,$M_\odot$). The outermost dashed line
indicates the gain radius, and the inner (bold) solid, dashed,
and dash-dotted lines are the neutrinospheres 
of $\nu_e$, $\bar\nu_e$, and $\nu_x$, respectively. 
{\em Right panel:} Neutrino luminosities and mean energies
from an ECSN for the infall, $\nu_e$ breakout-burst, accretion 
phase, and PNS cooling evolution
(from~\cite{Huedepohletal2010}). The average energies are
defined as the ratio of energy to number fluxes.
(The left panel is reproduced with permission; copyright: ESO.)
}}
\label{jankafig5}
\end{figure}

\begin{figure}
\centerline{\psfig{figure=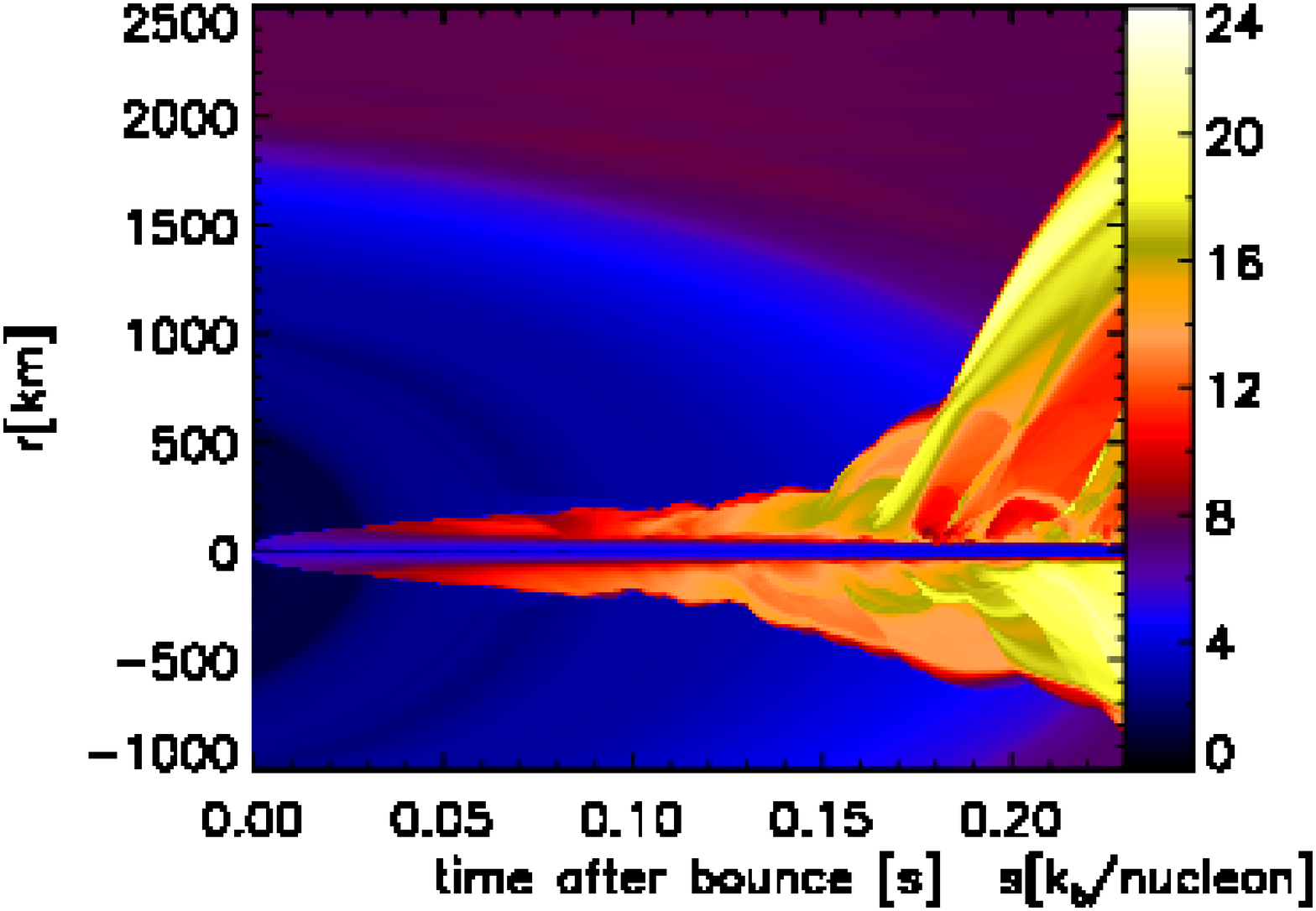,width=7truecm}
            \psfig{figure=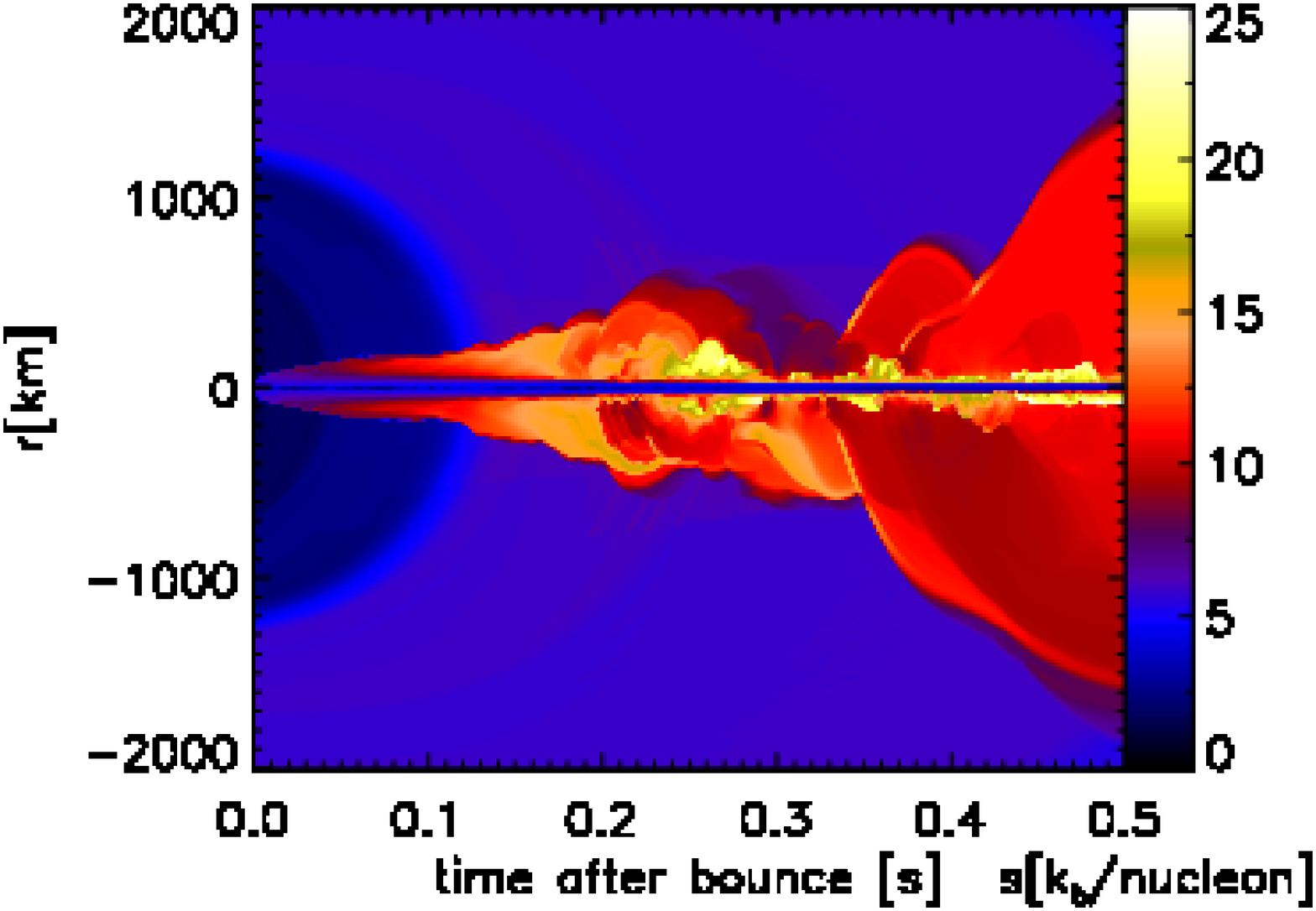,width=7truecm}}
\vspace{-0.3truecm}
\centerline{\psfig{figure=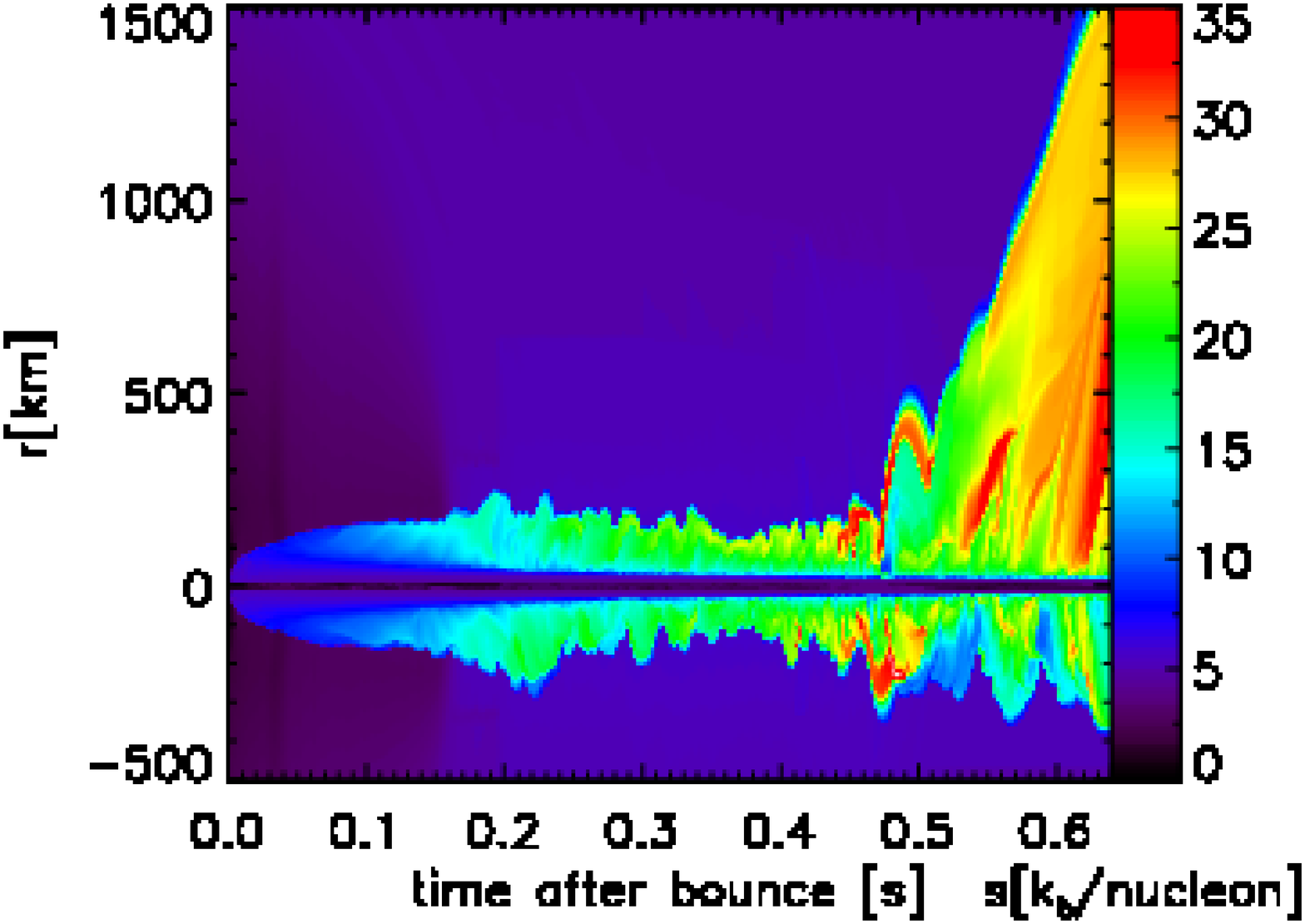,width=7truecm}
            \psfig{figure=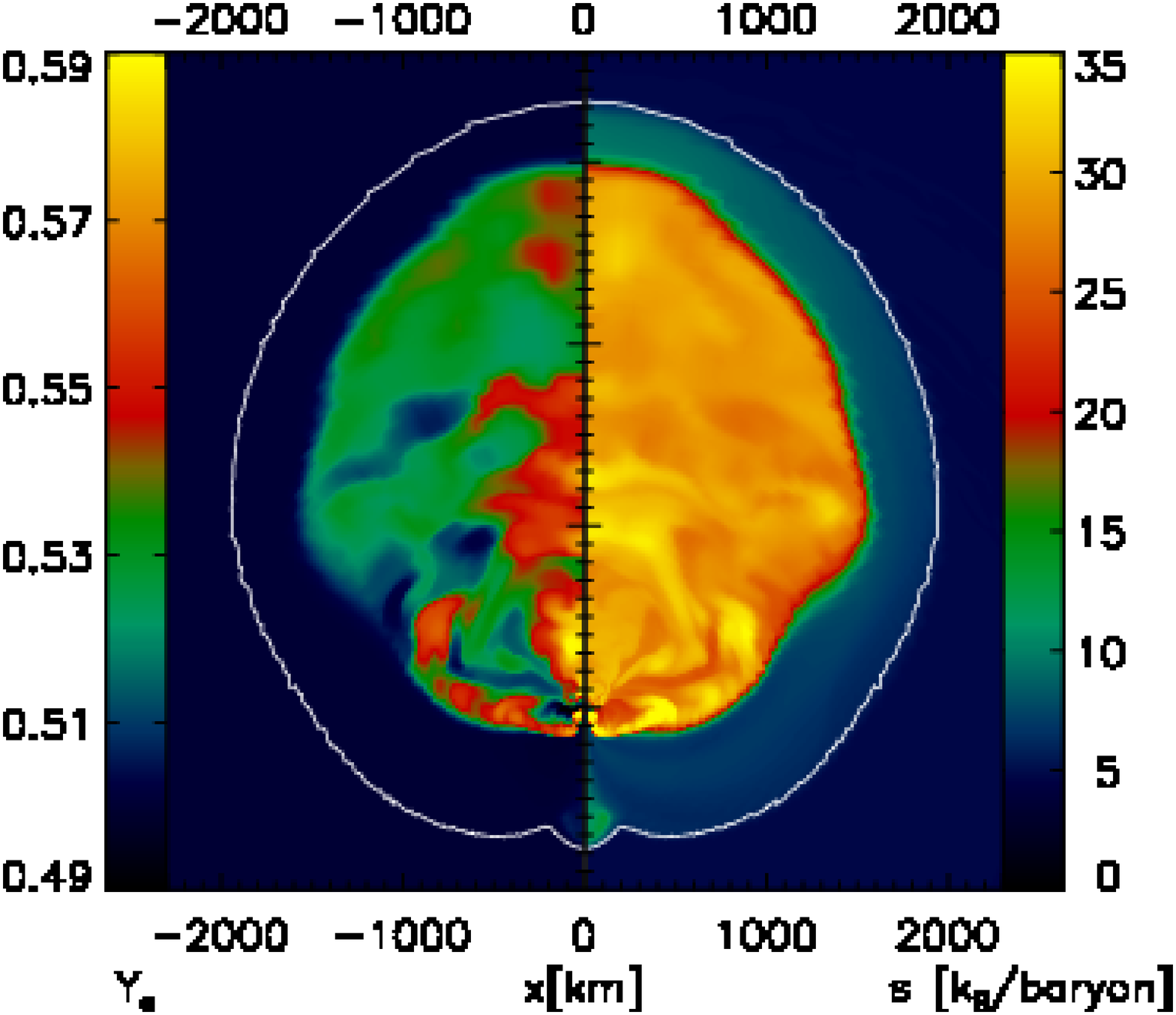,width=7truecm}}
\caption{{\small Neutrino-driven explosions of Fe-core 
progenitors~\cite{Muelleretal2011,MuellerJankaMarek2012}.
The {\em upper left, upper right, and lower left panels}
display the time evolution of color-coded entropy profiles in 
the north and south pole directions for 2D simulations of an
8.1\,$M_\odot$ ultra metal-poor ($10^{-4}$ solar metallicity)
star (A.~Heger, private communication), and 
11.2\,$M_\odot$~\cite{Woosleyetal2002} and
15\,$M_\odot$~\cite{WoosleyWeaver1995} solar-metallicity 
stars, respectively. The shock position is clearly visible
as a sharp boundary between high-entropy (yellow, red) and
low-entropy (blue, black) regions. Shock oscillations are
associated with violent convective activity in the 
neutrino-heating region and strong, bipolar SASI sloshing
motions of the whole postshock layer.
The explosions develop highly aspherically
in all cases. The {\em lower right panel} shows, for example,
an extreme dipole asymmetry of the cross-sectional distribution
of electron fraction ($Y_e$; {\em left}) 
and entropy at 775\,ms p.b.\ for 
the 15\,$M_\odot$ model, which explodes in a unipolar way.
The NS is located at the position of the 
lowermost long tickmark on the vertical axis, far away from
the geometrical center of the roundish shock contour (white
line).}}
\label{jankafig6}
\end{figure}

\section{EXPLOSION MECHANISMS}
\label{sec:mechanisms}

In this section we will review the mechanisms by which the 
gravitational binding energy of the collapsing stellar core
can be tapped to eject the outer stellar layers in a SN
blast. A particular problem in understanding the onset of 
the explosion of massive stars is connected to the need of 
reversing implosion to explosion by transferring energy from
the nascent NS to the overlying shells. This is different from
thermonuclear SNe (SNe~Ia) of white dwarfs (WDs), where 
the combustion (by deflagration or detonation) of carbon and
oxygen to nickel and silicon in an essentially hydrostatic 
object releases sufficient energy to unbind and destroy the
whole star. 

The typical energy scale of the explosion of a WD near its
Chandrasekhar mass limit is set by the release of nuclear 
binding energy associated with the conversion of
$\sim$1\,$M_\odot$ C+O to Si+Ni (roughly $2\times 10^{51}$\,erg)
minus the gravitational binding energy of the initial, highly 
degenerate WD (several $10^{50}$\,erg). But what sets the 
energy scale of CCSNe? Why does the large majority of 
``normal'' cases have explosion energies similar to SNe~Ia?
The answer to this question is connected to the initial state
of the dying star, in which the gravitationally unstable 
Fe-core is a configuration resembling a massive, degenerate
WD, surrounded by dense shells whose gravitational
binding energy is of the same order of magnitude,
i.e., around (1--15)$\times 10^{50}$\,erg.

Any self-regulated mechanism for powering the explosion
will deposit an energy in this range, possibly a few times
this value, before the energy transfer is turned off.
The neutrino-heating mechanism is such a self-regulated
process, because the matter particles absorbing energy 
from neutrinos will react by expanding away from the
heating region as soon as they have acquired an energy of
the order of their binding energy. This will evacuate the
heating region and diminish further energy input. But CC
events show a large diversity concerning their kinetic
energy, ranging from $\gtrsim$$10^{50}$\,erg to 
some $10^{51}$\,erg for SNe and up to several $10^{52}$\,erg
for HNe (Fig.~\ref{jankafig3}). Explosion energies far in
excess of the initial gravitational binding energy of the 
ejecta suggest a different driving mechanism than neutrino
heating, a process without the self-regulation described 
above. Magnetorotational explosions fulfill this requirement,
because the blast-wave energy is extracted from the huge 
reservoir of rotational energy of a rapidly spinning PNS
by magnetic fields and can be delivered in the form of
electromagnetic energy.

In the following the status of our present understanding of
both of these mechanisms will be summarized and also other,
more controversial suggestions will be addressed.

\subsection{Thermonuclear Mechanism}

While ignition of thermonuclear combustion in compression-heated,
free-falling shells was recognized not to be able to blow
matter outwards~\cite{ColgateWhite1966}, Russian 
authors~\cite{Gershtein1976,Chechetkinetal1976,Zmitrenkoetal1978,Chechetkinetal1980}
later proposed neutrino radiation from the collapsing stellar core
to heat the degenerate C+O shell of a low-mass progenitor star
at hydrostatic conditions and a density around 
$2\times 10^9$\,g/cm$^3$ and thus to ignite a thermonuclear 
burning front that explodes the star. The heating was considered
to be mainly by neutrino-electron scattering. 

Although this is a cute idea, neither the stellar 
nor the dynamical conditions assumed for this
scenario could be verified by detailed progenitor
and explosion models. In ONeMg-core progenitors, for example,
which define the low-mass limit of stars that undergo core 
collapse to radiate large neutrino luminosities, the C+O shell 
is initially located between roughly 500 and 1000\,km (at
densities $\lesssim 4\times 10^8$\,g/cm$^3$)
and falls dynamically inward (with compression-induced burning)
much before it is exposed to
a high fluence of neutrinos (see Fig.~\ref{jankafig5}).
If, in contrast, the O and C layers are farther out at 
$r > 1000\,$km as in more massive Fe-core 
progenitors (Fig.~\ref{jankafig2}), the
neutrino flux is diluted by the large distance from the source
and the electron densities (and degeneracy) there are much 
lower than adopted 
in~\cite{Gershtein1976,Chechetkinetal1976,Zmitrenkoetal1978,Chechetkinetal1980} 
so that neutrino-electron
scattering cannot raise the temperature to the ignition 
threshold. 

Presently PISNe are the only stellar CC events where the explosion
mechanism is known to be based on thermonuclear energy release
(Sect.~\ref{sec:PISNe}),
but a closer examination of the possibility of neutrino-triggered
burning in the significantly more compact low-metallicity stars
might be interesting.

\subsection{Bounce-Shock Mechanism}
\label{sec:bouncemechanism}

The purely hydrodynamical bounce-shock
mechanism~\cite{ColgateJohnson1960,ColgateGrasbergerWhite1961}, 
in which the shock wave launched at the moment of core bounce
(Sect.~\ref{sec:FeCSNe}) causes the ``prompt'' ejection of 
stellar mantle and envelope, has been a matter of intense 
research in the 1980's (for a review, see~\cite{Bethe1990}).
However, for more than 20 years now detailed analytical 
analysis of collapse and bounce 
physics~(e.g., \cite{Mazurek1982} and refs.\ therein) as well
as all modern CC simulations
---despite continuous improvements and significant quantitative
differences in details, mainly linked to important refinements
of electron captures on heavy nuclei and neutrino-electron
scattering during infall--- agree in the basic outcome:
The prompt mechanism cannot succeed in causing the explosion
of any progenitor star.

Upgrades of the microphysics turned out to disfavor
prompt explosions by decreasing the 
size of the homologously and subsonically collapsing ``inner 
core'', whose mass scales with the instantaneous Chandrasekhar 
mass, $M_\mathrm{Ch}(t)\propto Y_e^2(t)$, and whose edge
defines the location of shock formation at bounce. 
(The number fraction $Y_i$ of particles, here of electrons
($e$), is defined as the number of particles ($i$) per nucleon.)
With the presently most sophisticated treatment of neutrino
emission by electron captures on nuclei and free protons during 
core infall~\cite{langanke_03} the central electron fraction 
$Y_{e,\mathrm{c}}$ after neutrino trapping decreases to
0.25--0.27 (corresponding to a trapped
lepton fraction of 0.285--0.30), and the shock formation point
(defined by the location where the entropy first reaches a 
value of 3\,$k_\mathrm{B}$ per nucleon) lies at an 
enclosed mass of only
0.4--0.5\,$M_\odot$~\cite{Jankaetal2007,Hixetal2003,Burasetal2006B}. 
Moreover, because preferred nuclear EoSs are relatively stiff, 
the rebound of the inner core is too weak to transfer a large 
energy to the shock. The flow discontinuity, running into 
supersonically infalling material at densities below
$\sim$$10^{13}$\,g/cm$^3$, quickly loses its initial energy
by heating the plasma to entropies of several $k_\mathrm{B}$ 
per nucleon and thus disintegrating heavy nuclei to free
nucleons (which consumes roughly $1.7\times 10^{51}$\,erg per
0.1\,$M_\odot$). A short transient period of positive postshock
velocities therefore lasts only 1--2\,ms, after which the
velocity in the whole postshock region becomes negative again.
A negative postshock velocity  
defines the moment of shock stagnation, at which time
the shock has travelled through just 0.3--0.35\,$M_\odot$ of 
iron material and is still deep inside the stellar Fe-core.
Since the preshock density then is still above $10^{11}$\,g/cm$^3$,
shock stagnation happens well before shock breakout and thus
before the release of the prompt burst of $\nu_e$. Lepton number
and energy ($\sim$2$\times$10$^{51}$\,erg) drain by the escape
of the $\nu_e$ burst is therefore not causal for the shock 
stagnation.

Despite negative velocities and thus accretion flow to the
central NS in the downstream region of the shock, the latter
continues to propagate outward in mass as well as radius.
This motion of the shock stagnation radius is a response to 
the massive accretion of infalling matter (initially with
a rate $\dot M \gg 1\,M_\odot$/s; cf.\ Fig.~\ref{jankafig5}), 
which emits energy and lepton number in neutrinos and thus
settles onto the PNS only gradually, creating the 
postshock pressure that pushes
the shock position outwards. Finally, after reaching a 
maximum radius of typically 100--150\,km, the accretion 
shock retreats again in 1D models, following the contraction
of the nascent NS roughly according to the relation
\begin{equation}
R_\mathrm{s} \propto
\frac{(L_\nu\left\langle\epsilon_{\nu}^2\right\rangle)^{4/9}
R_\mathrm{ns}^{16/9}}{\dot M^{2/3}M_\mathrm{ns}^{1/3}}
\propto
\frac{R_\mathrm{ns}^{8/3} (k_\mathrm{B}T_\nu)^{8/3}}{\dot M^{2/3}
M_\mathrm{ns}^{1/3}} 
\propto
\frac{L_\nu^{4/3}}{\dot M^{2/3}M_\mathrm{ns}^{1/3}
(k_\mathrm{B}T_\nu)^{8/3}}
\, ,
\label{eq:shockradius}
\end{equation}
which can be derived by combining equations~(33, 39, 43, 44, 56, 63)
of ref.~\cite{Janka2001A} and assuming 
$R_\mathrm{g}\propto R_\mathrm{ns}$ for the ``gain radius''
$R_\mathrm{g}$ (see Sect.~\ref{sec:neutrinomechanism}) as well as 
$L_\nu\propto R_\mathrm{ns}^2 T_{\nu}^4$,
and $\left\langle\epsilon_{\nu}^2\right\rangle \propto
(k_\mathrm{B}T_{\nu})^2$ for neutrino ($\nu \in \{\nu_e,\bar\nu_e\}$)
luminosity and mean squared
energy, respectively. The radius of maximum shock expansion at
this stage is still well below the dissociation radius of iron,
for which the equality
$GM_\mathrm{ns} m_\mathrm{u}/R_\mathrm{diss} = 8.8\,$MeV ($m_\mathrm{u}$
is the atomic mass unit) yields $R_\mathrm{diss} \approx 200$\,km.
This means that the matter behind the shock is fully disintegrated into
neutrons and protons.

It is during this period of $\sim$100\,ms of slow shock expansion 
that a ``gain radius'' emerges, at which energy losses by neutrinos
for $r < R_\mathrm{g}$ change to neutrino heating for 
$r > R_\mathrm{g}$~\cite{BetheWilson1985}. Before this moment
neutrino losses are dominant in the whole postshock layer.
The onset of neutrino-energy deposition also allows convective
activity to develop behind the shock: Neutrino heating 
creates a negative entropy gradient ${\mathrm d}s/{\mathrm d}r$, 
which is unstable in the
strong gravitational field according to the Ledoux criterion,
\begin{equation}
C_\mathrm{L} = \left (\frac{\partial \rho}{\partial s}\right )_{Y_e,P}
            \frac{\mathrm{d}s}{\mathrm{d}r} +
            \left (\frac{\partial \rho}{\partial Y_e}\right )_{s,P}
            \frac{\mathrm{d}Y_e}{\mathrm{d}r}  >  0  \,.
\label{eq:ledoux}
\end{equation}
This criterion, however, defines growth conditions for convection
and Rayleigh-Taylor structures exactly only for a static layer, but 
needs to be generalized for the infalling flow in the postshock
region~\cite{Foglizzoetal2006,Burasetal2006B,Schecketal2008}.
Convective activity can take place there only when
the inward advection timescale $t_\mathrm{adv}\sim r/|v_r|$
for radial velocity $v_r$
is sufficiently longer than the convective growth 
timescale of perturbations (inverse Brunt-V\"ais\"al\"a or
buoyancy frequency), $t_\mathrm{conv}\sim
\left (g_\mathrm{grav}|C_\mathrm{L}|/\rho\right )^{-1/2}$,
or than the buoyancy acceleration timescale, $t_\mathrm{buoy}\sim
\left (g_\mathrm{grav}(\delta\rho/\rho)/r\right )^{-1/2}$,
of blobs with density contrast $\delta\rho/\rho$ in a local
gravitational field with acceleration $g_\mathrm{grav} = GM/r^2$
\cite{Janka2001B,Foglizzoetal2006}.

The breaking of spherical symmetry by hydrodynamic instability,
manifesting itself in the growth of initially small, random seed
perturbations to large-scale structures,
turned out to be generic to the shock stagnation
phase in collapsing stellar cores and to be decisive for the
success of the neutrino-heating mechanism and for the further
destiny of the stalled accretion shock.

\subsection{Neutrino-Heating Mechanism}
\label{sec:neutrinomechanism}

The development of a neutrino-heating layer is a natural 
consequence of the contraction of the PNS and associated
compactification of its surrounding accretion layer 
during the postbounce accretion phase. The contraction of the 
PNS leads to increasing
neutrinospheric temperatures and therefore growing mean 
energies of the radiated neutrinos (Fig.~\ref{jankafig7}).
More energetic neutrino emission
together with the decreasing postshock temperature
at larger shock radii allows for the appearance of a 
gain radius: Since the temperature in the postshock layer
drops roughly like $r^{-1}$ (this is well fulfilled for
convectively mixed, isentropic conditions, while in 1D
the gradient is even steeper), the neutrino-cooling rate
per nucleon by captures
of (nondegenerate) $e^-$ and $e^+$ on protons and neutrons
drops with $q_\nu^- \propto T^6\propto r^{-6}$. In
contrast, the neutrino-heating rate per nucleon 
(which is largely dominated by $\nu_e$, $\bar\nu_e$
absorption on free $n$, $p$, respectively) scales with
$q_\nu^+ \propto L_\nu\langle\epsilon_\nu^2\rangle r^{-2}$
and thus decreases less steeply with $r$
than $q_\nu^-$, enabling a crossing point $R_\mathrm{g}$
to occur~\cite{BetheWilson1985}, where 
$T_\mathrm{g}^3R_\mathrm{g}\propto 
\sqrt{L_\nu\langle\epsilon_\nu^2\rangle}$.

\subsubsection{Heating Efficiency and Energetics}
\label{sec:heatingeffiency}
With a density profile $\rho \propto r^{-3}$ between gain radius
$R_\mathrm{g}$ and shock $R_\mathrm{s}$ 
(see, e.g.,~\cite{Janka2001A}) and the preshock 
(free-fall) velocity $v_0 = -\sqrt{2GM_\mathrm{ns}/R_\mathrm{s}}$, 
mass infall rate $\dot M = 4\pi R_\mathrm{s}^2|v_0|\rho_0$ 
of the progenitor star, and density jump at the shock,
$\beta = \rho_1/\rho_0 \sim 10$, the optical depth for 
$\nu_e$ and $\bar\nu_e$ absorption in the gain layer can be
estimated as
\begin{equation}
\tau \approx 0.026\,\left ( 
\frac{k_\mathrm{B}T_\nu}{4\,\mathrm{MeV}}\right )^{\! 2}
\left (\frac{\dot M}{0.1\,M_\odot\,\mathrm{s}^{-1}}\right )
\left ( \frac{R_\mathrm{s}}{200\,\mathrm{km}} \right )^{\! 3/2}
\left ( \frac{R_\mathrm{g}}{100\,\mathrm{km}} \right )^{\! -2}
\left (\frac{M_\mathrm{ns}}{1.5\,M_\odot} \right )^{\! -1/2} \, ,
\label{eq:optdepth}
\end{equation}
where $Y_n \approx Y_p \approx 0.5$ and 
$\langle \sigma_\mathrm{abs}\rangle \approx 3.26\times 10^{-41}
[k_\mathrm{B}T_\nu/(4\,\mathrm{MeV})]^2$\,cm$^2$ for the average
absorption cross section of a blackbody neutrino spectrum with
temperature $T_\nu$ (therefore $\langle\epsilon_\nu^2\rangle \approx
21 (k_\mathrm{B}T_\nu)^2$) was used. Equation~\ref{eq:optdepth} 
suggests that for typical 
accretion rates, $\dot M = 0.1$--0.5\,$M_\odot$/s, several percent 
of the neutrino luminosity from the neutrinosphere can be absorbed 
in the gain layer, thus accounting for a neutrino-heating rate
$Q_\nu^+ = \tau\,(L_{\nu_e} + L_{\bar\nu_e})\approx
10^{51}$--$10^{52}$\,erg/s for $\nu_e$ and $\bar\nu_e$ luminosities, 
$L_\nu$, of some $10^{52}$\,erg/s 
during the postbounce accretion phase (Fig.~\ref{jankafig7}).

However, in a dynamical situation as in the gain layer, where 
the matter is not at rest, the optical depth (which determines
the interaction probability of a crossing neutrino) is not a perfectly
appropriate measure for the heating efficiency.
This holds in particular in the multi-dimensional case, where
accretion funnels carry cold (low-entropy) matter from the shock
towards the NS, while neutrino-heated matter expands outward in 
high-entropy bubbles. At such conditions the residence time of the
matter in the gain layer accounts for the duration of its exposure
to neutrino heating. While in the 1D case the advection time 
$t_\mathrm{adv} \sim (R_\mathrm{s}-R_\mathrm{g})/|v_1|$ with 
$v_1 = v_0/\beta$ measures how long the accretion flow needs
from $R_\mathrm{s}$ to $R_\mathrm{g}$~\cite{Janka1998,Janka2001B},
the dwell time in the gain region is better captured in the multi-D 
situation by the more general 
expression~\cite{Burasetal2006B,MarekJanka2009}
\begin{equation}
t_\mathrm{dwell}\,\approx \,\frac{M_\mathrm{g}}{\dot M} \,,
\label{eq:dwelltime}
\end{equation}
which relates the mass in the gain layer $M_\mathrm{g}$ with
the mass accretion rate $\dot M$ through the shock and (for
conditions near steady state) through the gain radius. With an 
energy-transfer rate per nucleon by neutrino absorption of $q_\nu^+ =
n_\nu \langle\epsilon_\nu\rangle\langle \sigma_\mathrm{abs}\rangle c$, 
where $n_\nu\langle\epsilon_\nu\rangle = L_\nu/(4\pi r^2c)$, each
nucleon absorbs an energy of $q_\nu^+\,t_\mathrm{dwell} \sim 50\,$MeV
for time $t_\mathrm{dwell}\sim 0.1$\,s when $k_\mathrm{B}T_\nu = 4$\,MeV,
$L_\nu = 3\times 10^{52}$\,erg/s ($\nu \in \{\nu_e, \bar\nu_e\}$),
and $r = R_\mathrm{g} \sim 100$\,km are
assumed. This corresponds to a temperature of $\sim$3\,MeV and an entropy
of $\sim$20\,$k_\mathrm{B}$ per nucleon of an $e^\pm$-photon-dominated
plasma at $\rho\sim 10^9$\,g/cm$^3$ (cf.\ Figs.~\ref{jankafig4} 
and \ref{jankafig6}). The total energy deposition rate by
$\nu_e$ plus $\bar\nu_e$ absorption in the gain layer thus becomes
\begin{eqnarray}
Q_\nu^+ \!&=&\! q_\nu^+\,\frac{M_\mathrm{g}}{m_\mathrm{u}} \cr
\! &\sim& \!  9.4\times 10^{51}\,\frac{\mathrm{erg}}{\mathrm{s}}\,
\left ( \frac{k_\mathrm{B}T_\nu}{4\,\,\mathrm{MeV}} \right )^{\! 2} \!
\left ( \frac{L_\nu}{3\cdot 10^{52}\,\mathrm{erg/s}} \right ) \!
\left ( \frac{M_\mathrm{g}}{0.01\,M_\odot} \right ) \!
\left ( \frac{R_\mathrm{g}}{100\,\mathrm{km}} \right )^{\! -2} \! .
\label{eq:heatingrate}
\end{eqnarray}
Equation~(\ref{eq:heatingrate}) corresponds to a heating efficiency of
\begin{equation}
\frac{Q_\nu^+}{L_{\nu_e}+L_{\bar\nu_e}} \,\sim\,
0.16
\left ( \frac{k_\mathrm{B}T_\nu}{4\,\,\mathrm{MeV}} \right )^{\! 2} \!
\left ( \frac{M_\mathrm{g}}{0.01\,M_\odot} \right )  \!
\left ( \frac{R_\mathrm{g}}{100\,\mathrm{km}} \right )^{\! -2} 
\label{eq:heatingefficiency}
\end{equation}
and an integral energy deposition of
\begin{eqnarray}
E_N  &\sim& Q_\nu^+\,t_\mathrm{dwell}       \cr 
     &\sim& 9.4\times 10^{50}\,\mathrm{erg} 
\left ( \frac{k_\mathrm{B}T_\nu}{4\,\,\mathrm{MeV}} \right )^{\! 2} \!
\left ( \frac{L_\nu}{3\cdot 10^{52}\,\mathrm{erg/s}} \right )\!\times \cr
     &\phantom{\sim}& \phantom{9.4\times 10^{50}\,\mathrm{erg}}
\left ( \frac{M_\mathrm{g}}{0.01\,M_\odot} \right )^{\! 2} \!
\left ( \frac{\dot M}{0.1\,M_\odot\,\mathrm{s}^{-1}} \right )^{\! -1} \!
\left ( \frac{R_\mathrm{g}}{100\,\mathrm{km}} \right )^{\! -2}  \,.
\label{eq:totheating}
\end{eqnarray}
These numbers, reduced by 20--30\% for neutrino-cooling losses 
in the gain layer, are well compatible with results of detailed
simulations~\cite{MarekJanka2009,MuellerJankaMarek2012}.

\subsubsection{Hydrodynamical Explosion Models}
\label{sec:hydromodels}
Neutrino-driven explosions can be found in 1D simulations only
for ECSNe of low-mass progenitors
(\cite{Kitauraetal2006,Jankaetal2008,Fischeretal2010},
considering an 8.8\,$M_\odot$ star with an ONeMg 
core~\cite{Nomoto1984,Nomoto1987}). Because of the
extremely steep density gradient at the edge of the
ONeMg core and the corresponding rapid decrease of $\dot M$,
the radius of the accretion shock grows continuously
(cf.\ Eq.~\ref{eq:shockradius}), thus creating ideal conditions
for neutrino-energy transfer (Fig.~\ref{jankafig5}). The latter 
drives a baryonic outflow, which carries the energy for the 
explosion. With the most sophisticated available treatment of 
neutrino-matter interactions 
(Table~\ref{jankatab1}; Sect.~\ref{sec:neutrinotransport}) an
explosion energy $E_\mathrm{ONeMg} \approx 10^{50}$\,erg was
obtained~\cite{Kitauraetal2006,Jankaetal2008}, 
which is enhanced at most by $\sim$10\% in 2D models due to a 
brief phase ($\sim$50--200\,ms p.b.) of convective overturn
behind the rapidly expanding
shock~\cite{Jankaetal2008,WanajoJankaMueller2011}.
The low explosion energy and little nickel ejection (several
$10^{-3}\,M_\odot$; \cite{WanajoJankaMueller2011}) 
are compatible with estimates for the
Crab SN (\cite{Nomotoetal1982} and refs.\ therein).

For more massive Fe-core progenitors nonradial hydrodynamic
instabilities ---convective 
overturn~\cite{Herantetal1994,Burrowsetal1995,JankaMueller1996,FryerWarren2002} 
in combination with SASI activity~\cite{Blondinetal2003}---
turned out to be decisive for the success of the neutrino-heating
mechanism~\cite{Burasetal2006B,MarekJanka2009}. While 1D models
did not explode, the Garching group found neutrino-driven, though
weak, explosions for 11.2 and 15\,$M_\odot$ stars in 2D
simulations~\cite{Burasetal2006B,MarekJanka2009}.
Recently, these results with the 
Prometheus-VERTEX program~\cite{RamppJanka2002,Burasetal2006A}
could be confirmed by general relativistic
2D simulations~\cite{MuellerJankaMarek2012} based on the newly 
developed CoCoNut-VERTEX code~\cite{MuellerJankaDimmelmeier2010},
which also produced explosions for solar-metallicity 
27\,$M_\odot$ and 
metal-poor ($10^{-4}$ solar metallicity) 
8.1\,$M_\odot$ progenitors with Fe cores
(see Fig.~\ref{jankafig6} and \cite{MuellerJankaHeger2012}).

Neutrino-driven explosions for a variety of stars were also
seen in 2D and 3D simulations of other 
groups with different multi-group treatments of neutrino
transport~\cite{Bruennetal2009,Suwaetal2010,Liebendoerferetal2010,TakiwakiKotakeSuwa2011},
whereas the Arizona-Jerusalem-Princeton (AJP) collaboration 
could not observe any success of the neutrino-heating 
mechanism~\cite{Burrowsetal2006,Burrowsetal2007}.
This underlines the sensitivity of the outcome qualitatively
and quantitatively to details of the input and methods.
While there are many differences between the modeling approaches
concerning numerics and microphysics, the Garching 2D models in 
particular include the full set of neutrino-matter interactions of
Table~\ref{jankatab1} and a careful implementation of all fluid-motion
dependent terms as well as GR effects in the transport. All
were recognized to be 
significant~\cite{Lentzetal2011,Burasetal2006A} 
but were simplified or ignored
in the AJP calculations due to the omission of energy-bin coupling
there (see also Sect.~\ref{sec:neutrinotransport}). 
Instead of attributing conflicting results to transport
differences, it has been repeatedly stated that the success of the 
Garching simulations is disputable
because the unacceptably soft LS180-EoS was 
used (e.g.,~\cite{OttOConnorDasgupta2011}). This criticism, however,
misses foundation because of the arguments given in
Sect.~\ref{sec:microphysics} and the fact that the 11.2\,$M_\odot$ 
explosion could be reproduced with the stiffer 
STOS-EoS (Fig.~\ref{jankafig4}),
which did not lead to an explosion in the 11.2\,$M_\odot$ run of the
AJP collaboration~\cite{Burrowsetal2007}.

Self-induced neutrino-flavor conversions in the SN core, which
could swap lower-energetic $\nu_e$ and $\bar\nu_e$ spectra with
hotter $\nu_x$ and $\bar\nu_x$ spectra and thus might enhance 
the neutrino heating behind the shock and strengthen the SN explosion,
have recently been shown not to have an impact during the postbounce
accretion phase. Because detailed SN models yield electron densities 
that are higher than the neutrino densities 
(mostly even $n_e\gg n_\nu$), the matter background dominates
and thus suppresses collective neutrino oscillations by dephasing
the flavor evolution of neutrinos travelling on different
trajectories~\cite{Chakrabortyetal2011A,Chakrabortyetal2011B,Dasguptaetal2011,Sarikasetal2011}.

\subsubsection{Effects of Nonspherical Flows}
\label{sec:nonsphericaleffects}
Nonradial, turbulent flows increase the residence time of matter
in the gain 
layer~\cite{Burasetal2006B,MurphyBurrows2008B,TakiwakiKotakeSuwa2011}
and thus the mass $M_\mathrm{g}$ in this region 
(for given $\dot M$; Eq.~\ref{eq:dwelltime}).
This leads to a higher total energy deposition rate by neutrinos, 
$Q_\nu^+$ (Eq.~\ref{eq:heatingrate}), and to an enhanced integral 
energy transfer $E_N$ (Eq.~\ref{eq:totheating}).

Rayleigh-Taylor fingers, for example, which develop in a convectively 
unstable situation (Eq.~\ref{eq:ledoux}) channel cool, freshly
accreted material from immediately downstream of the shock towards
the PNS and thus closer to the gain radius, where neutrino heating
is strongest. At the same time, expanding bubbles of buoyant,
high-entropy gas allow freshly heated matter 
to rise away from the gain
radius instead of being accreted inward to the cooling layer. This
reduces energy losses by the reemission of neutrinos, which can
have important dynamical consequences because cooling for 
$r < R_\mathrm{g}$ is usually much larger than net heating in the
gain layer. The combination of all such effects leads to an
increase of the temperature and pressure in the gain layer,
which in turn pushes the shock farther out. A positive feedback cycle
is the consequence, which for sufficiently strong neutrino heating
enables an explosion in the multi-D case even when the
neutrino-driven mechanism fails in 1D.

SASI activity can basically have the similar effects. It is not
only associated with shock expansion and nonradial mass flows, 
thus allowing for a larger efficiency of neutrino-energy deposition,
it also leads to secondary shocks that dissipate kinetic energy and
produce extra heating and higher entropies, strengthening the
convective activity and giving additional push to the
shock~\cite{Schecketal2008,MarekJanka2009}.

After the onset of the explosion, the nonspherical situation
permits simultaneous
shock expansion and ongoing accretion. This maintains higher
neutrino fluxes and stronger neutrino heating for a longer
time compared to the 
1D case~\cite{MarekJanka2009}, where the accretion luminosity 
decays as soon as shock expansion quenches the mass infall to 
the PNS.

While hydrodynamical simulations clearly demonstrate that violent
convective and SASI activity are crucial for the success of the 
neutrino-heating mechanism~\cite{MuellerJankaHeger2012}, 
the nature of the SASI and the exact
role of hydrodynamic instabilities and turbulent motions for the
onset of the explosion are still a matter of intense research.
The SASI, whose amplitude grows from small pressure and/or
entropy perturbations in an oscillatory way with 
highest growth rates for the lowest-order spherical harmonics 
(dipolar $\ell = 1$ and quadrupolar $\ell = 2$)
components~\cite{Blondinetal2003,BlondinMezzacappa2006,Ohnishietal2006,Foglizzoetal2007},
seems to be an ideal candidate to explain not only the global
asymmetries found in the SN core in simulations, but also
the large asphericities and mixing effects
that are observed in most SNe and SN 
remnants and that are probably linked to the measured high kick
velocities of many young pulsars (cf.~Sect.~\ref{sec:remnants}).
Linear growth analysis, numerical toy models for the linear 
and nonlinear regimes~\cite{GuiletFoglizzo2011,Schecketal2008},
and laboratory shallow-water experiments~\cite{Foglizzoetal2011}
yield evidence that the underlying instability mechanism is
an advective-acoustic cycle~\cite{Foglizzo2001,Foglizzo2002}
rather than a purely acoustic one~\cite{BlondinMezzacappa2006}.

\subsubsection{Runaway Threshold}
\label{sec:runaway}
Burrows \& Goshy~\cite{BurrowsGoshy1993} conjectured the transition
to the explosion to be a global instability of the postshock
layer. Considering steady-state accretion conditions in 1D, they 
determined a critical neutrino luminosity $L_{\nu,\mathrm{c}}(\dot M)$
as a function of the mass accretion rate that damps shock expansion
(cf.~Eq.~\ref{eq:shockradius}), above which they could not find 
accretion solutions and thus expected the onset of the explosion. 
Their reasoning is supported by subsequent similar analyses of
stationary accretion
flows~\cite{YamasakiYamada2005,YamasakiYamada2006,PejchaThompson2011}
as well as hydrodynamical
studies~\cite{JankaMueller1996,MurphyBurrows2008B,Nordhausetal2010,Hankeetal2011,Fernandez2011}.

The existence of a functional relation $L_{\nu,\mathrm{c}}(\dot M)$
as threshold condition to an explosion can be qualitatively
understood by simple analytic considerations. Numerical 
simulations~\cite{Thompsonetal2005,Burasetal2006B,MarekJanka2009,MurphyBurrows2008B,Fernandez2011}
have shown that the necessary condition for runaway expansion is
given by $t_\mathrm{adv}/t_\mathrm{heat} > 1$, i.e., the 
explosion can set in when the advection timescale
of the mass flow through the gain layer,
\begin{equation}
t_\mathrm{adv} = \int_{R_\mathrm{g}}^{R_\mathrm{s}}
\frac{\mathrm{d}r}{|v_r|} \sim \beta\,\frac{R_\mathrm{s}}{|v_0|}
\sim \beta\,\frac{R_\mathrm{s}^{3/2}}{\sqrt{2GM_\mathrm{ns}}} \, ,
\label{eq:advectiontime}
\end{equation}
exceeds the heating timescale for neutrinos to deposit enough
energy to lift matter from a gravitationally bound to an unbound
state. In this context the total energy of the gas is the relevant
quantity, i.e., the sum of internal, gravitational, and kinetic
energies, which in a bound state is negative. Making the assumption
that this energy scales roughly with the gravitational 
energy near the gain radius, which for a nucleon is 
$\epsilon_\mathrm{g} = -GM_\mathrm{ns}m_\mathrm{u}/R_\mathrm{g}$,
one obtains
\begin{equation}
t_\mathrm{heat} \sim \frac{|\epsilon_\mathrm{g}|}{q_\nu^+}
\propto 
\frac{M_\mathrm{ns}R_\mathrm{g}}{L_\nu\langle\epsilon_\nu^2\rangle}
\, .
\label{eq:heatingtime}
\end{equation}
Setting both timescales equal, $t_\mathrm{adv} = t_\mathrm{heat}$,
and using Eq.~(\ref{eq:shockradius})
for $R_\mathrm{s}$ and the fact that $R_\mathrm{g}$ follows 
approximately $R_\mathrm{ns}$, i.e., $R_\mathrm{g}\propto
R_\mathrm{ns} \propto L_\nu^{1/2}/(k_\mathrm{B}T_\nu)^2$ (which 
was also applied in deriving Eq.~\ref{eq:shockradius}),
leads to
\begin{equation}
L_{\nu,\mathrm{c}}(\dot M) \propto 
\beta^{-2/5}\,\dot M^{2/5}\,M_\mathrm{ns}^{4/5} \,.
\label{eq:critlum}
\end{equation}
This relation reproduces the functional behavior found
in~\cite{BurrowsGoshy1993} very well (the numerical factor
of the scaling relation becomes (5--6)$\times 10^{52}$\,erg/s 
for $\beta \sim 10$, $\dot M = 1\,M_\odot$/s, and 
$M_\mathrm{ns} = 1.5\,M_\odot$, slightly varying with the 
choice of other involved parameters).
It must be pointed out, however, that the limiting
luminosity for steady-state accretion solutions as derived 
in~\cite{BurrowsGoshy1993} was shown to be usually close to, 
but not identical with the runaway threshold at
$t_\mathrm{adv} > t_\mathrm{heat}$. The latter roughly 
coincides with the time when the fluid behind the 
shock begins to develop positive total specific energy
(see the excellent study of~\cite{Fernandez2011}).

Numerous studies for both stationary accretion
flows~\cite{YamasakiYamada2005,YamasakiYamada2006} 
and time-dependent conditions in collapsing stellar 
cores~\cite{JankaMueller1996,MurphyBurrows2008B,Nordhausetal2010,Hankeetal2011}
have demonstrated that the critical neutrino luminosity 
for fixed mass accretion rate is significantly lowered 
in the multi-dimensional case, typically by several 10\%. 
The possible (or combined) reasons for this improvement were
discussed in Sect.~\ref{sec:nonsphericaleffects}, but many
aspects are still unsettled. For example, the properties
and consequences of neutrino-driven turbulence (e.g., 
convective energy transport and pressure) are a matter of
ongoing research~\cite{MurphyMeakin2011} and the effects of
3D hydrodynamics have not been clarified yet. While there
is hope that these might make the runaway easier than in
2D~\cite{Nordhausetal2010,TakiwakiKotakeSuwa2011}
and thus lead to earlier and more powerful explosions,
not all studies revealed a significant reduction of the 
threshold luminosity in 3D relative to 2D~\cite{Hankeetal2011}.
The 2D/3D comparison obviously depends on subtle differences of
the background flow, neutrino source terms or even numerics, and 
requires further exploration. A sophisticated neutrino transport
seems necessary for reliable answers.

\subsubsection{Modes of Global Instability}

The results of SN simulations and analytic studies
suggest that the onset of the explosion is connected to 
a global runaway instability of the postshock accretion 
layer~\cite{BurrowsGoshy1993,Jankaetal2005NuclPhysA758}
fueled by neutrino energy deposition above a certain 
threshold (see Sect.~\ref{sec:runaway}).
An important question concerns the type 
of mode that grows fastest to trigger the 
runaway~\cite{Fernandez2011}. Unstable oscillatory and
non-adiabatic radial modes were observed in many time-dependent
1D simulations ---in agreement with linear stability
analysis~\cite{YamasakiYamada2007}---
for neutrino luminosities intermediate between those that
are too low to drive explosions and those that suffice to
trigger an explosion by the instability of a nonoscillatory
radial mode~\cite{JankaMueller1996,Burasetal2006A,Ohnishietal2006,MurphyBurrows2008B,Nordhausetal2010,Hankeetal2011,FernandezThompson2009,Fernandez2011}.
But what happens in the multi-dimensional case? Is the
runaway there caused by an unstable radial oscillatory or
nonoscillatory mode, whose development is affected by 
turbulence altering the conditions of the background flow?
Or is a nonradial nonoscillatory (possibly convective) or
oscillatory (SASI) mode decisive? Exploration of the growth
conditions has only begun, suggesting that unstable large-scale,
nonoscillatory modes require the highest driving
luminosities~\cite{YamasakiYamada2007}, but their growth may 
strongly depend on the conditions in the SN 
core~\cite{MuellerJankaHeger2012}, the
dimensionality of the problem, and even a modest rate of
rotation~\cite{YamasakiFoglizzo2008}. While first 3D 
simulations~\cite{Iwakamietal2008,Liebendoerferetal2010,Nordhausetal2010,Wongwathanaratetal2010,Hankeetal2011,TakiwakiKotakeSuwa2011}
show strongly damped or no radial oscillations, suggesting
that SASI modes are less strong in 3D and the explosion
might be connected to unstable nonoscillatory 
modes~\cite{BurrowsDolenceMurphy2012,MurphyDolenceBurrows2012}, 
none of these simulations was performed with a combination of
sufficiently sophisticated neutrino transport, high enough
numerical resolution, and a consistent inclusion of all 
dissipative processes (such a the decay of the
NS core luminosity, changes of the accretion luminosity, and
the shrinking of the nascent NS, all of which provide a 
negative feedback) included and combined consistently.
Final answers will require well-resolved, full-scale 3D 
radiation-hydrodynamics simulations with reliable neutrino
treatment.

\subsection{Magnetohydrodynamic Mechanisms}
\label{sec:magneticmechanism}

MHD phenomena, in particular the magnetorotational mechanism 
(MRM) proposed in \cite{Bisnovatyi-Kogan1970,OstrikerGunn1971}, have
been discussed as potential drivers of SN explosions already in
the 1970's (e.g., \cite{Meieretal1976,Bisnovatyi-Koganetal1976}) and
were explored by first axi-symmetric simulations with approximate 
microphysics and artificially imposed stellar core rotation and 
magnetic field configurations in~\cite{LeBlancWilson1970,Symbalisty1984}. 
These and a flood of subsequent 2D calculations, which either
ignored or radically simplified the neutrino physics
(e.g., \cite{Kotakeetal2004,Sawaietal2005,Obergaulingeretal2006,Moiseenkoetal2006}
and refs.\ therein), or more recently used neutrino transport
by MGFLD (assuming, inappropriately, the stellar medium to be at
rest;~\cite{Burrowsetal2007}), have confirmed that MHD processes 
and especially the MRM are able viable agents to extract energy 
from a highly magnetized NS and to violently expel the outer
stellar layers.

Because of the extremely low resistivity of SN matter, magnetic field
lines are frozen in the flow. Magnetic flux conservation therefore 
leads to compressional amplification of the average strength of the
nonradial field during CC,
$B \propto R_\mathrm{core}^{-2}\propto \rho_\mathrm{core}^{2/3}$,
and a corresponding growth of the energy density of the magnetic fields
($\propto B^2$). Initial fields as expected in stellar cores at the
onset of gravitational instability, 
i.e., several $10^9$\,G for the dominant toroidal 
component~\cite{HegerWoosleySpruit2005}, can thus grow by a factor
$>$1000 but cannot gain dynamically relevant strength, for which the
magnetic pressure has to reach a fair fraction of the gas pressure.

Therefore secondary amplification mechanisms are crucial to further
boost the magnetic energy density to values close to equipartition
with the stellar plasma. In the MRM such an increase in energy density
can be achieved by tapping the huge reservoir of rotational energy,
$E_\mathrm{rot} \propto J_\mathrm{core}^2/(M_\mathrm{core}R_\mathrm{core}^2)$,
that builds up at the expense of gravitational energy due to
angular momentum ($J_\mathrm{core}$) conservation during the infall.
The rotational energy in a rapidly spinning
PNS with average revolution period $P_\mathrm{ns}$ thus becomes
\begin{equation}
E_\mathrm{rot}\,\sim\, 2.4\times 10^{52}\,\mathrm{erg}\,
\left ( \frac{M_\mathrm{ns}}{1.5\,M_\odot} \right )
\left ( \frac{R_\mathrm{ns}}{10\,\mathrm{km}} \right )^{\! 2}
\left ( \frac{1\,\mathrm{ms}}{P_\mathrm{ns}} \right )^{\! 2} \,.
\label{eq:rotenergy}
\end{equation}
The amplification can either happen through the wrapping of
an (initially present or convectively created) poloidal field,
stretching it into a toroidal one, which leads to a linear increase
with the number of windings. Or it can take place
by exponential amplification with characteristic
timescale of order $4\pi|\mathrm{d}\Omega/\mathrm{d}\,\ln r|^{-1}$ 
(with $\Omega(r) = 2\pi/P_\mathrm{rot}(r)$ being the angular 
frequency for local spin period $P_\mathrm{rot}$) through the
magnetorotational instability (MRI; \cite{BalbusHawley1998,Akiyamaetal2003}),
whose growth conditions in SN cores were studied in detail
in~\cite{Obergaulingeretal2009}. Both processes require differential
rotation, which naturally develops during infall. Saturation fields 
of order
\begin{equation}
B^2\,\sim\, 4\pi \rho r^2 \Omega^2 
\left ( \frac{\mathrm{d}\ln\Omega}{\mathrm{d}\ln r}\right )^{\! 2} 
\label{eq:saturationfield}
\end{equation}
can be expected in an MRI-unstable environment, in which 
$\mathrm{d}\Omega/\mathrm{d}\,\ln r < 0$ must hold to enable
the growth of long-wavelength, slow-magnetosonic waves. For 
sufficiently large angular velocities, fields of order
$10^{15}$--$10^{16}$\,G were estimated~\cite{Akiyamaetal2003}.

The ejection of matter can be driven by magnetic pressure and
hoop stresses, magnetic buoyancy, or gas heating due to the
dissipation of rotational energy through turbulent magnetic
viscosity~\cite{Meieretal1976,Akiyamaetal2003,Thompsonetal2005}.
Well collimated, bipolar outflows or jets along the rotation axis 
with characteristic power 
\begin{equation}
\dot E_\mathrm{MHD}\,\sim\, 10^{52}\,\frac{\mathrm{erg}}{\mathrm{s}}\,
\left ( \frac{B}{10^{15}\,\mathrm{G}}\right )^{\! 2} 
\left ( \frac{r}{30\,\mathrm{km}} \right )^{\! 3} 
\left ( \frac{\Omega}{10^3\,\mathrm{rad}\,\mathrm{s}^{-1}} \right ) 
\label{eq:jetpower}
\end{equation}
may be 
generic~\cite{WheelerMeierWilson2002,Akiyamaetal2003,Burrowsetal2007B}.

Since the MRM can tap only the free energy of differential 
rotation in the PNS, $E_\mathrm{rot}^\mathrm{free} \ll E_\mathrm{rot}$,
angular velocities near the Keplerian rate
of the progenitor core ($P_\mathrm{core} \sim 1\,$s) are required
for magnetic fields to grow to dynamical significance. 
SN simulations~\cite{Thompsonetal2005,Burrowsetal2007B}
suggest that the spin period must be 
$P_\mathrm{core}\lesssim 2$--5\,s, leading to NSs rotation
periods of 
$P_\mathrm{ns} \sim (R_\mathrm{ns}/R_\mathrm{core})^2\,P_\mathrm{core}$
under the assumption of strict angular momentum conservation.
Present stellar evolution models that include angular momentum
loss through magnetic breaking, however, yield typical core-rotation
periods of $P_\mathrm{core}\gtrsim 100$\,s 
before collapse (cf.\ Sect.~\ref{sec:FeCSNe}). 
Such slowly rotating stellar cores are
consistent with observed spin rates of 
newly born white dwarfs~\cite{Charpinetetal2009}
and with the estimated spin periods of new-born pulsars of
$\sim$10\,ms~\cite{HegerWoosleySpruit2005,MeynetEggenbergerMaeder2011}, 
but they are much too slow to provide
the rotational energy reservoir for powering SNe 
through the MRM (see Eq.~\ref{eq:rotenergy}).

A variety of mechanisms have also been proposed for
magnetic field amplification in collapsing cores with
no or slow rotation,
e.g.\ by a convective dynamo~\cite{ThompsonDuncan1993},
turbulent SASI motions in the postshock region~\cite{Endeveetal2010}
or exponential steepening of Alfv{\'e}n waves created by fluid
perturbations at Alfv{\'e}n points in the accretion flow of the 
PNS~\cite{Guiletetal2011}. Moreover, Alfv{\'e}n waves emitted from 
the convective layer inside the PNS (thus extracting energy from the
rich reservoir of gravitational binding energy of
the contracting remnant) were suggested to provide extra
energy to the stalled SN shock by dissipative heating~\cite{Suzukietal2008}
similar to the heating of the solar corona by Alfv{\'e}n waves emerging
from the surface of the Sun. Recent 2D CC simulations with neutrino
transport~\cite{ObergaulingerJanka2011}, however, find that these
mechanisms are either inefficient or able to amplify the fields only 
locally. Large-scale fields with dynamical importance must reach 
magnetar strength (at least $10^{14}$--$10^{15}$\,G) but in the 
absence of magnetorotational processes seem to require pre-collapse
fields 100 times stronger than predicted by stellar evolution models.

Magnetic fields are therefore likely to play some role in all SN cores,
but at the moment it seems only certain that they are crucial for the
explosion of very rapidly spinning stars, which are probably
linked to long GRBs and HNe (Sect.~\ref{sec:GRBSNe}).
MHD mechanisms have the advantage of not being strongly coupled to
the mass in the gain layer and
the mass-accretion rate through the stalled shock, which determine
the explosion energy of SNe powered by neutrino
heating (Eq.~\ref{eq:totheating}). MHD-driven explosions
can therefore be considerably more energetic than neutrino-driven 
SNe, where blast-wave energies of $\sim$(1--2)$\times 10^{51}$\,erg
may be the upper limit (see Sect.~\ref{sec:NSBH}). 
Large global deformation and
well collimated jets must be expected to be generic to 
MHD explosions of very rapidly rotating stellar cores and seem
to be characteristic of most hyperenergetic SNe~Ib/c.

Reliable and predictive multi-dimensional simulations of such 
phenomena are hampered by the fact that the true nature of
MHD phenomena can only be treated in 3D, and such models should
also include reasonably realistic neutrino transport. Another
problem arises from the extreme dependence
of the dynamical evolution on the initial conditions, in
particular the rotation rate and profile of the stellar core
(e.g., \cite{Kotakeetal2004,Sawaietal2005,Obergaulingeretal2006})
and the initial strength and geometry of the magnetic field
(e.g.,~\cite{Moiseenkoetal2006} and refs.\ therein). Moreover,
a large variety of MHD instabilities, among them the MRI, demand 
high numerical resolution, which is particularly hard to achieve
in 3D models, and which adds to the computational demands that 
result from long evolution times on the one hand and severe
time-step constraints set by high Alfv{\'e}n speeds,
$v_\mathrm{A} = B/\sqrt{4\pi\rho}$, on the other. 
The exploration of magnetorotational explosions
will therefore remain a challenging task for the coming years.

\subsection{Acoustic Mechanism}

A new CCSN mechanism was envisioned 
in~\cite{Burrowsetal2006,Burrowsetal2007} based on results
of 2D hydrodynamic simulations, which did not yield explosions
by neutrino-energy deposition. 
At late times after bounce ($\gtrsim$1\,s), large-amplitude 
dipole ($\ell = 1$) gravity-mode oscillations of the PNS core were
found to be excited by SASI sloshing motions of the postshock 
layer and by anisotropic accretion downdrafts.
The PNS vibrations (with amplitudes of several km) were damped
by sending strong sound waves into the surrounding medium.
Running down the density gradient away from the PNS the waves
steepened into secondary shocks. The dissipation of the latter 
helped to heat the postshock region. Thus robust explosions were
obtained for a variety of progenitors. For the conversion
rate of accretion power, 
\begin{equation}
\dot E_\mathrm{acc}\,=\,
\frac{G M_\mathrm{ns}\dot M}{R_\mathrm{ns}}
\,\sim\, 1.3\times 10^{52}\,\frac{\mathrm{erg}}{\mathrm{s}}\,
\left (\frac{M_\mathrm{ns}}{1.5\,M_\odot}\right )\,
\left (\frac{\dot M}{0.1\,M_\odot\,\mathrm{s}^{-1}}\right )\,
\left (\frac{30\,\mathrm{km}}{R_\mathrm{ns}}\right )\, ,
\label{eq:accretionpower}
\end{equation}
into acoustic power, one can estimate~\cite{Burrowsetal2006,Burrowsetal2007C}:
\begin{eqnarray}
\dot E_\mathrm{sound}\! &\sim&\! \frac{\pi\rho}{2}
\left (g_\mathrm{ns}R_0\right )^{3/2}\,H_0^2  \cr
\! &\sim& \! 0.5\!\times\! 10^{51}\,
\frac{\mathrm{erg}}{\mathrm{s}}\,\,\rho_{11}\,
g_{\mathrm{ns},13}^{3/2}\,
\left (\frac{R_0}{10\,\mathrm{km}} \right )^{\! 3/2}\!
\left (\frac{H_0}{3\,\mathrm{km}}\right )^{\! 2} \,,
\label{eq:sonicpower}
\end{eqnarray}
(see also eq.~1 in \cite{Burrowsetal2006}).
Here, $R_0$ is the accretion-stream radius, $H_0$ the
wave height, $\rho_{11} = \rho/(10^{11}$\,g/cm$^3$) the 
average density at the ``surface'' of the PNS core, and 
$g_{\mathrm{ns},13} = g_\mathrm{ns}/(10^{13}$\,cm/s$^2$)
the average gravitational acceleration
($g_\mathrm{ns} = G M_\mathrm{ns}/R_\mathrm{ns}$)
at the PNS ``surface''. The reference value of
$\dot E_\mathrm{sound}$ of Eq.~(\ref{eq:sonicpower}) is 
suggestive. It exceeded the neutrino-energy deposition rate
($\sim\tau\,(L_{\nu_e}+L_{\bar\nu_e})$; 
Sect.~\ref{sec:heatingeffiency})
at late times in the numerical models. The violently vibrating
PNS thus acted as a transducer channelling accretion energy
efficiently into sound.

The fraction of the accretion power that is converted into
core g-mode activity of the PNS could not be extracted 
reliably from the numerical calculations 
of~\cite{Burrowsetal2006,Burrowsetal2007}, thus leaving the
value of $H_0$ uncertain. Also final numbers for the explosion
energies could not be determined, but the 2D explosions occurred
very late, implying large NS masses and tending to be
low-energetic. 
Fundamental questions about the excitation efficiency of the
large-amplitude, low-order g-modes in the PNS remain to be
answered, in particular whether 3D simulations would yield this
phenomenon as well. So far other groups have 
not been able to reproduce the results
(e.g.,~\cite{MarekJanka2009}), maybe because their models
either were not evolved to sufficiently late times or exploded
by neutrino heating before. A serious counterargument to the 
proposed scenario was made in~\cite{WeinbergQuataert2008}. 
Employing pertubation analysis the authors concluded
that non-linear coupling between the low-order
primary modes and pairs of high-order g-modes, whose small
wavelengths cannot be resolved
in hydrodynamical simulations, damps the 
low-order mode amplitudes to dynamically insignificant size.
The thermalized pulsational energy is lost by neutrino 
emission.

\subsection{Phase-Transition Mechanism}
\label{sec:phasetransitmechanism}

A first-order hadron-to-quark matter phase transition that 
occurs at a sufficiently low density can
have dynamical consequences during the postbounce accretion
phase of the collapsing stellar core. This was discovered 
by~\cite{Sagertetal2009,Fischeretal2011} using a hybrid EoS
with a mixed phase that was softer than the hadronic phase
and the pure quark phase. The latter was described by
suitable choices of the parameters in the MIT bag model for
strange (u,d,s) quark matter. Different from laboratory
(heavy-ion collision) conditions, where the proton fraction $Y_p$
is close to 0.5, the mixed phase appears at
subnuclear densities for SN matter with $Y_p \lesssim 0.3$,
and for all proton-to-baryon ratios shows a decrease 
of the transition density with higher temperatures.
This is in contrast to other models for the hadron-quark phase
transition, which predict an increase of the phase-transition
density with increasing temperature
(e.g., \cite{Prakashetal1995,Fischeretal2011B}).

The very special properties of the hybrid EoS lead to gravitational
instability of the PNS when it has accreted enough matter and has
heated up during its contraction, entering the transition to
quark matter in a growing dense-core region. The decrease of the
effective adiabatic index there below the critical value for 
stability triggers a second, supersonic implosion to the denser 
pure quark phase, where the EoS suddenly stiffens again. This leads 
to considerable release of gravitational binding energy and the
formation of a strong, second bounce shock, which catches up with
the stalled primary shock to cause a SN explosion even in 1D 
models. When the second shock breaks out of the neutrinospheres,
$e^+$ captures by neutrons in the shock-heated matter emit
a $\bar\nu_e$ burst that may be detectable for a Galactic
SN~\cite{Dasguptaetal2010}.

Though this is an interesting, new scenario, the fine tuning of
the QCD phase transition is problematic. In particular, all
EoS versions that lead to explosions so far are not compatible
with the 1.97$\pm$0.04\,$M_\odot$ NS mass limit of
PSR~J1614-2230~\cite{Demorestetal2010}. Changing the EoS parameters
to reduce this inconsistency leads to a larger radius of the
hybrid star and a less extreme density difference between hadronic
and pure quark phases~\cite{Fischeretal2011}. Whether SN explosions 
can be obtained with deconfinement scenarios compatible with 
PSR~J1614-2230 still needs to be shown.

\begin{figure}
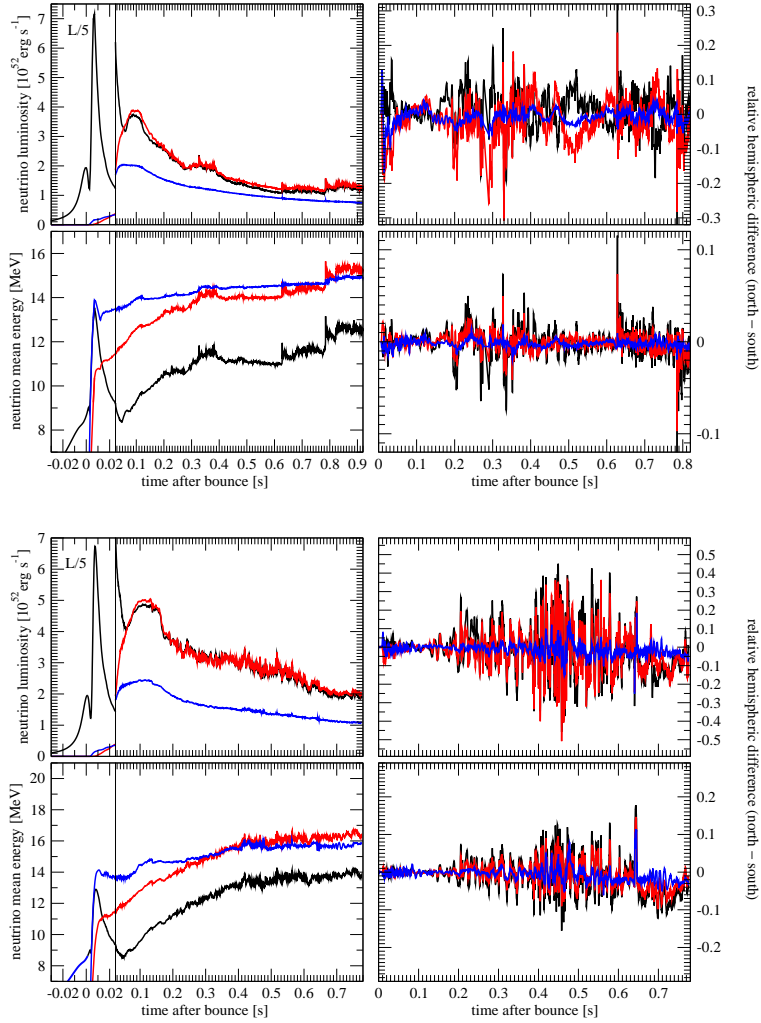

\centerline{\psfig{figure=JankaFig7a.eps,width=10truecm}}
\vspace{0.5truecm}
\centerline{\psfig{figure=JankaFig7b.eps,width=10truecm}}
\caption{{\small 
Neutrino signals from general relativistic 2D
simulations of core collapse and explosion of 11.2\,$M_\odot$
({\em upper plot}) and 15\,$M_\odot$ ({\em lower plot})
stars shown in Fig.~\ref{jankafig6}~\cite{Muelleretal2011}.
The {\em left panels} of each plot show luminosities (i.e.,
total neutrino-energy loss rates of the PNS; {\em upper panels})
and mean energies (defined by the ratio of total neutrino
energy-loss rate to number-loss rate, $\dot E_\nu/\dot N_\nu$; 
{\em lower panels})
with black lines for $\nu_e$, red for $\bar\nu_e$, and blue for
one kind of heavy-lepton neutrino $\nu_x$. The {\rm right panels}
display the corresponding relative hemispheric differences
after core bounce (the infall remains spherical).
All quantities are measured in the lab frame at large distance.
Note that the fluctuations, sudden jumps, and north-south 
differences at $t > 300$\,ms in the upper plot
are caused by violent, time-dependent, anisotropic downflows
and corresponding changes of the accretion rate of the PNS.}}
\label{jankafig7}
\end{figure}

\section{OBSERVABLE SIGNALS FROM THE SUPERNOVA CORE}
\label{sec:coresignals}

Neutrinos, gravitational waves, and
heavy-element formation in the neutrino-heated ejecta are
direct probes of the processes in the SN core. 
Because of the gain in sophistication of the models,
in particular in the neutrino transport and by the inclusion
of relativistic effects (recently also in 2D simulations),
and because of the growing understanding of hydrodynamic
instabilities during the postbounce accretion phase, 
interesting new aspects were discovered and are fit to
change our notion of some of the accompanying phenomena
and implications of CCSNe fundamentally.

\subsection{Neutrinos}
\label{sec:neutrinos}

Neutrinos and antineutrinos of all flavors radiated from the SN 
core (see production processes in Table~\ref{jankatab1})
carry information of the thermodynamic conditions
(temperature, degeneracy) there, but also reflect dynamical
processes associated with NS formation. A measurement of 
a neutrino signal from a future Galactic event could thus 
help to unravel the explosion mechanism. 

The shock-breakout burst of $\nu_e$ is a well known 
example of such a dynamical feature. It emerges when
a flood of neutrinos freshly produced in shock-heated
matter can suddenly escape when the bounce-shock 
reaches the neutrino-transparent regime at sufficiently low 
densities. Another example, though more on the exotic
side, is the neutrino flash connected to the
hadron-quark phase transition described in 
Sect.~\ref{sec:phasetransitmechanism}. Also the pronounced
rise of the mean energy of heavy-lepton neutrinos, $\nu_x$,
when a PNS approaches collapse to a 
BH~\cite{Sumiyoshietal2008,Fischeretal2009}
and possible ---so far unclarified--- signatures of 
magnetohydrodynamics can be mentioned here.

Moreover, the
large-amplitude radial oscillations~\cite{Burasetal2006A}
or low-multipole (dipolar, quadrupolar) nonradial
oscillations of the postshock layer (due to the SASI 
or due to convective activity for higher-multipole 
asymmetries) lead to quasi-periodic variations
of the accretion flow to the PNS and to corresponding 
fluctuations of the accretion luminosity and mean
energies of radiated neutrinos~\cite{MarekJankaMueller2009}. 
The effect is particularly strong for $\nu_e$ and
$\bar\nu_e$, for which a significant part of the
luminosity (amounting to a fair fraction of the
progenitor-specific
accretion power, Eq.~\ref{eq:accretionpower}) originates
from $e^\pm$~captures on free nucleons in the hot accretion 
layer. In Fig.~\ref{jankafig7} these fluctuations are
visible at $t \gtrsim 100$\,ms p.b. 
Note that the accretion luminosities depend strongly 
on the progenitor, and in both of the 11.2 and 15\,$M_\odot$
models, accretion continues until the end of the 
simulated evolution simultaneously with the accelerating 
expansion of the shock (Fig.~\ref{jankafig6}) and with the 
development of positive ejecta energy. Ongoing accretion is 
obvious because the $\nu_e$ and $\bar\nu_e$ luminosities are
still considerably higher than those of $\nu_x$, whereas
after accretion has ceased all luminosities 
become nearly equal (Fig.~\ref{jankafig5};
\cite{Huedepohletal2010,Fischeretal2010,MarekJanka2009}).
The ``luminosities'' in the left panels of Fig.~\ref{jankafig7}
are energy-loss rates, $\dot E_{\nu}$, of the PNS
($\nu\in \{\nu_e,\bar\nu_e,\nu_\mu,\bar\nu_\mu,\nu_\tau,
\bar\nu_\tau\}$) and not observable fluxes, the variation
amplitudes are therefore damped by the integration over all 
directions, and their true magnitude (percents to tens of percents) 
can be better read off the right
panel, where relative hemispheric differences are displayed.

The SASI and convective modulation of the neutrino
emission was not only seen in 2D simulations with RbR
neutrino transport, but also with multi-angle treatment
(\cite{Brandtetal2011}; cf.~Sect.~\ref{sec:neutrinotransport}
for a discussion of methods). It may be detectable
for a Galactic SN at a fiducial distance of 10\,kpc
with IceCube or future megaton-class 
instruments~\cite{Lundetal2010}. First 3D simulations with
approximative neutrino transport suggest that the 
variation amplitudes could be smaller than in  
2D~\cite{MuellerJankaWongwathanarat2011},
but more and better 3D models with multi-group transport
are needed for reliable information.

Another remarkable property of the neutrino signals in
Figs.~\ref{jankafig5} and \ref{jankafig7} is the close 
similarity and even crossing of the mean energies of
$\bar\nu_e$ and $\nu_x$~\cite{MarekJankaMueller2009,Huedepohletal2010}.
Instead of the previous notion that $\nu_x$ are
significantly more energetic than $\nu_e$ and $\bar\nu_e$,
i.e., instead of a neutrino-energy hierarchy of 
$\langle\epsilon_{\nu_e}\rangle 
<\langle\epsilon_{\bar\nu_e}\rangle <\langle\epsilon_{\nu_x}\rangle$
with typically 
$\langle\epsilon_{\nu_x}\rangle \gtrsim 18$--20\,MeV,
state-of-the-art models now yield
$\langle\epsilon_{\nu_e}\rangle <
\langle\epsilon_{\bar\nu_e}\rangle \sim \langle\epsilon_{\nu_x}\rangle$
and $\langle\epsilon_{\nu_x}\rangle\lesssim 13$--16\,MeV
(Figs.~\ref{jankafig5}, \ref{jankafig7}, and
ref.~\cite{MarekJankaMueller2009}; also ref.~\cite{Lentzetal2011},
where rms energies are given, however).
The exact value depends on the time and EoS: A softer EoS
lead to a more compact and hotter PNS and thus higher mean
energies~\cite{MarekJankaMueller2009}.

While during the later accretion phase the order of 
$\langle\epsilon_{\bar\nu_e}\rangle$ 
and $\langle\epsilon_{\nu_x}\rangle$ can be reversed
(Fig.~\ref{jankafig7}), one obtains a mild hierarchy 
$\langle\epsilon_{\bar\nu_e}\rangle < \langle\epsilon_{\nu_x}\rangle$
during the PNS cooling phase (Fig.~\ref{jankafig5}), which, however,
diminishes to insignificant differences at very late times (seconds
after bounce), where the mean energies of all neutrino kinds
become nearly identical,
$\langle\epsilon_{\nu_e}\rangle \approx
\langle\epsilon_{\bar\nu_e}\rangle\approx\langle\epsilon_{\nu_x}\rangle$
(Fig.~\ref{jankafig5}, \cite{Huedepohletal2010,Fischeretal2011C}).
The late behavior can be understood by the flat temperature 
profile inside the PNS during the late cooling stage and the close 
proximity of the neutrinospheric positions of all neutrinos
then. This proximity is caused by the strong dominance of neutral-current
scatterings in the effective opacity (i.e., inverse mean free path)
for thermal coupling between neutrinos and stellar medium, 
$\kappa_\mathrm{eff} \equiv
\sqrt{\kappa_\mathrm{e}(\kappa_\mathrm{s}+\kappa_\mathrm{e})}$
($\kappa_\mathrm{e}$ and $\kappa_\mathrm{s}$ being the opacities
for neutrino-production processes and nucleon scatterings, 
respectively) at conditions where $e^-$ are strongly degenerate
and neutrons start to become 
degenerate~\cite{Fischeretal2011C}\footnote{For the influence
of the EoS-specific nucleon potential energies in dense NS matter,
which affect the $\beta$-processes of $\nu_e$ and $\bar\nu_e$ but
were ignored in the models of Figs.~\ref{jankafig5} and \ref{jankafig7},
see refs.~\cite{Roberts2012,RobertsReddy2012,MartinezPinedoetal2012}.}.
The close similarity of $\langle\epsilon_{\nu_x}\rangle$ and
$\langle\epsilon_{\bar\nu_e}\rangle$ during the early postbounce
and accretion phases is fostered by nucleon-nucleon-bremsstrahlung
as main $\nu_x$ production channel~\cite{Thompsonetal2000}, 
because it shifts the energy-sphere of $\nu_x$ to lower
temperature~\cite{Keiletal2003}. However, the effect
is considerably enhanced (compare the two cases discussed
in~\cite{Huedepohletal2010}) by energy losses of $\nu_x$ in
collisions with free nucleons $N = n,\,p$ (``inelastic''
---better ``non-conservative''--- nucleon recoil;
Table~\ref{jankatab1}). Such losses occur when the neutrinos
diffuse out through the (optically) thick scattering layer
between energy- and transport-spheres~\cite{Raffelt2001,Keiletal2003}.
The small
but very frequent energy transfers with an average value per
collision of $\langle\Delta\epsilon_\nu\rangle_{\nu N} \sim 
\epsilon_\nu(6\,k_\mathrm{B}T-\epsilon_\nu)/(m_Nc^2)$~\cite{Tubbs1979}
can force the $\nu_x$-spectrum to become even softer than that
of $\bar\nu_e$, whose production in a hot accretion layer by
$e^+$ captures on neutrons is very efficient.
  
The close similarity of the spectra of all neutrinos
and the corresponding relevance of non-conservative 
nucleon recoils, which is still widely ignored, 
have a bearing on the consequences of neutrino-flavor
conversions, e.g., the rise time of the detectable 
$\bar\nu_e$ signal~\cite{Chakrabortyetal2011C}, and on
neutrino-induced or flavor-oscillation dependent
r-process nucleosynthesis in 
SNe~\cite{Banerjeeetal2011,Duanetal2011}.
Also the steep rise of $\langle\epsilon_{\nu_x}\rangle$ 
before BH formation, which was found without taking 
non-conservative nucleon recoils into 
account~\cite{Sumiyoshietal2008,Fischeretal2009}, 
may be affected.

\begin{figure}
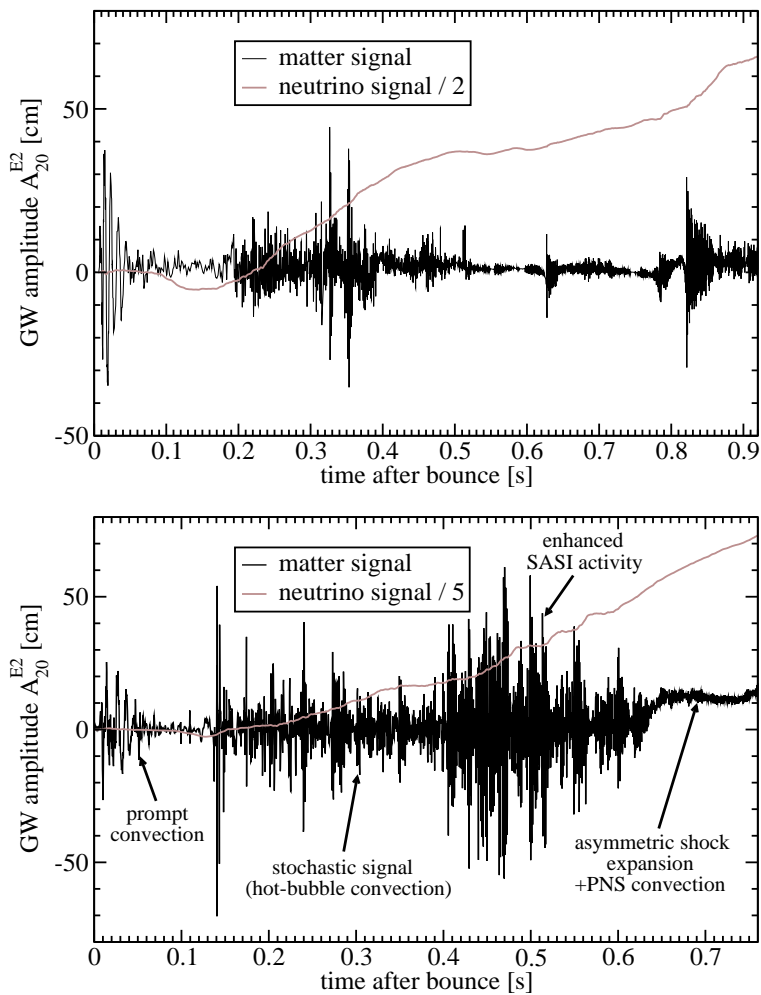

\centerline{\psfig{figure=JankaFig8a.eps,width=10truecm}}
\vspace{0.3truecm}
\centerline{\psfig{figure=JankaFig8b.eps,width=10truecm}}
\caption{{\small 
Amplitudes of gravitational waves (GWs) from the general
relativistic 2D simulations of core collapse and explosion
of 11.2\,$M_\odot$ ({\em upper plot}) and 15\,$M_\odot$
({\em lower plot}) stars shown in
Fig.~\ref{jankafig6}~\cite{Muelleretal2011}.
The light brown lines (scaled down by factors of two and
five in the upper and lower panel, respectively) display the 
growing amplitude connected with the asymmetric neutrino
emission. The matter signal (solid black line)
exhibits activity phases associated with
strong, prompt postbounce convection
(for $t_\mathrm{pb} \lesssim 50$\,ms), increasingly
violent convective and SASI mass motions in the postshock layer
before the explosion sets in (between $\sim$100\,ms and
350\,ms (500\,ms) in the 11.2\,$M_\odot$ (15\,$M_\odot$) case),
and the continued impact of asymmetric
accretion downdrafts on the PNS after the launch of the
explosion. The non-zero value of the matter signal at late
times is a consequence of the aspherical expansion
of the shocked ejecta.}}
\label{jankafig8}
\end{figure}

\subsection{Gravitational Waves}

Any nonspherical, accelerated mass motions in the SN core 
act as source of GWs, whose amplitude at a source distance
$D$ scales with the second derivative of the mass-quadruple moment,
$h\sim (2G/c^4)(\ddot Q/D)$. The GW signal reflects the activity
phases partly visible also in the neutrino-luminosity 
variations (compare Figs.~\ref{jankafig7} and \ref{jankafig8}).
Convective overturn caused by negative entropy gradients 
in the deceleration region of the bounce shock and in the 
neutrino-heating layer, the violent SASI sloshing of the whole
postshock volume, and the impact of accretion downdrafts, which
occur even after the onset of the explosion, induce
surface g-modes in the outer layers
of the PNS~\cite{MarekJankaMueller2009}, whose frequency determines 
that of the GW emission~\cite{MurphyOttBurrows2009}. Because the
buoyancy (Brunt-V\"ais\"al\"a) frequency connected to the gravity 
waves depends on the compactness of the PNS (see text following 
Eq.~\ref{eq:ledoux}), not only the stiffness of the EoS has a big
influence~\cite{MarekJankaMueller2009} but also GR gravity is
crucial to predict the GW spectrum, which for relativistic 
models peaks around 800--1000\,Hz while Newtonian simulations
yield significantly lower frequencies~\cite{Muelleretal2011}.

GWs are expected to carry characteristic signatures of
the explosion mechanism. While wave components associated with 
long-lasting convective and SASI activity and a broad-band
signal seem typical of
neutrino-driven explosions, the large-amplitude g-mode
oscillations of the PNS core, which are essential for the
acoustic mechanism, would lead to a dramatic rise of the GW
amplitude shortly before the blast sets in~\cite{Ottetal2006}. 
In contrast, rapid stellar core rotation
as required for MHD explosions would produce a powerful
GW burst at core bounce, possibly followed by postbounce
emission due to triaxial instabilities~\cite{Ott2009}.

The GW features and activity phases seen in 2D models are
also found in 3D simulations, though the amplitudes may be 
different. Without symmetry constraints, however, the 
detailed signal structure varies strongly with the 
observer direction and does not possess any template
character~\cite{Kotakeetal2009,Scheideggeretal2010,Muelleretal2011,Kotake2011}.

\subsection{Heavy Elements}
\label{sec:heavyelements}

Besides reprocessing shock-heated stellar layers
by explosive burning, nucleosynthesis takes place in the 
neutrino-heated ejecta from the close NS vicinity. The ejecta
consist of two components and have great
potential for diagnostics of the SN mechanism.

The first component consists of the early ejecta
from the phase of shock revival. Its composition at freeze-out
depends on the expansion timescale, which is intrinsically linked
to the blast dynamics and thus to the explosion mechanism,
but it also depends on the neutron-to-proton ratio set by the
competition of $e^\pm$~captures on nucleons and the inverse
$\nu_e$ and $\bar\nu_e$ captures (top two beta-processes in
Table~\ref{jankatab1}). 

A good example for the relevance of these effects are recent
2D results of ONeMg-core explosions, where acceleration by 
convective buoyancy expulses early ejecta so rapidly that this
material is able to retain a neutron excess sufficient for weak
r-processing, in contrast to 1D models where beta-reactions
in the more slowly ejected plasma lift $Y_e$ close to 0.5 and
above~\cite{WanajoJankaMueller2011}. It will be interesting to
explore the combination of composition and asymmetry 
differences of the early ejecta in magnetorotational
explosions compared to neutrino-driven ones when eventually
self-consistent, well resolved multi-dimensional MHD models
with sophisticated neutrino transport (instead of no or highly
simplified neutrino treatment) become available.

The second component is the neutrino-driven wind blown off the
surface of the hot PNS by neutrino-energy deposition above the 
neutrinosphere after the explosion has been launched. The 
properties of this ---in the absence of 
rotation and/or strong magnetic fields--- essentially 
spherically symmetric outflow depend on the strong
gravity field of the NS and on the properties (luminosities 
and spectra) of the radiated neutrinos, which determine the
strength of the heating~\cite{QianWoosley1996,Thompsonetal2001}.
Again the beta-processes of free nucleons (Table~\ref{jankatab1})
set the $n/p$ ratio of this environment. For sufficiently high
entropy and sufficiently large neutron excess this wind may
provide an interesting site for r-process 
nucleosynthesis~\cite{Hoffmanetal1997}.

However, besides the long-standing problem of insufficient 
entropy~\cite{Takahashietal1994,Robertsetal2010},
sophisticated hydrodynamic models find the wind to be
proton-rich~\cite{Huedepohletal2010,Fischeretal2010}. This 
is a consequence of the close similarity of the spectra
and luminosities of $\nu_e$ and $\bar\nu_e$ during the
PNS cooling phase, $L_{\bar\nu_e}\approx L_{\nu_e}$
and $\langle\epsilon_{\bar\nu_e}\rangle\approx 
\langle\epsilon_{\nu_e}\rangle$
(Fig.~\ref{jankafig5} and Sect.~\ref{sec:neutrinos}),
which enforces $Y_e > 0.5$. Since~\cite{QianWoosley1996} 
\begin{equation}
Y_e \sim \left [ 1 + 
\frac{L_{\bar\nu_e}(\varepsilon_{\bar\nu_e}-2\Delta)}{L_{\nu_e}
(\varepsilon_{\nu_e}+2\Delta)}\right ]^{-1}
\label{eq:windye}
\end{equation}
with $\varepsilon_\nu = 
\langle\epsilon_\nu^2\rangle/\langle\epsilon_\nu\rangle$
and $\Delta = (m_n-m_p)c^2 \approx 1.29$\,MeV, values of
$Y_e < 0.5$ require considerably more energetic 
$\bar\nu_e$ than $\nu_e$, satisfying $\varepsilon_{\bar\nu_e} -
\varepsilon_{\nu_e} > 4\Delta$. Recently it was shown that the
nucleon potential energies in dense NS matter, connected to the
nuclear symmetry energy, may cause sufficiently large spectral
differences of $\nu_e$ and $\bar\nu_e$ to bring the wind $Y_e$
slightly below 
0.5~\cite{Roberts2012,RobertsReddy2012,MartinezPinedoetal2012}.
But it still needs to be seen whether this reduction allows for 
an r-process.

A dominance of protons prevents r-processing
but might enable a
$\nu$$p$-process \cite{Froehlichetal2006,Pruetetal2006}.
Active-sterile $\nu_e$-$\nu_\mathrm{s}$ conversions involving 
a possible sterile neutrino $\nu_\mathrm{s}$ in the eV-mass
range, as suggested by an anomaly of reactor $\bar\nu_e$
spectra and their distance and energy variation, can 
decrease the proton excess by removing $\nu_e$ and thus 
suppressing their absorption on neutrons. A recent 
investigation based on an ECSN model, however, revealed 
only a modest effect, insufficient for an
r-process~\cite{Tamborraetal2012}.
But the results depend in a complex way on the interplay
between $\nu_e$-$\nu_\mathrm{s}$ MSW matter oscillations
and collective $\nu\bar\nu$ flavor conversion, which 
strongly reduces the pure matter effect. More exploration
seems necessary.

\begin{figure}
\centerline{\psfig{figure=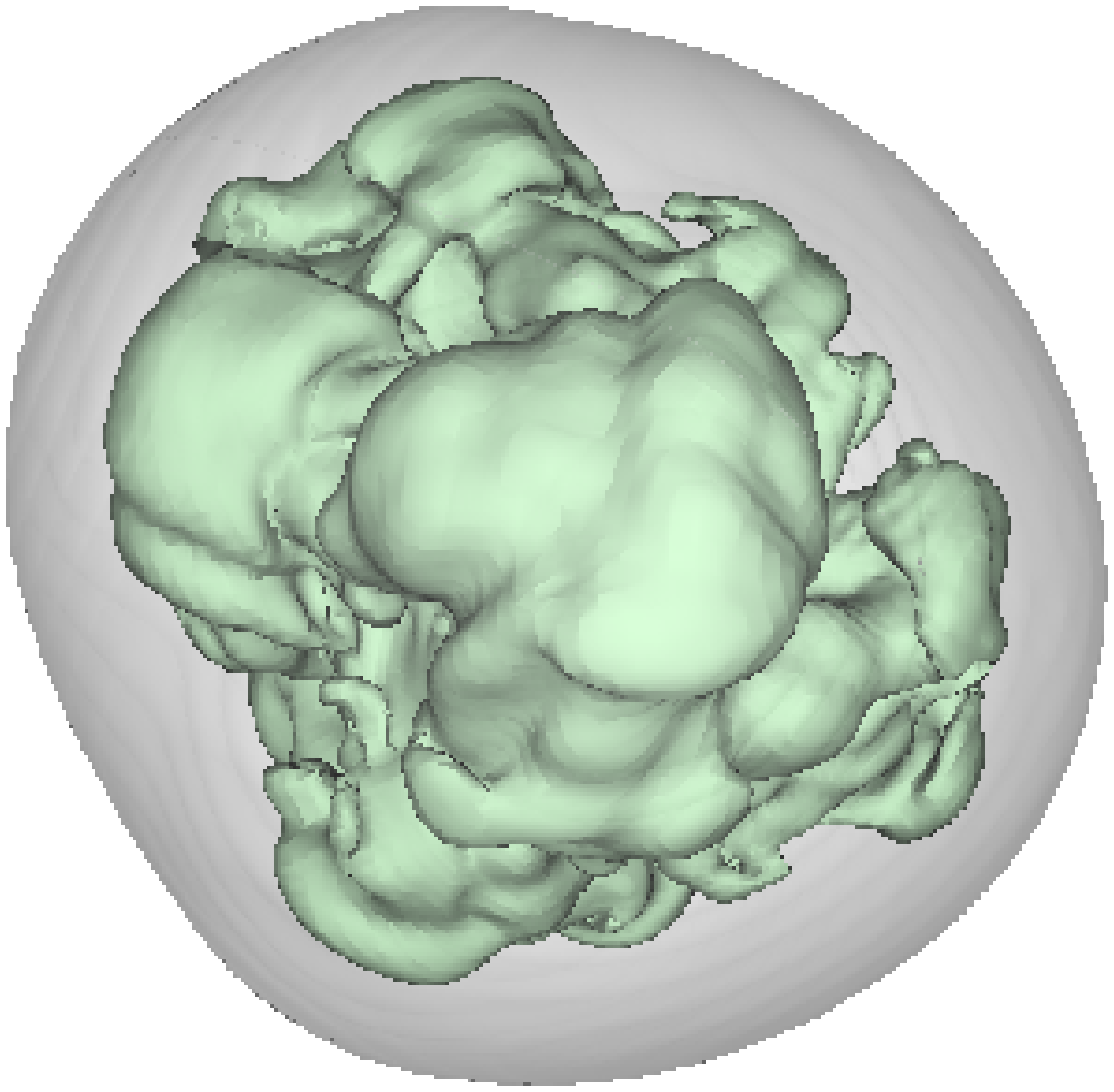,width=4.5truecm}\hspace{0.6truecm}
            \psfig{figure=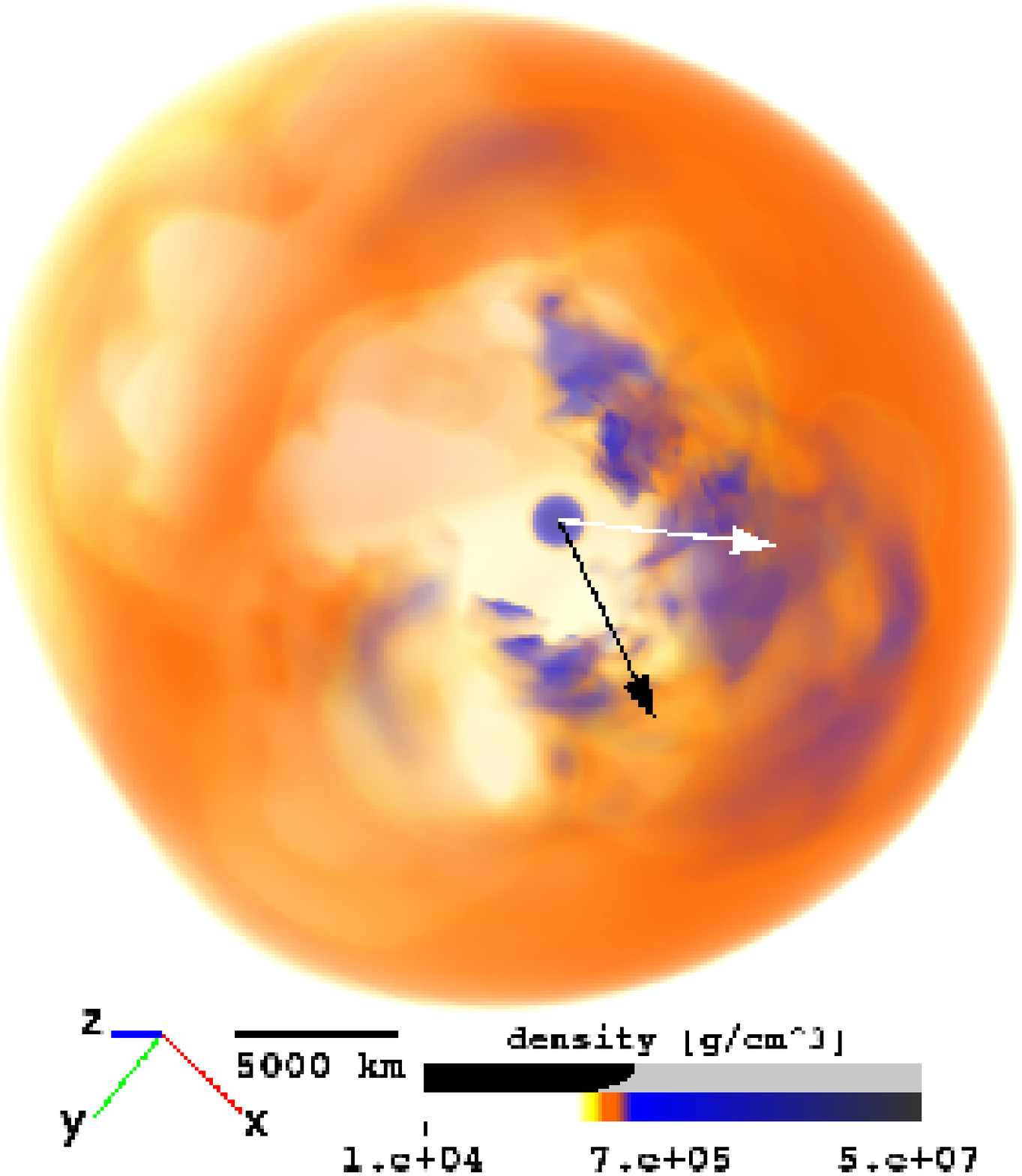,width=4.5truecm}\hspace{0.6truecm}
            \psfig{figure=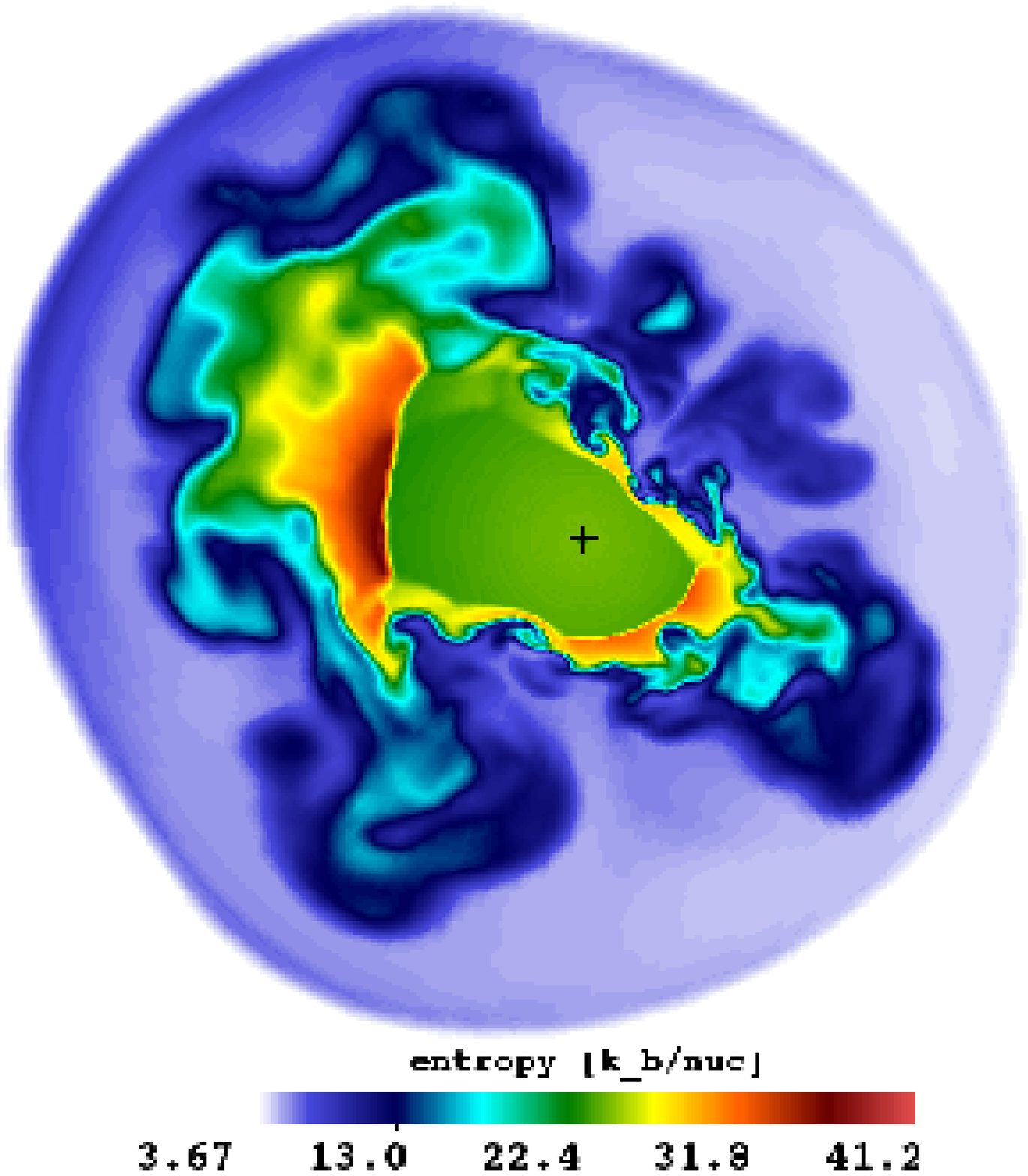,width=4.5truecm}}
\vspace{0.5truecm}
\centerline{\psfig{figure=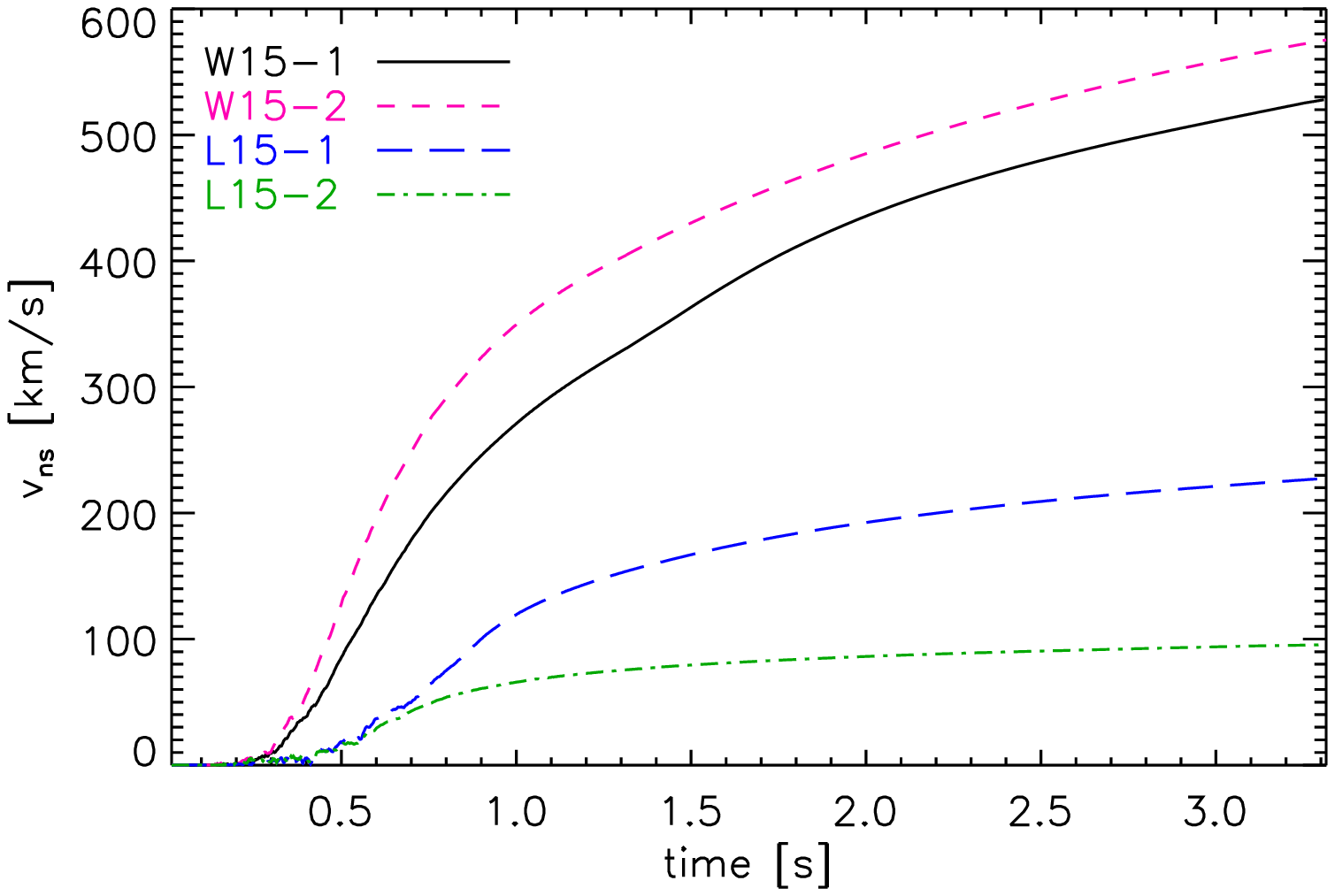,width=7truecm}\hspace{0.6truecm}
            \psfig{figure=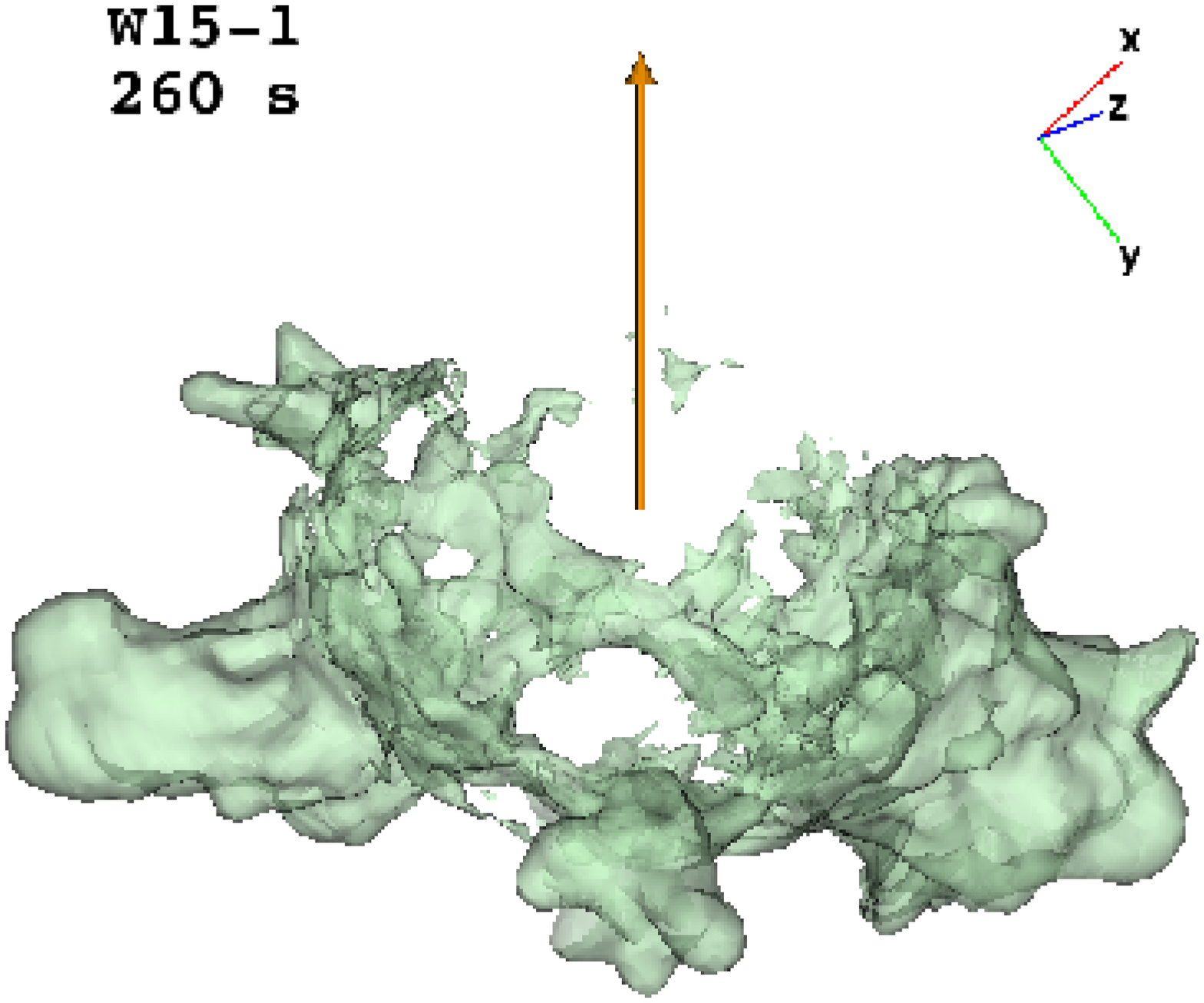,width=7truecm}}
\caption{{\small 
NS kicks and anisotropic Ni ejection for asymmetric SN explosions in 3D
simulations~\cite{Wongwathanaratetal2010,Wongwathanaratetal2012}.
{\em Upper panels:} Entropy isosurfaces of SN shock and convective
bubbles ({\em left}) and ray-casting image of the density ({\em middle})
at $t=1.3$\,s after bounce. The deformed boundary
is the shock, the viewing direction is normal to the 
plane of NS kick and spin vectors (white and black arrows), which 
define the plane of the entropy distribution ({\em right}). 
The NS (black cross) is clearly displaced from the geometrical 
center of the expanding shock towards the side of weaker explosion.
It is accelerated mainly by the asymmetric gravitational
attraction of less rapidly expanding, dense ejecta clumps 
(intense reddish and bluish in middle image).
{\em Left lower panel:} Recoil velocity of the NS vs.\ time
for four 3D explosion simulations of different stars. The acceleration
continues even later than 3\,s and kicks of $>$600\,km/s are reached.
{\em Right lower panel:} Anisotropic production of radioactive
$^{56}$Ni by explosive nuclear burning behind the expanding shock.
For large NS kicks nickel is ejected preferentially in the direction
where the shock is stronger, i.e., opposite to the NS motion
(red arrow).}}
\label{jankafig9}
\end{figure}

\begin{figure}
\centerline{\psfig{figure=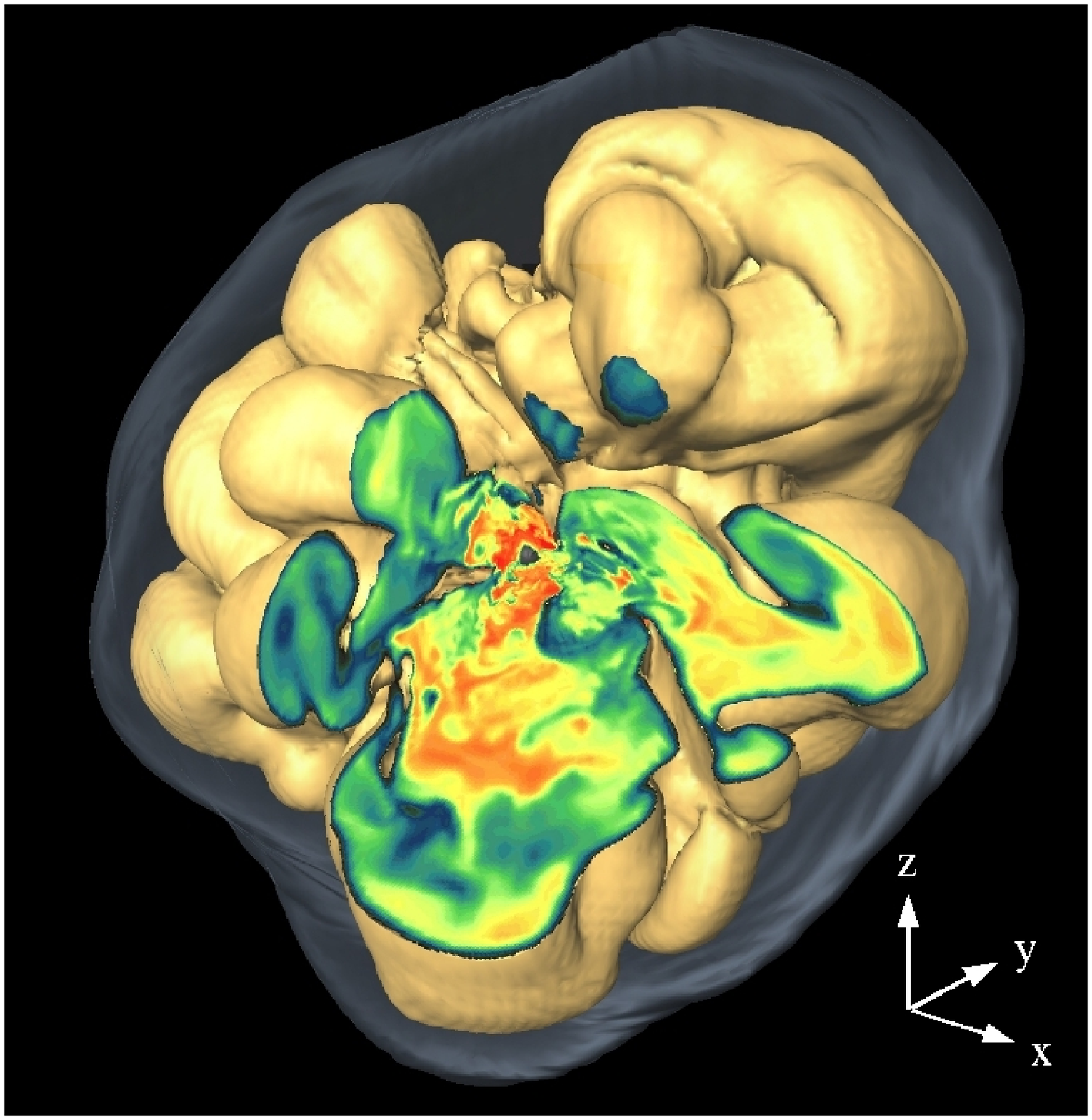,height=7truecm}
            \psfig{figure=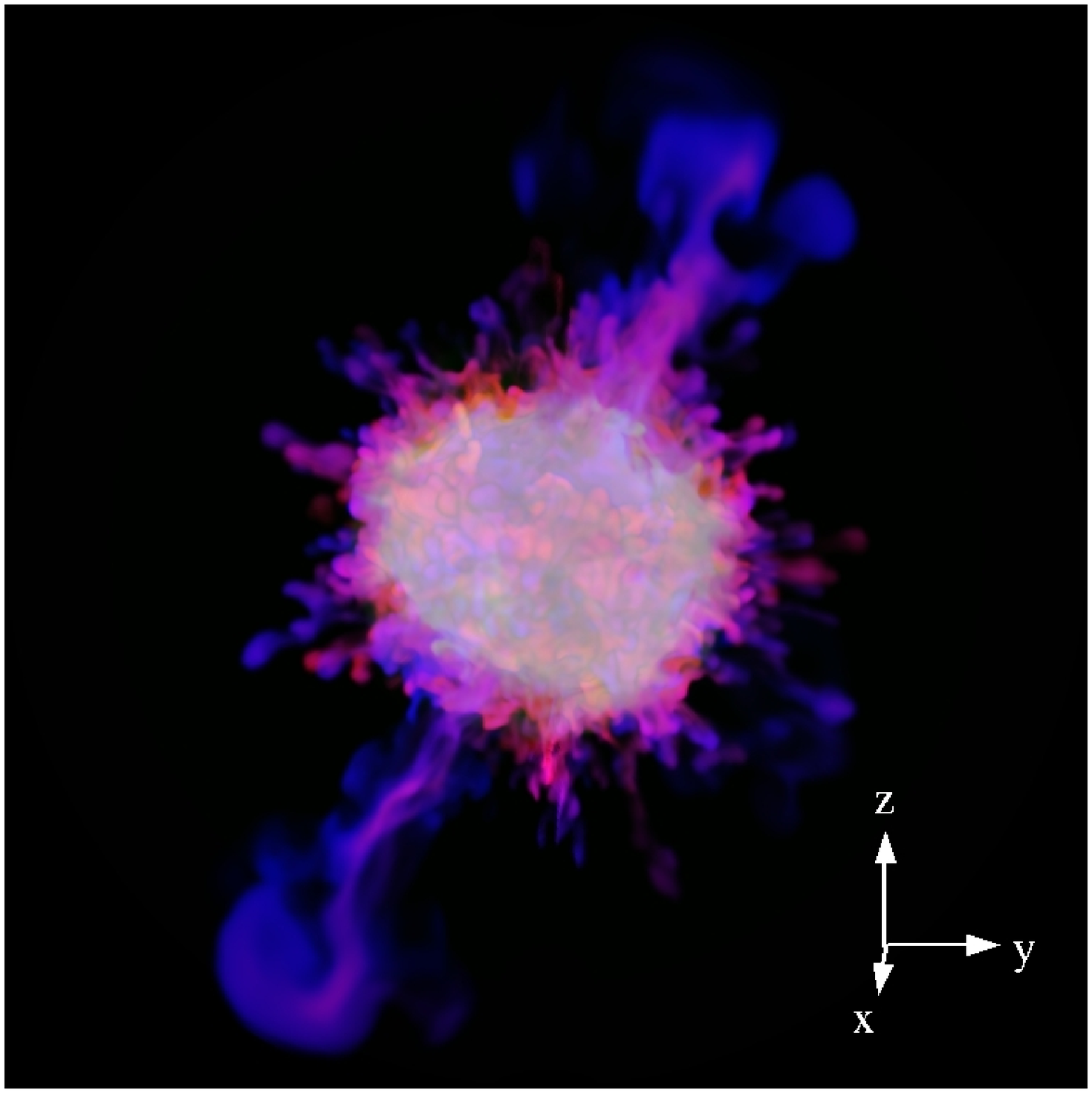,height=7truecm}}
\caption{{\small 
{\em Left panel:} Asymmetric shock front (outer bluish, nearly
transparent surface) and mushroom-like, high-entropy bubbles of 
neutrino-heated plasma around the central NS (dark grey surface near
the middle) at 0.5\,s after bounce in a 3D SN 
simulation~\cite{HammerJankaMueller2010}. The shock front has a 
diameter of $\sim$4000\,km.
An octant is cut out to show the entropy distribution (color-coded
between $\sim$10 and 21\,$k_\mathrm{B}$ from blue to yellow to red) 
in the expanding Rayleigh-Taylor mushrooms surrounded by cooler
accretion downdrafts. All visible structures have grown from 
tiny, random seed perturbations by hydrodynamic instabilities.
{\em Right panel:} Asymmetric ejection of different chemical elements 
during the explosion of the left image, but $\sim$9000\,s later into 
the SN evolution~\cite{HammerJankaMueller2010}. 
The side length of the displayed volume is about $7.5\times 10^7\,$km.
The largest bubbles on the left image have seeded the growth of the
most prominent Rayleigh-Taylor fingers in the right picture, which 
expand with up to 4500\,km/s. They are surrounded by the helium and
hydrogen of the outer stellar shells (not visible). Together with the
smaller features they thus 
carry heavier chemical elements from deep stellar layers far into
more slowly expanding, lighter SN material. Blue filaments contain
dominantly nickel, red fingers mostly oxygen, and green is associated
with carbon. A mix of nickel and oxygen appears in pink. 
The whitish glow results from a contamination with other colors as a 
consequence of the volume rendering for the visualization.}}
\label{jankafig10}
\end{figure}

\begin{figure}
\centerline{\psfig{figure=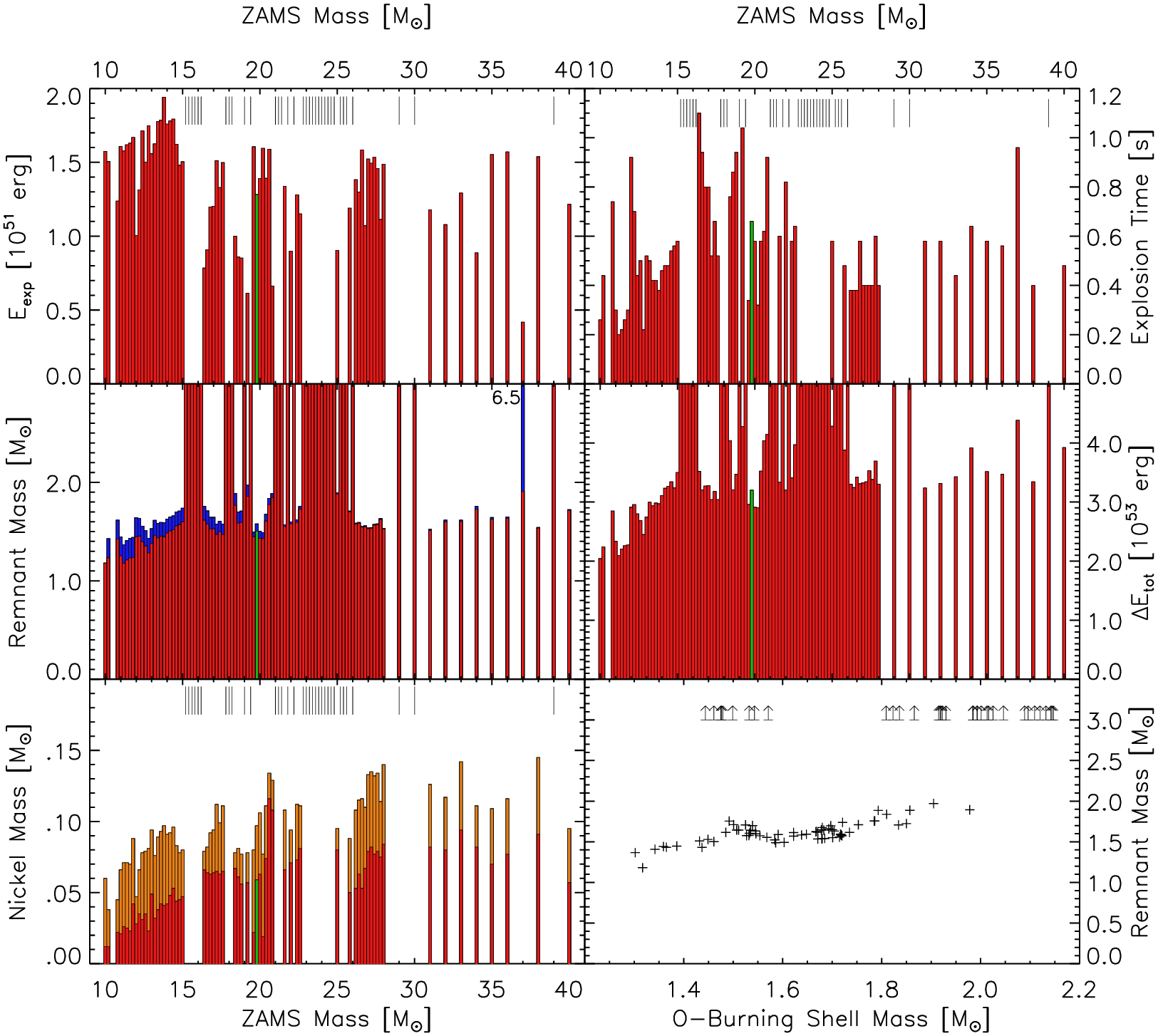,width=12truecm}}
\caption{{\small 
Explosion and remnant properties predicted
by parametrized 1D neutrino-driven SN 
simulations~\cite{Uglianoetal2012} of a large set of 
progenitor stars from~\cite{Woosleyetal2002}. Explosion
energy ({\em top left}), time of onset ({\em top right}),
baryonic remnant mass ({\em middle left}), neutrino-energy
release by the compact remnant ({\em middle right}), and
ejected Ni mass ({\em bottom left}) are shown as functions
of stellar birth (ZAMS) mass. The lower right plot gives
the compact remnant mass vs.\ the enclosed mass at the base
of the O-burning shell of the
progenitor. Neutrino cooling of the dense NS core was 
prescribed such that the properties of SN~1987A were
roughly reproduced for $\sim$20\,$M_\odot$ stars
(green histogram bar). Accretion neutrino luminosity was
self-consistently computed by approximate neutrino 
transport. The ticks in some panels mark masses
where computed models did not explode. Bars of remnant
masses reaching to the upper panel edge 
(3\,$M_\odot$) and arrows in the bottom right panel
signal formation of a BH containing the whole mass of 
the star at collapse. The only exception is the 
37\,$M_\odot$ progenitor, where the 
explosion ejects $\sim$3.2\,$M_\odot$ while fallback
creates a BH with 6.5\,$M_\odot$. Blue segments
indicate fallback masses and orange segments Ni-mass
uncertainties because of unclear Ni abundance in the
$\nu$-heated ejecta.
}}
\label{jankafig11}
\end{figure}

\section{EXPLOSION PROPERTIES AND COMPACT REMNANTS}
\label{sec:remnants}

The explosion mechanism establishes the link between
progenitors stars and SNe and their remnants. It is
therefore natural to seek for observables that might
provide indirect evidence of the processes triggering
the onset of the blast. In the following the implications
and limitations of neutrino-driven explosions will be
briefly discussed with respect to pulsar kicks, SN
asymmetries, and progenitor-dependent explosion and
remnant properties.

\subsection{Pulsar Kicks and Spins}
\label{sec:pulsarkicks}

Young pulsars are observed to possess average space velocities
around 400\,km/s, some of them even more than 
1000\,km/s~\cite{Hobbsetal2005}. This is too high to be understood
by the breakup of binary systems in SN explosions, and natal 
kicks of the NSs are required~\cite{Laietal2001}.

An asymmetric initiation of the explosion can naturally impart
a recoil velocity to the NS due to linear momentum conservation.
Impulsive momentum transfer by hydrodynamic forces of anisotropically 
expelled gas during the convective launch phase of the explosion, 
however, is hardly able to account for more than $\sim$200\,km/s
even in the most optimistic case~\cite{JankaMueller1994}.
Therefore a strong unipolar asymmetry in the progenitor star
prior to collapse ---in contrast to the higher-multipole
asymmetries usually resulting
from the stochastic and chaotic growth of hydrodynamical 
instabilities seeded by small, random perturbations 
(Sect.~\ref{sec:nonsphericaleffects})---
has been hypothesized to define a preferred direction
in which the SN blast could break out with the highest
velocities~\cite{BurrowsHayes1996,ArnettMeakin2011}. 
However, stellar
evolution models, self-consistently evolved in 3D through all
advanced burning stages up to gravitational instability, which
could lend convincing support to the existence of such global, 
low-multipole pre-collapse asymmetries, do not exist yet.

Anisotropic neutrino emission is another potential NS kick mechanism
by which the NS could achieve a recoil velocity of
$v_\mathrm{ns} \sim \xi_\nu\cdot 33000\,\mathrm{km/s}\,
(E_\nu/3\times 10^{53}\,\mathrm{erg})(M_\mathrm{ns}/1.5\,M_\odot)^{-1}$.
An asymmetry $\xi_\nu$ of 1\% of the total neutrino energy loss
would thus kick the NS to more than 300\,km/s. The asymmetric 
neutrino emission associated both with postbounce 
accretion~\cite{Schecketal2006,Wongwathanaratetal2010} and
with the convective activity during the PNS cooling 
evolution~\cite{Janka2001B}, however, is highly time variable and
nonstationary in space and time; therefore, it can hardly account 
for recoil velocities of more than 10\,km/s.
Exotic mechanisms, invoking ultrastrong NS magnetic dipole fields
and non-standard, still uncertain neutrino properties have therefore 
been suggested as speculative solution
(e.g.,~\cite{Kusenkoetal2008} and refs.\ therein).

Probably the most plausible origin of the NS velocities was
proposed on the basis of 2D explosion simulations by Scheck et 
al.~\cite{Schecketal2004,Schecketal2006}, 
whose results received recent confirmation
by 2D~\cite{Nordhausetal2010B,Nordhausetal2011} and 3D
models~\cite{Wongwathanaratetal2010,Wongwathanaratetal2012}.
Scheck et al.\ showed that the asymmetric expulsion of gas does 
not only exert ``contact forces'' during a few 100\,ms, 
in which the explosion is lauched and ejecta and PNS interact
hydrodynamically, but that the asymmetric ejecta
exert a long-time, anisotropic gravitational pull, which can 
accelerate the PNS over seconds to velocities of many 100\,km/s
(Fig.~\ref{jankafig9}). For particularly large asphericity
of the ejecta a NS velocity of
$v_\mathrm{ns} > 1000$\,km/s was obtained~\cite{Schecketal2006}.
A hemispheric asymmetry of the mass distribution of only
$\Delta m = \pm 10^{-3}\,M_\odot$ in a shell expanding away from
the NS from an initial radius $r_\mathrm{i} = 100$\,km with
$v_\mathrm{s} = 3000$\,km\,s$^{-1}$ can tug the NS to a velocity
of $v_\mathrm{ns} \approx 2G\Delta m/(r_\mathrm{i}v_\mathrm{s})
\approx 900$\,km\,s$^{-1}$~\cite{Wongwathanaratetal2010,Wongwathanaratetal2012}.

Gravitational forces of anisotropically ejected gas can thus 
mediate an efficient, long-lasting acceleration of the NS, 
transferring momentum from the anisotropically ejected matter
to the compact remnant.
Since the NS is gravitationally pulled by the slower, usually 
denser ejecta associated with a weaker explosion shock,
Wongwathanarat et al.~\cite{Wongwathanaratetal2012}
expect the bulk of the iron-group nuclei and of other elements
heavier than $^{28}$Si, which are explosively produced in the
shock-heated ejecta, to be expulsed preferentially in the 
direction opposite to the NS motion. They predict a very strong 
asymmetry of the nickel ejection in the case of large NS kicks
(see Fig.~\ref{jankafig9}),
which could be an observationally accessible, characteristic 
feature of the hydrodynamical-gravitational kick mechanism.

Asymmetrical convective downdrafts and rising bubbles as well as
violent, low-multipole SASI sloshing modes, which have spiral 
components in 3D, can establish angular momentum separation between 
PNS and ejecta and thus may cause considerable PNS rotation 
even if the stellar core did not rotate before
collapse (\cite{BlondinMezzacappa2007,Fernandez2010};~because of 
the use of an inner boundary condition, however, these results were
questioned by~\cite{Rantsiouetal2011}). Naturally, any anisotropic 
mass infall that hits the accretor not exactly head-on can exert a 
torque and spin up the PNS. A mass $\Delta m = 10^{-3}\,M_\odot$ 
that has an impact velocity $v_\mathrm{imp} \sim
\sqrt{2GM_\mathrm{ns}/R_\mathrm{ns}}\sim 10^{10}\,$cm/s
and an impact parameter $d \equiv \zeta R_\mathrm{ns}\sim 30$\,km 
when colliding with the NS
transfers an angular momentum of $\Delta J_\mathrm{ns} =
\Delta m\,v_\mathrm{imp} d\sim 6\times 10^{46}$\,g\,cm$^2$/s,
corresponding to a NS spin period of 
$T_\mathrm{ns} = 2\pi I_\mathrm{ns}/\Delta J_\mathrm{ns}
\sim 0.2$\,s for a typical value of the NS moment of inertia 
of $I_\mathrm{ns} \sim 2\times 10^{45}$\,g\,cm$^2$.
Indeed, 3D explosion simulations yield $T_\mathrm{ns}$ in
the range of hundreds of milliseconds to 
seconds~\cite{Wongwathanaratetal2010,Wongwathanaratetal2012}.
Nevertheless, angular momentum transferred to the PNS by 
hydrodynamical flows during the development of the explosion 
and in the post-explosion accretion phase is unlikely to be
sufficient to account for the estimated NS birth spin periods
of order $\sim$10\,ms, which seems to require rotation of the 
collapsing stellar core~\cite{HegerWoosleySpruit2005,Ottetal2006B}.
Explaining a possible spin-kick correlation of observed
NSs remains a challenge for any discussed kick mechanism
connected to explosion asymmetries of progenitor stars with or 
without rotation.

\subsection{Supernova Asymmetries}
\label{sec:snasymmetries}

The large asymmetries imprinted on the ejecta by the violent,
nonradial mass motions in the SN core, which precede and 
accompany the neutrino-driven revival of the blast wave, 
seed the growth of secondary Rayleigh-Taylor instability
in the shock-accelerated outer shells of the 
exploding star~\cite{Kifonidisetal2003}.
Since the developing Rayleigh-Taylor mushrooms are denser
than the surrounding gas, they are less decelerated than
their environment and can penetrate the composition
interfaces of the progenitor, retaining high velocities
as the SN ejecta expand. Thus they carry freshly synthesized
radioactive nickel and other heavy elements from the vicinity
of the nascent NS into the outer stellar layers. Significant
amounts of the initially innermost ejecta can be mixed deep 
into the helium shell and even the hydrogen layer of the
disrupted star~\cite{Kifonidisetal2006}, destroying the
well-stratified onion-shell structure of the progenitor.

In 3D simulations of a SN~1987A progenitor model, large 
nickel-dominated clumps (containing up to several 
$10^{-3}\,M_\odot$ of $^{56}$Ni) were found to speed through
the stellar hydrogen envelope with up to 4500\,km/s
(Fig.~\ref{jankafig10};~\cite{HammerJankaMueller2010}).
This can explain mixing phenomena and asymmetries observed in
SN~1987A, e.g.\ the detection of X-rays and $\gamma$-rays
from the radioactive nickel decay much earlier than predicted
by 1D explosion models~\cite{Arnettetal1989}. The outward
mixing of radioactive nickel and inward displacement of
hydrogen can well account for the shape and width of the
lightcurve maximum of SN~1987A 
(V.~Utrobin, private communication).

However, it is still unclear whether explosion asymmetries 
associated with the development of hydrodynamic instabilities
in the SN core and the subsequent growth of mixing 
instabilities in the stellar envelope are able to explain the
prolate shape of the SN~1987A ejecta cloud. It will
also have to be seen whether they can account for the 
extremely fast ``jet'' structures observed ahead of the 
explosion shock in the Cassiopeia~A SN remnant. Moreover,
the global asphericity of most SNe~Ib/c might require larger
nonradial deformation than the asymmetric structures that can 
stochastically grow from initially small random perturbations.

\subsection{Neutron Stars and Black Holes} 
\label{sec:NSBH}

Remnant masses and explosion properties (energy, ejected
$^{56}$Ni mass) and their systematics with the progenitor
mass also carry information on the explosion mechanism.
The observational basis of determined or constrained NS
and BH masses \cite{LattimerPrakash2010,Casares2007,Ziolkowski2010},
SN-progenitor connections \cite{Smartt2009,Smarttetal2009},
and estimated explosion parameters 
(e.g., \cite{UtrobinChugai2011,Tanakaetal2009}; Fig.~\ref{jankafig3}) 
is rapidly growing.

From the observed mass distribution of compact remnants and its
possible gap between $\sim$2 and 5\,$M_\odot$ at the boundary
between NSs and BHs, it was inferred that the
SN engine must launch the (neutrino-powered) explosion within 
100--200\,ms after bounce in order not to overproduce remnants
in the gap \cite{Belczynskietal2011,Fryeretal2011}. This was
considered as argument that the mechanism is supported by 
Rayleigh-Taylor (convective) rather than SASI instability. 
However, in the SN core both of these nonradial instabilities
occur simultaneously~\cite{Schecketal2008,MuellerJankaHeger2012} 
and cannot be separated just on the basis of a timescale argument.
Moreover, the population evolution models
of \cite{Belczynskietal2011,Fryeretal2011} used very
simple theoretical considerations to determine the 
explosion energy for early and late explosions and to
estimate the fallback mass of matter that initially moves
outward but ultimately fails to escape because of 
insufficient blast-wave energy.
The analytic theory ignores, for example,
dynamical effects and the nonnegligible additional 
power carried by the early neutrino-driven wind 
(cf.~\cite{Schecketal2006}).
Other approaches to predict mass distributions of NSs and/or
BHs were based either on piston-driven explosions with 
predefined mass cut and explosion energy 
(e.g.,~\cite{Zhangetal2008}), or on a single-parameter 
criterion to distinguish progenitors that are likely to
explode or not~\cite{OConnorOtt2011}.

In~\cite{Uglianoetal2012} an alternative approach was
adopted. Hydrodynamical simulations in 1D were performed
for a large set (roughly 100) of solar-metallicity progenitors 
of~\cite{Woosleyetal2002} using an analytic, time-dependent
two-zone model of the cooling, contracting PNS, whose free 
parameters were calibrated such that the explosion energy
and $^{56}$Ni mass of SN~1987A were reproduced for stars
with ZAMS mass around 20\,$M_\odot$. The 
effects of accretion luminosity were taken into account 
by simplified neutrino transport~\cite{Schecketal2006}.
With this prescription all stellar collapses and possible
explosions were simulated for at least 15\,s beyond core
bounce and were followed after the PNS cooling for hours
to days later until the fallback mass was determined.

Results of the calculations are shown in Fig.~\ref{jankafig11}
and reveal a number of interesting insights, which, of course, 
depend on the considered progenitor set:
\begin{itemize}
\item
Because the stellar structure varies nonmonotonically,
the SN properties depend on the progenitor mass
in a complex way. Large differences of the explosion
characteristics are possible for small mass differences.
\item
Failed explosions with BH formation seem possible for 
progenitors below 20\,$M_\odot$, and successful SNe
with NS formation are found also between 20 and
40\,$M_\odot$.
\item
Neutrino-driven explosions with energies in excess of
$2\times 10^{51}$\,erg and $^{56}$Ni production of 
significantly more than $\sim$0.1\,$M_\odot$ seem 
unlikely.
\item
The time of the onset of the SN blast (measured by
the moment the shock passes 500\,km) varies
between $\sim$0.1\,s and 1.1\,s, so it includes ``early''
and ``late'' cases. Later explosions tend to be
less energetic because less mass is available
for being heated by neutrinos.
\item
The NS baryonic masses are in the range of 
$\sim$1.2--2\,$M_\odot$. The smallest BH, formed by
fallback, contains 6.5\,$M_\odot$, all other BHs 
originate from failed explosions and contain all the
mass of the progenitor at collapse ($>$8.5\,$M_\odot$).
The possible gap of the observed remnant distribution
is clearly reproduced.
\item
Fallback is larger for the lower-mass progenitors 
where an extended hydrogen envelope
leads to a stronger reverse shock. The result of little
fallback in solar-metallicity progenitors is compatible
with conclusions drawn from an analysis of observed
double NS systems~\cite{PejchaThompsonKochanek2012}.
\item
Although the remnant mass is an almost monotonic function 
of the enclosed mass at the base of the oxygen-burning
shell, the latter is no reliable indicator
for the fate of the star because some models with relatively
small Si-cores do not explode.
\item
Neutrino-driven explosions are fostered by
big ``jumps'' in the stellar density and entropy 
profiles (cf.\ Fig.~\ref{jankafig2}), 
reducing the mass-infall rate (and ram pressure) and
allowing the shock to expand (cf.\ Eq.~\ref{eq:shockradius}).
\end{itemize}
Certainly these results are based on 1D simulations and
many approximations were made. Therefore they can be only
a very first step, but nevertheless are enlightening
concerning the implications of neutrino-powered
explosions. They challenge a number of paradigms
for the progenitor-explosion and progenitor-remnant 
connections. In particular the limited blast-wave energy
and nickel production
support arguments in favor of another explosion mechanism
for HNe. These events are likely to be triggered by
magnetorotational processes. More research, observationally
and theoretically, will have to clarify whether there
is a continuous transition between both, associated with
a varied degree of progenitor rotation and leading to
a continuous spectrum of explosion energies that reach
from the neutrino-powered regime of 
$E_\mathrm{exp}\lesssim 2\times 10^{51}$\,erg
to the hyperenergetic regime of 
$E_\mathrm{exp} > 10^{52}$\,erg as suggested by
some phemomenological studies (see Fig.~\ref{jankafig3}).

\section{SUMMARY, CONCLUSIONS, OUTLOOK}

Supernova theory has made remarkable progress over the 
past decade, promoted by common interests of the astro-,
particle (neutrino), nuclear, and gravitational physics 
communities and by an increasing number of active (young)
researchers in the field. A deeper understanding of the
physical mechanisms that initiate and 
fuel SN and HN explosions of massive 
stars is of crucial importance not only for establishing
the progenitor-remnant connection but also for predicting 
the properties of stellar explosions, their nucleosynthetic 
output, and the characteristics of their gravitational-wave
and neutrino signals. 

The most sophisticated present simulations demonstate
that neutrino-energy deposition can power ECSNe
(even in spherical models) of $\sim$9\,$M_\odot$ stars 
with ONeMg-cores near the lower mass limit for
SN progenitors (Fig.~\ref{jankafig5}).
Overall, the features of such explosions, e.g., low energy
and little nickel production, seem to be compatible with 
observational candidates like the Crab SN and some faint 
transients. Multi-dimensional simulations suggest these
explosions to be potential sources of light r-process nuclei
up to silver and palladium (Sect.~\ref{sec:heavyelements}).
Several groups have also reported successful neutrino-driven
explosions (with multi-group neutrino transport) for Fe-core
progenitors above 10\,$M_\odot$ (Sect.~\ref{sec:hydromodels};
Figs.~\ref{jankafig4}, \ref{jankafig5}). Ultimate confirmation
of the viability of this mechanism for a wider range of
progenitor masses therefore seems to be in reach. 

The onset of the explosion can be understood as a global
runaway instability of the accretion layer, whose initiation
depends on the power of neutrino-energy deposition.
While the exact mode of the runaway is still a matter 
of exploration and debate (e.g., low-multipole SASI or 
higher-multipole convective, oscillatory or nonoscillatory?),
its threshold in terms of the driving 
neutrino luminosity is lowered by nonradial fluid motions in
the neutrino-heating layer. Such flows play a supportive role
because they stretch the residence time of matter in the gain
region and thus decrease the heating timescale and increase 
the efficiency of neutrino-energy deposition, leading to
successful explosions even when sophisticated spherical models
fail (Sect.~\ref{sec:nonsphericaleffects}). The efficiency
of neutrino-energy transfer, the growth conditions
and growth rates of different hydrodynamic
instabilities, and the critical luminosity threshold
for an explosion may not only depend on the dimension and
thus will ultimately require simulations in 3D, but 
have been shown to depend also on putative ``details'' of
the physics ingredients like approximations for
the energy and velocity dependence of the neutrino transport,
the neutrino-interaction rates, general relativity, and the
contraction of the nascent NS in response to the nuclear EoS
(Sect.~\ref{sec:numericalmodeling}; Fig.~\ref{jankafig4}).
Moreover, the outcome of the complex neutrino-hydrodynamical
simulations can be sensitive to the numerical resolution,
which naturally is subject to limitations in full-scale,
multi-dimensional SN-core models.

While detailed modeling of the processes in collapsing 
stars now pushes forward from the second to the third dimension,
facing considerable computational challenges and demands
mainly for the neutrino transport,
a growing host of studies begins to explore the 
observational consequences of neutrino-driven explosions. 
In view of existing and upcoming 
big detection facilities, in particular neutrino and
gravitational-wave signals (Figs.~\ref{jankafig5},
\ref{jankafig7}, \ref{jankafig8}) are of relevance for 
SN-core diagnostics targeting a future Galactic SN. The
former even have the potential to yield valuable information
on particle properties of the neutrinos provided the
characteristics of the SN emission are sufficiently
well understood (e.g., 
\cite{Kachelriessetal2005,Chakrabortyetal2011C,Ellisetal2011,Ellisetal2012}). 
Sophisticated neutrino transport and interaction
treatments have revealed interesting signal features like
an amazing robustness of the neutronization $\nu_e$
burst~\cite{Kachelriessetal2005}, characteristic differences
of the rise time of the $\bar\nu_e$ and $\nu_x$ emission
after bounce~\cite{Chakrabortyetal2011C}, luminosity 
variations associated with nonsteady flows in the accretion
layer~\cite{MarekJankaMueller2009,Brandtetal2011}, 
and a close
similarity of the luminosities and spectra of neutrinos and
antineutrinos of all flavors during the PNS cooling 
phase (Sect.~\ref{sec:neutrinos}, Fig.~\ref{jankafig5};
\cite{Huedepohletal2010,Fischeretal2011C}) with important
consequences for SN nucleosynthesis 
(Sect.~\ref{sec:heavyelements}).

While a Galactic SN in the near future is a realistic
possibility, it will be a unique event and might not provide
evidence of wider validity. Photometric and spectroscopic
diagnostics of extragalactic SNe and of gaseous, young
SN remnants, which reveal information on explosion energies,
$^{56}$Ni production, ejecta masses, asymmetries, and
composition, as well as progenitor constraints
(cf.\ Fig.~\ref{jankafig3}) are therefore extremely valuable, 
and more is desirable. First-principle explosion models 
begin to become mature enough to be linked to such 
observations, a possibility that defines a fruitful 
territory for future reseach. Neutrino-driven explosion
models also begin to allow for predictions of compact
remnant (NSs and BHs) masses, kicks, and spins. 

Nonradial hydrodynamic instabilities in the 
collapsing stellar core, which can grow from small,
random initial perturbations before neutrino
heating revives the stalled shock, lead to low-multipole
asymmetries that trigger anisotropic and inhomogeneous 
expulsion of matter. Hydrodynamic instabilities in the SN
core therefore do not only yield a natural
explanation of the origin of pulsar kicks up to more
than 1000\,km/s (Sect.~\ref{sec:pulsarkicks}, 
Fig.~\ref{jankafig9}); they also seed large-scale mixing
processes in the exploding star, accounting for the
penetration of high-velocity clumps of inner-core material 
into the hydrogen and helium ejecta of
well observed SN explosions (Sect.~\ref{sec:snasymmetries},
Fig.~\ref{jankafig10}).

First results of a systematic exploration of the
progenitor-supernova connection based on the neutrino-heating 
mechanism show strong sensitivity of the explosion properties 
on the stellar structure and, for the employed set of 
stellar models~\cite{Woosleyetal2002}, 
large variations even within narrow 
progenitor-mass intervals (Sect.~\ref{sec:NSBH}; 
Fig.~\ref{jankafig11}). The explosion models can reproduce
fundamental properties of the empirical remnant-mass
distribution but reveal that neutrino-driven explosions
are unlikely to explain SN energies above
$\sim 2\times 10^{51}$\,erg and nickel masses 
significantly higher than 0.1\,$M_\odot$.
This underlines the need for an alternative engine that
powers stellar blast waves with energies from several
$10^{51}$\,erg up to more than $10^{52}$\,erg.
Such hyperenergetic events, which typically also exhibit  
unusually large nickel ejection (Fig.~\ref{jankafig3})
and deformation, are most probably energized 
by magnetorotational effects.

Many questions remain to be answered in this context 
and require more observations and theoretical work.
What discriminates progenitors of ``normal'' SNe from those
of HNe? Is rapid rotation of the progenitors the crucial
parameter? Is it connected to binary evolution?
Is there a continuous spectrum of stellar explosions
connecting the SN and HN regimes? Is a mixed mechanism,
neutrino-heating in combination with magnetorotational
energy transfer, at work in such events?

On the theory side the mission of clarifying the 
SN engines is severely handicapped by the unavailability 
of multi-dimensional stellar evolution models with the
quality to reduce the major uncertainties of the stellar
structure, rotation, and magnetic fields at the onset of
core collapse. It is clear that reliable theoretical
predictions of the progenitor-remnant connection and
of explosion properties ---energies, nucleosynthetic
yields, asymmetries, remnant masses, and neutrino and 
GW signals--- heavily depend on a firm knowledge of the 
stellar conditions at the time the gravitational instability
is reached.

\begin{acknowledgments}
Helpful discussions with P.~Mazzali and 
A.~Wei\ss\ are acknowledged. The author is very grateful to John Eldridge
and Stephen Smartt for providing panel~d of Fig.~\ref{jankafig3},
to Mrs.\ Rosmarie Mayr-Ihbe for preparing Fig.~\ref{jankafig1},
and to A.~Marek, B.~M\"uller, M.~Ugliano, and A.~Wongwathanarat for
providing figures of their results from publications in preparation.
Data from simulations by the Garching group are accessible either
openly or upon request at {\tt http://www.mpa-garching.mpg.de/\-ccsnarchive/}.
This work was supported by the Deutsche Forschungsgemeinschaft through
Son\-der\-for\-schungs\-be\-reich/\-Trans\-re\-gio~27 
``Neutrinos and Beyond'',
Son\-der\-for\-schungs\-be\-reich/\-Trans\-re\-gio~7
``Grav\-i\-tational-\-Wave Astronomy'',
and the Cluster of Excellence EXC~153 ``Origin and Structure of the
Universe''. Computing time at the John von Neumann Institute for
Computing (NIC) in J\"ulich, the H\"ochstleistungsrechenzentrum 
(HLRS) of the University Stuttgart, the Rechenzentrum (RZG) Garching,
and through DECI-5 and DECI-6 grants of the DEISA initiative of the EU
FP7 are acknowledged.\\
\\
Posted with permission from the Annual Review of Nuclear and 
Particle Science, Volume 62 \copyright\ 2012 by Annual Reviews,
{\tt http://www.annualreviews.org}. 
\end{acknowledgments}

\bibliography{JankaReferences-file}

\end{document}